\documentclass[final,3p,times,authoryear]{elsarticle}
\usepackage{numcompress}
\usepackage{natbib}
\usepackage{amssymb}
\usepackage{color}
\newcommand\apjl{{\@eapj@ApJLetters}}%     % Astrophysical Journal, Letters 

\begin{document}

\journal{Physics Reports}

\begin{frontmatter}

\title{Gravitational Lenses as High-Resolution Telescopes}

\author[ad1,ad2]{Anna Barnacka}
%\thanksref{corr}}
%\author[ad2]{Author Two}
\address[ad1]{Harvard-Smithsonian Center for Astrophysics, 60 Garden St, MS-20, Cambridge, MA 02138, USA}
\address[ad2]{Astronomical Observatory, Jagiellonian University, Cracow, Poland}
%\author{}
\address{ E-mail: {\tt abarnacka@cfa.harvard.edu}.}

\begin{abstract}
The inner regions of active galaxies host the most extreme and energetic phenomena in the universe 
including, relativistic jets, supermassive black hole binaries, and recoiling supermassive black holes.
However, many of these sources can not be resolved with direct observations.
I review how strong gravitational lensing can be used to elucidate the structures of these sources from radio frequencies up to very high energy gamma rays.  
The deep gravitational potentials surrounding galaxies act as natural gravitational lenses.
These gravitational lenses split background sources into multiple images,
each with a gravitationally-induced time delay.
These time delays and positions of lensed images depend on the source location, and thus, 
can be used to infer the spatial origins of the emission. 
%
%I review application of  these methods to bright gravitationally lensed quasars with relativistic jets.
For example, using gravitationally-induced time delays improves angular resolution of modern gamma-ray instruments
by six orders of magnitude ($\times 10^6$), 
and provides evidence that gamma-ray outbursts can be produced at even thousands of light years from a supermassive black hole, 
and that the compact radio emission does not always trace the position of the supermassive black hole. 
These findings provide unique physical information about the central structure of active galaxies, 
force us to revise our models of operating particle acceleration mechanisms,
and challenge our assumptions about the origin of compact radio emission. 
Future surveys, including LSST,  SKA, and Euclid, will provide observations for hundreds of thousands of gravitationally lensed sources, 
which will allow us to apply strong gravitational lensing to study the multi-wavelength structure for large ensembles of sources.
This large ensemble of gravitationally lensed active galaxies will allow us to elucidate the physical origins of multi-wavelength emissions, 
their connections to supermassive black holes, and their cosmic evolution.

\end{abstract}

\begin{keyword}
Strong Gravitational Lensing \sep Active Galaxies \sep Radio Loud Quasars \sep Supermassive Black Holes 

\end{keyword}

\end{frontmatter}

%\linenumbers

%\newpage
\tableofcontents

\newpage
%%%%%%%%%%%%%%%%%%%%%%%%%%%%%%%%%%%%%%%%%%%%%%%%%%%%%%
\section{Introduction}
%%%%%%%%%%%%%%%%%%%%%%%%%%%%%%%%%%%%%%%%%%%%%%%%%%%%%%

The historical label 'quasi-stellar objects' (QSOs) branded belief that quasars are point sources. 
Recent decades have provided an enormous advancement in technology allowing us to resolve emission from radio to X-ray frequencies. 
These multi-wavelength observations reveal complexities at all scales, from regions as small as the event horizon of supermassive black holes to hundreds of kpcs. 

Future surveys will be able to detect fainter sources and will cover a larger fraction of the sky. 
However, future facilities will not provide significant improvement in angular resolution -
leaving numerous assumptions concerning the spatial structures of sources untested. 

Here, I review the use of strong gravitational lensing to overcome technological shortcomings of current and future facilities 
to uncover the complex structure of active galaxies. 
Gravitational lensing is a powerful tool to examine the geometry, content, and forces at work in the universe. 
Spacetime curved by baryonic and dark matter act as a lens 
that magnifies and distorts emission from distant astrophysical sources. 
In this review, I will focus on gravitationally-induced time delays 
and positions of lensed images as a proxy to reveal the spatial structures of sources. 

Currently, we have hundreds of lensed quasars discovered. 
The next decade will provide discoveries of hundreds of thousands lensed quasars more,
opening an era of strong gravitational lensing, which will lead to discoveries of the physics of sources. 

The methods presented in this review have been applied to a handful of available objects to date.
However, the results have already provided profound insights into the nature of relativistic jets and their connections to supermassive black holes (SMBHs). 
This review acts as a complementary guide to useing strong gravitational lensing to resolve the origins of emission for large ensembles of sources from future surveys  
and points to new approaches to answer the most challenging problems in astrophysics, including
\begin{itemize}
\item the spatial origins of variable emission, including the sites of gamma-ray flares, which is linked directly to our understanding of particle acceleration mechanisms,
\item discovering and resolving a population of SMBH binaries, linked to galaxy evolution and the gravitational wave background from their coalescence, 
\item discovering and resolving the most distant quasars, which will provide further insights into their formation and seeds of supermassive black holes.
\end{itemize}

{\bf Scope of the review}. 
Sections~\ref{sec:Theory} and~\ref{sec:Sources} present a brief overview of theory of gravitational lensing and the physical nature of the lensed sources.  
Section~\ref{sec:Resolving} presents an introduction to resolving sources using strong gravitational lensing. 
Section~\ref{sec:TDA} reviews use of gravitationally-induced time delays and their application to test the origin of gamma-ray emission.  
Section~\ref{sec:HPT} reviews the Hubble parameter tuning approach and demonstrates its application to reveal origin of gamma-ray flares in respect to compact radio emission. 
Section~\ref{sec:Caustics} reviews use of caustics of lensing galaxies as non-linear amplifiers to illuminate offsets of emitting regions at different energies 
and presents an overview of applications. 
Section~\ref{sec:Esemble} discusses the application of gravitational telescopes  as a tool to investigate the universe in the coming era of large surveys.

%%%%%%%%%%%%%%%%%%%%%%%%%%%%%%%%%%%%%%%%%%%%%%%%%%%%%%
\begin{figure}[ht!]
 % \vspace{5mm}
  \centering
  %,bb=0 0 651 365
  \includegraphics[width=16cm,angle=0]{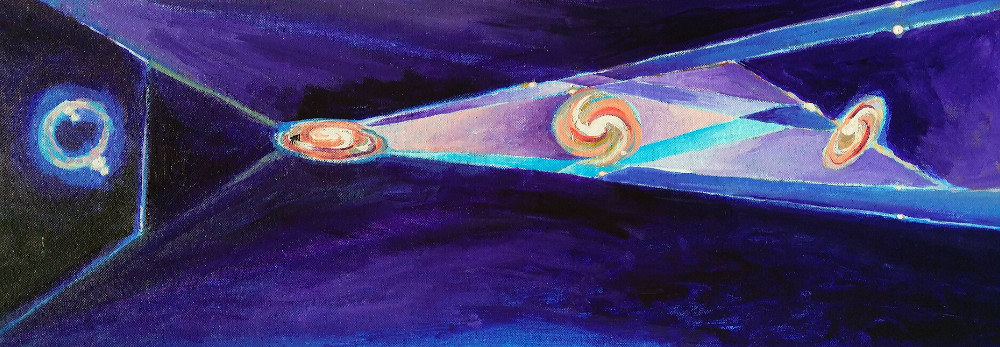}
  \caption{
  		{\it Gravitational Lenses as High-Resolution Telescopes} (Marlena Bocian Hewitt, 2018).
		Strong gravitational lensing phenomena depicted using cubism and futurism.
  		The painting captures the complexity of a source and propagation of its emission in spacetime as curved by a lensing galaxy.
		  }
  \label{fig:GL}
 \end{figure}
%%%%%%%%%%%%%%%%%%%%%%%%%%%%%%%%%%%%%%%%%%%%%%%%%%%%%%
\section{Strong Gravitational Lensing Theory \label{sec:Theory}}
%%%%%%%%%%%%%%%%%%%%%%%%%%%%%%%%%%%%%%%%%%%%%%%%%%%%%%

The mathematician Henri Poincare's work on geometry had a profound influence on science as well as the arts. 
In science, it led to the geometrization of space and time in Einstein's general relativity theory of 1915, 
with predictions on light bending by curved spacetime.
This is known today as gravitational lensing.  
In the arts, it inspired the early offshoots into  cubism, which abandoned a single viewpoint 
and used geometric shapes and interlocking planes to represent reality.  

{\it Gravitational Lenses as High-Resolution Telescopes}, by contemporary artist  Marlena Bocian Hewitt\footnote{http://www.marlenabocian.com},
captures both cubism and futurism elements in its depiction of gravitational lensing. 
%Painting~\ref{fig:GL} shows gravitational lensing phenomena depicted using cubism and futurism by 
%
Gravitational lensing phenomena occur when a massive object, such as a galaxy (lens), 
is located close to the line-of-sight between an observer and a source. 
Figure~\ref{fig:GL} shows multiple planes in gravitational lensing. 
Starting on the right,  the source plane shows an emission originating from an active galaxy. 
Next, the lens plane illustrates a lensing galaxy and its effects of bending the light of the source. 
Toward the left, the observer plane represents the observer's galaxy 
and converging emission of the source focused by the lensing galaxy.  
On the left, the image plane shows the observed image of the lensed source.  

One of the first predictions of general relativity proposed by Einstein was the deflection of light by the Sun. 
Einstein calculated the deflection angle $\alpha$  of light passing at the distance $R_\odot$ from the Sun to be
\begin{equation} 
\alpha = \frac{4GM_{\odot}}{c^2} \frac{1}{R_\odot}=1.75 \, \mbox{arcsec}\,,
\end{equation}
where $M_\odot$  and $R_\odot$  are the Sun's mass and radius, respectively. 
 Einstein's solution predicted the deflection by the Sun twice larger than the Newtonian estimation for a slow particle.
The deflection of light by the Sun can be tested by observing the change in the positions of stars as they pass near the Sun during an eclipse. 
Such observations were performed by Sir Arthur Eddington and his collaborators during the total solar eclipse of May 29, 1919. 
%The first measurement of the deflection angle was performed by Eddington during a total solar eclipse. 
The Eddington expedition provided the first experimental confirmation of Einstein's general relativity, 
demonstrating that  light bending is a result of curved spacetime, as opposed to gravitational force acting on a particle. 

Observations of the gravitational lensing effect in Eddington's experiment were limited to measuring small changes in positions of the background stars. 
The gravitational lensing phenomena can result in producing multiple images of a source. 
The shape of lensed images depends on the geometry of curved spacetime and the alignment of the source, lens, and observer. 
When the source is aligned with the lens center,
deflection is defined as the Einstein angle 
 \begin{equation} 
 \label{eq:EA}
\theta_E = \sqrt{\frac{4GM}{c^2} \frac{1}{D}}\,,
\end{equation}
where $M$ is the mass of the lens and $D$ is a ratio of angular distances
\begin{equation}
 D\equiv \frac{D_{OL}D_{LS} }{D_{OS}}\,,
  \label{eq:D}
 \end{equation}
with distances from the observer to the lens $D_{OL}$,  from the observer to the source $D_{OS}$, 
and from the lens to the source $D_{LS}$.
If the sources are aligned within the Einstein angle, then deflection enables multiple light paths to reach the observer
and  as a result multiple images of the source are observed. 

The Einstein angle of a star behind the Sun would be $\sim40\,$arcsecond, 
which is $\sim50$ times smaller than the Sun's angular size seen from the Earth.
The multiple images produced when the Sun acts as a lens will be occulted  by the Sun itself. 
Thus, multiple images produced when the Sun is acting as a lens cannot be observed  from the Earth.  

In principle, the emission of multiple images would not be occulted if another star acted as a lens. 
The angular size of a star observed from the Earth is significantly smaller than the Einstein angle for such  a configuration. 
However, observing the lensing of stars by stars was considered technologically impossible due to
the very small separation between lensed images on the order of milli-arcseconds for Galactic stellar lensing. 
In the case of extra-galactic stellar lensing in which the lens is located at a cosmological distance ($\sim 1\, Gpc$) from the Earth
the separation of the lensed images is of the order of micro-arcseconds, which dubbed  this phenomena ``microlensing".
  
As an alternative, \cite{1937ApJ....86..217Z,1937PhRv...51..290Z} was the first to point out that instead of lensing stars by stars, 
galaxies are likely to be gravitationally lensed and the image separation, 1~arcsecond, would be detectable.
 Zwicky also noticed that the lensing phenomena would provide a way to measure the mass of a galaxy by 
using the Einstein angle (Equation~\ref{eq:EA}).

The multiple paths that light can travel in curved spacetime result in multiple images of the source.
The difference in path lengths and traversed gravitational potentials result in a time delay between the lensed images. 
The time delay can be predicted assuming a model of the lens.
However, conversion of the predicted time delay into physical units of time 
requires knowing cosmological distances.
Thus, time delays can be used  to infer cosmological parameters.  

The cosmological application of lensing phenomena was pointed out in 1964 by \cite{1964MNRAS.128..307R},
who proposed the possibility of estimating the expansion rate of the universe (the Hubble constant) using  
time delay between images of a lensed supernova. 
 Refsdal's method and the possibility of observing lensed sources  with multiple images started gaining momentum thanks to the discovery that quasi-stellar radio stars (known today as quasars) are at cosmological distances  \citep{1968ApJ...151..393S}.  
Bright and distant quasars become the best candidates to exhibit detectable gravitational lensing effects. 
In 1979, \cite{1979Natur.279..381W} discovered the first gravitationally lensed quasar QSO 0957+561A,B
with two images separated by $6\,$arcseconds. 
The similarity in morphology and spectra of the two images and detection of the foreground galaxy
provided evidence that these twin images, A and B, were the lensed images of a background quasar.

When the source, lens, and observer are well aligned, light follows multiple paths forming a complete ring with the radius equal to the Einstein angle.  
A typical galaxy with a mass of  $1.25\times 10^9\,$ M$_{\odot}$ enclosed within the Einstein angular scale
acting as a lens produces an Einstein ring with an angular size of $\sim 0.5\,$arcseconds.
The first Einstein ring was discovered by \cite{1988Natur.333..537H} using radio observations and provided the determination of the mass of the lensing galaxy. 
The Einstein radius provides one of the most straightforward ways to estimate the mass of a galaxy including both baryonic and dark matter. 

The scale of the Einstein radius separates strong and weak lensing regimes. 
In the strong lensing regime, the source is aligned within one Einstein radius from the lens center, 
and multiple lensed images of the source are observed. 
Strong gravitational lensing is a powerful tool for exploring the universe \citep{1992grle.book.....S}. 
It magnifies distant objects and provides a way to observe their structure and detailed properties 
(e.g., \citep{2012ApJ...759...66Y,2012ApJS..199...25P,2012A&A...542L..31L}). 

Extended sources can be considered as multiple point sources deflected by a lens depending on their position. 
As a result, images of extended sources can form arcs. 
Arcs are commonly observed when a lens is a galaxy cluster, and the source is extended. 
The interpretation of cosmological arcs as background galaxies strongly distorted and elongated by a foreground cluster was proposed by  \cite{1987Natur.325..572P}.

It was also  \cite{1986ApJ...304....1P} who revised the idea of microlensing. 
The lensed images produced by the lensing of stars by stars cannot be separated, 
but an effect of changing magnification on timescales from hours to years caused by a background star being lensed by an object of a mass  between $10^{-6}$M$_\odot$ and $10^2$M$_\odot$ would be detectable.
Paczynski estimated that at any given time, one in a million stars in the Large Magellanic Cloud (LMC) might 
be measurably magnified by the gravitational lens effect of an intervening star in the halo of our Galaxy \citep{1996astro.ph..6001N}. 
The advance in charge coupled device (CCD) technology in the late 1980s allowed monitoring the light curves of millions of stars for the first time, 
making the microlensing effect a standard tool in astrophysics \citep{2006astro.ph..4278W,2012RAA....12..947M,2015IJMPD..2430020R}.

Weak lensing is a regime in which multiple images do not arise. 
It occurs usually when the projected position of the source is larger than one Einstein radius from the lens mass center. 
As a result, sources experience relatively weak magnification accompanied by an increase in size and stretch in tangential direction around the foreground mass.
The weak lensing effect is impossible to detect for individual sources due to the unknown intrinsic luminosity and shape of a lensed galaxy. 
Thus, it is difficult to distinguish the effects of lensing from the intrinsic morphology of a galaxy. 
However, effects of weak lensing can be detected statistically, by averaging over a large number of galaxies,
which can reveal the tangential stretch  of galaxies caused  by weak lensing effects. 
Weak lensing is a powerful tool to map large massive structures across the universe and provides insights into the structure formation and evolution 
\citep{2014MNRAS.442.1326K,2014MNRAS.441.2725F,2013MNRAS.432.2433H,2017arXiv171003235M}.

This review focuses on strong gravitational lensing. 
The strong gravitational lensing effect has been widely used to probe the mass distribution of lenses 
\citep[see][]{2013SSRv..177...75H,1993ApJ...404..441K,2012ApJS..199...25P,2016ApJ...819..114K,1998ApJ...498L.107T}. 
This review focuses on using strong gravitational lensing to study  sources.
Thus, here, I will continue the story of using strong gravitational lensing to resolve spatially complex structures of active galaxies.  

%%%%%%%%%%%%%%%%%%%%%%%%%%%%%%%%%%%%%%%%%%%%%%%%
\begin{figure}[ht!]
 % \vspace{5mm}
  \centering
  %,bb=0 0 651 365
  %https://docs.google.com/document/d/11ysqpCyaOyWrU2l-pCZXVS4uwztkPnDzbxO0BqkMjeI/edit
  \includegraphics[width=10cm,angle=0]{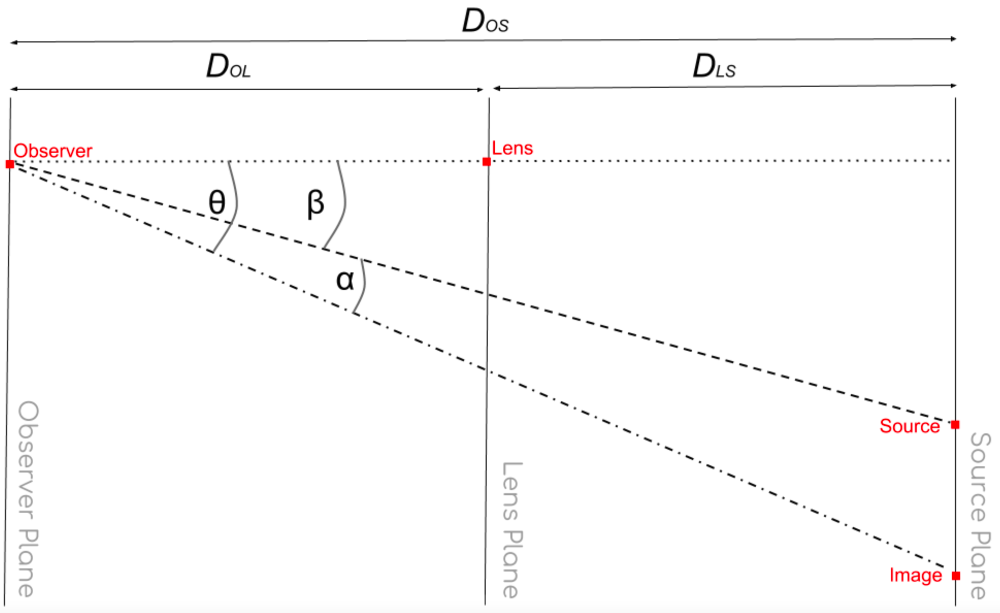}
  \caption{Sketch of a typical gravitational lensing system.  
  		Diagram shows positions of the observer, lens, source and one of the lensed images of the source.
		The distances observer-source, observer-lens, and lens-source are shown on the top. 
		The deflection angle $\alpha$ and the position angles $\beta$ of the source and $\theta$ of the lensed image are defined as well. 
		The lens equation can be derived from this geometrical arrangement if $\alpha,\beta,\theta \ll 1$.
		  }
  \label{fig:GLSketch}
 \end{figure}
%%%%%%%%%%%%%%%%%%%%%%%%%%%%%%%%%%%%%%%%%%%%%%%%
\subsection{Lens Equation\label{sec:LensEquation}}
%%%%%%%%%%%%%%%%%%%%%%%%%%%%%%%%%%%%%%%%%%%%%%%%

Figure~\ref{fig:GLSketch} illustrates angles and distances in the lens sytem related as $\theta D_{OS} = \beta D_{OS} - \hat{\alpha} D_{LS}$.
Using the reduced deflection angle
\begin{equation}
\vec{\alpha} = \frac{D_{LS}}{D_{OS}}\vec{\hat{\alpha}},
\end{equation}
the positions of the source, ${\beta}$,  and its lensed image, ${\theta}$ are related through the lens equation 
\begin{equation}
\vec{\beta} = \vec{\theta} -\vec{\alpha}(\vec{\theta} ) \,,
\end{equation}
%
%where $\alpha$ is the deflection angle introduced in Section~\ref{sec:DeflectionAngle}. 
where the deflection angle $\alpha$ is the gradient of the effective gravitational potential $\psi$
\begin{equation}
\label{eq:alpha}
\vec{\alpha} = \vec{\triangledown}_\theta\psi \,.
\end{equation}

The lens equation is nonlinear, resulting in the possibility of creating multiple images for a single source position. 
The formation of lensed images follows the extrema (maxima, minima, and saddle points) of the Fermat surface \citep{1986ApJ...310..568B}. 
The Fermat surface can be calculated using Eq.~(61) from \citep{1996astro.ph..6001N}
\begin{equation}
(\vec{\theta} - \vec{\beta}) - \vec{\nabla}_\theta \psi = 0 \,.
\end{equation}
Light deflection can be calculated by studying geodesic curves and  can equivalently be described by the FermatÕs principle, as in geometrical optics.
The astrophysical applications discussed in this review are all well described by geometrical lensing.

%%%%%%%%%%%%%%%%%%%%%%%%%%%%%%%%%%%%%%%%%%%%%%%%
\subsection{Time Delays \label{sec:TimeDelay}}
%%%%%%%%%%%%%%%%%%%%%%%%%%%%%%%%%%%%%%%%%%%%%%%%

The deflection of light by a lens results in a delay in the time between the emitting of radiation by the source and the reception by the observer.
Lensed images of a source have similar temporal evolution (light curves), except for a shift in  time due to this time delay.
This time delay has two components. 
The first component is the geometrical time delay $\Delta t_{geom}$ caused by the differences in the length of geometrical paths of the deflected light rays compared to the unperturbed ones.
The second component is the gravitational time delay $\Delta t_{grav}$, called the Shapiro delay \citep{1964PhRvL..13..789S}.
The Shapiro effect is due to ``clocks" slowing down in gravitational fields.
As a result, light rays are delayed relative to their travel time in unperturbed spacetime. 
Thus, the gravitational time delay comes from the slowing down of photons traveling through the gravitational field of the lens and is  induced by the gravitational potential of the lens.

The total time delay introduced by the gravitational lensing of a source at the position $\vec{\beta}$ and a lens at redshift $z_L$ is  \citep{1996astro.ph..6001N} 
\begin{equation}
\frac{c\, \Delta t(\vec{\theta})}{(z_L+1)} = \frac{c (\Delta t_{geom}+\Delta t_{grav})}{(z_L+1)} =
\frac{D_{OS}D_{OL}}{D_{LS}}\left [  \frac{1}{2}(\vec{\theta} -\vec{\beta})^2 -\psi(\vec{\theta})\right ]  \,.
 \label{dt3}
\end{equation}
%
%where $\theta=r/D_{OL}$,  $\beta=r_S/D_{OL}$  and $\psi(\theta)$ is the effective 
%gravitational potential of the lens. 
%
The total time delay $\Delta t$ depends on the source position $\beta$,  the gravitational potential of the lens $\psi$,
and the distance ratio $D$. 
The time delay is proportional to the square of the angular offset between $\theta$ and $\beta$,
resulting  from the difference in light travel time for two images.
As a consequence, the time delay increases with distance of the source from the lens center. 

The representation of time delay effects is illustrated in Figure~\ref{fig:GL},
where two outbursts of emission are split by a lensing galaxy into two paths. 
The first outburst originated from the central engine, 
and the second from the bright knot along the jet. 
The emission passing closer to the center of the lens experiences a stronger gravitational potential.
As a result, the image closer to the lens center arrives delayed despite a shorter path in respect to the brighter image. 
The artistic representation of gravitational lensing illustrated by Painting~\ref{fig:GL}
also preserves the magnification of the lensed images.

%%%%%%%%%%%%%%%%%%%%%%%%%%%%%%%%%%%%%%%%%%%%%%%%
\subsection{Magnification \label{sec:magnification}}
%%%%%%%%%%%%%%%%%%%%%%%%%%%%%%%%%%%%%%%%%%%%%%%%

The properties of the lens mapping from the source to the lens plane are described by the Jacobian matrix A
\begin{equation}
A \equiv \frac{\partial \vec{\beta}}{\partial \vec{\theta}} = \left( \delta_{ij} - \frac{\partial \alpha_i(\vec{\theta}) }{\partial \theta_j} \right) \,.
\end{equation} 

The Jacobian A is in general a function of position $\vec{\theta}$, 
and is used to calculate the magnification, $\mu$, as an inverse of the determinant of A
\begin{equation} 
\mu = \frac{1}{\mbox{det}A} \,. 
\end{equation}

In geometrical optics approximation, the determinant of A, $\mbox{det}A=0$, corresponds to infinite magnification. 
In the lens plane, points where the magnification goes to infinity are called critical curves. 
These critical curves define regions where lensed images merge or are created. 
The critical curves mapped to the source plane are called caustics. 

In optics, a caustic is the envelope of light rays reflected or refracted by a curved surface or object, 
or the projection of that envelope of rays on another surface\footnote{$https://en.wikipedia.org/wiki/Caustic\_(optics)\#cite_note-WEI69-2$}. 
In gravitational lensing, a caustic is the envelope of rays refracted by spacetime curved by a gravitational lens
and projected into the source plane.

The number of lensed images changes when a source crosses a caustic curve. 
The caustic degenerates into a point for a single-point lens. 
For a spherically symmetric mass distribution, the critical curves are circles. 
For elliptical lenses or spherically symmetric lenses plus external shear, the caustics can consist of cusps and folds 
\citep{1997MNRAS.292..863W,1986ApJ...310..568B,2006MNRAS.372.1692A}.
The shape of a caustic can be the distorted by substructure in the lensing galaxy \citep{2006glsw.conf.....M}.
Mathematically, the approximate of the caustic is best described by catastrophe theory and Morse theory \citep{1993A&A...268..453E}.

%%%%%%%%%%%%%%%%%%%%%%%%%%%%%%%%%%%%%%%%%%%%%%%%
\subsection{Lens Model: Singular Isothermal Sphere} 
%%%%%%%%%%%%%%%%%%%%%%%%%%%%%%%%%%%%%%%%%%%%%%%%
\label{sec:SIS}

The singular isothermal sphere (SIS) profile is the simplest parameterization of 
the spatial distribution of matter for the inner density profile of galaxies and clusters of galaxies.
In general, the mass distribution of the lens composed of stellar and dark matter 
is well represented by an isothermal model over many orders of magnitude in radius.
It can reproduce the flat rotation curves of spiral galaxies
and deviate significantly only far outside the Einstein radius. 

%Galaxies which are responsible for large time delays between the images are 
%extended lenses. 
%A simple model of extended lenses is a singular isothermal sphere model (SIS).
In the SIS model, the mass increases proportionally to the radius $r$ and the force is proportional to $1/r$. 
The SIS model is a first approximation  model for the gravitational field of galaxies and  clusters of galaxies \citep{1988ApJ...333..522R},
and is consistent with the results of the Sloan Lens ACS Survey \citep{2007ApJ...667..176G,2010ApJ...724..511A}.

The density profile of the SIS model can be derived assuming that 
the matter content of the lens behaves as an ideal gas confined by 
a spherically symmetric gravitational potential. 
This gas is taken to be in thermal and hydrostatic equilibrium.
The three-dimensional density distribution of SIS is given by
\begin{equation}
\rho(r) = \frac{\sigma_\nu^2}{2\pi G}\frac{1}{r^2} \, ,
\end{equation}
where $\sigma_\nu$ is the one-dimensional velocity dispersion of stars in the galaxy modeled as ``gas" particles 
and $r$ is the distance from the lens center.
The corresponding surface mass distribution is obtained by projecting the three-dimensional density along the line-of-sight as
 \begin{equation}
\Sigma (\xi) = \frac{\sigma^2_\nu}{2 G}\frac{1}{\xi} \, .
\end{equation}
The SIS density profile has a singularity at $\xi=0$, where theoretically the density goes to infinity.
The projected mass enclosed within a cylinder of radius $r$ is given by
\begin{equation}
M(r)=\int_{0}^{r}\Sigma (\xi)2 \pi \xi d\xi \, .
\label{eq:M}
\end{equation} 
Using Equations~(\ref{eq:alpha})  and (\ref{eq:M}) one obtains the deflection angle 
\begin{equation}
\hat{\alpha}(r)=4 \pi \frac{\sigma_\nu^2}{c^2} \, .
\label{eq:ar}
\end{equation} 
Equation (\ref{eq:ar}) shows that the deflection angle for SIS is independent of $r$. 
Thus, Equation (\ref{eq:ar}) can be simplified to
\begin{equation}
\alpha=1.15 \left(\frac{\sigma_{\nu}}{200\,km\,s^{-1}}\right)^2 \, {\rm arcsec}\, .
\end{equation} 
The Einstein angle derived using the SIS model is
\begin{equation}
\theta_E = 4 \pi \frac{\sigma_\nu^2}{c^2} \frac{D_{LS}}{D_{OS}} = \hat{\alpha}\frac{D_{LS}}{D_{OS}}=\alpha  \,.
\end{equation}
The lensing potential of the SIS can be reduced to
\begin{equation}
\label{eq:SISLP}
\psi(\theta)=\theta_E|\theta| \,.
\end{equation}
The SIS model is circularly symmetric, which reduces the lens equation into one-dimension. 
Multiple lensed images of the source are created only if the source is inside the Einstein ring.  
There is also a third image, but it is hidden by the central singularity of the $\ln(r)$ potential.  
When $\beta < \theta_E$, the positions of the lensed images $\theta_A$ and $\theta_B$ can be determined using the deflection defined by Equation~(\ref{eq:alpha}) 
and the lensing potential of the SIS (Equation~\ref{eq:SISLP}) 
\begin{equation}
\label{eq:theta}
\theta_{A,B} = \beta\pm\theta_E \,.
\end{equation}
For the sources located outside the Einstein ring, $\beta > \theta_E$, 
there is only one image $\theta = \theta_A = \beta+\theta_E$. 

The magnification factor $\mu$ for a circularly symmetric lens is 
\begin{equation}
\mu=\frac{\theta}{\beta} \frac{d\theta}{d\beta} \,.
\end{equation}
Thus, the magnification for two images, $A$ and $B$ using the SIS model are
\begin{equation}
\mu_{A,B} = \frac{\theta_{A,B} }{\beta} \,,
\end{equation}
and the time delay for $\beta>0$ is
\begin{equation}
\label{eq:tdSIS}
\frac{2c\Delta t}{(1+z_L)} =  D (\theta_B^2 - \theta_A^2) \,.
\end{equation}

The SIS model allows for a simple analytical solution of the lens equation. 
%The reviewed the time delay method and the Hubble parameter tuning approach was originally evaluated using the SIS model. 
In the SIS model, the position of the source can be reconstructed using the positions or magnifications of the lensed images. 
The SIS model in the time delay formula reduces to the positions of the lensed images, which trace the position of the source. 
Thus, the time delay can be used to find the source location. 

%%%%%%%%%%%%%%%%%%%%%%%%%%%%%%%%%%%%%%
\begin{figure}
\begin{center}
\includegraphics[width=5.5cm,angle=0]{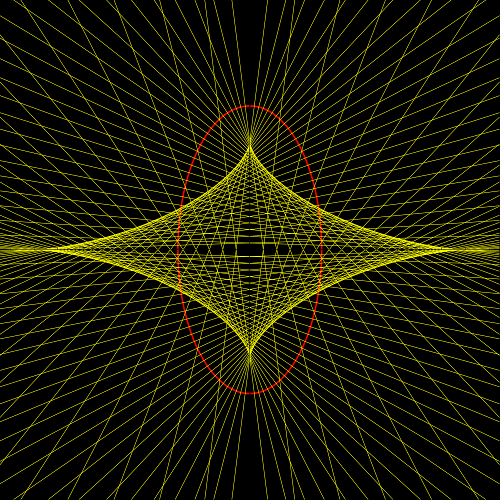}
\includegraphics[width=7.5cm,angle=0]{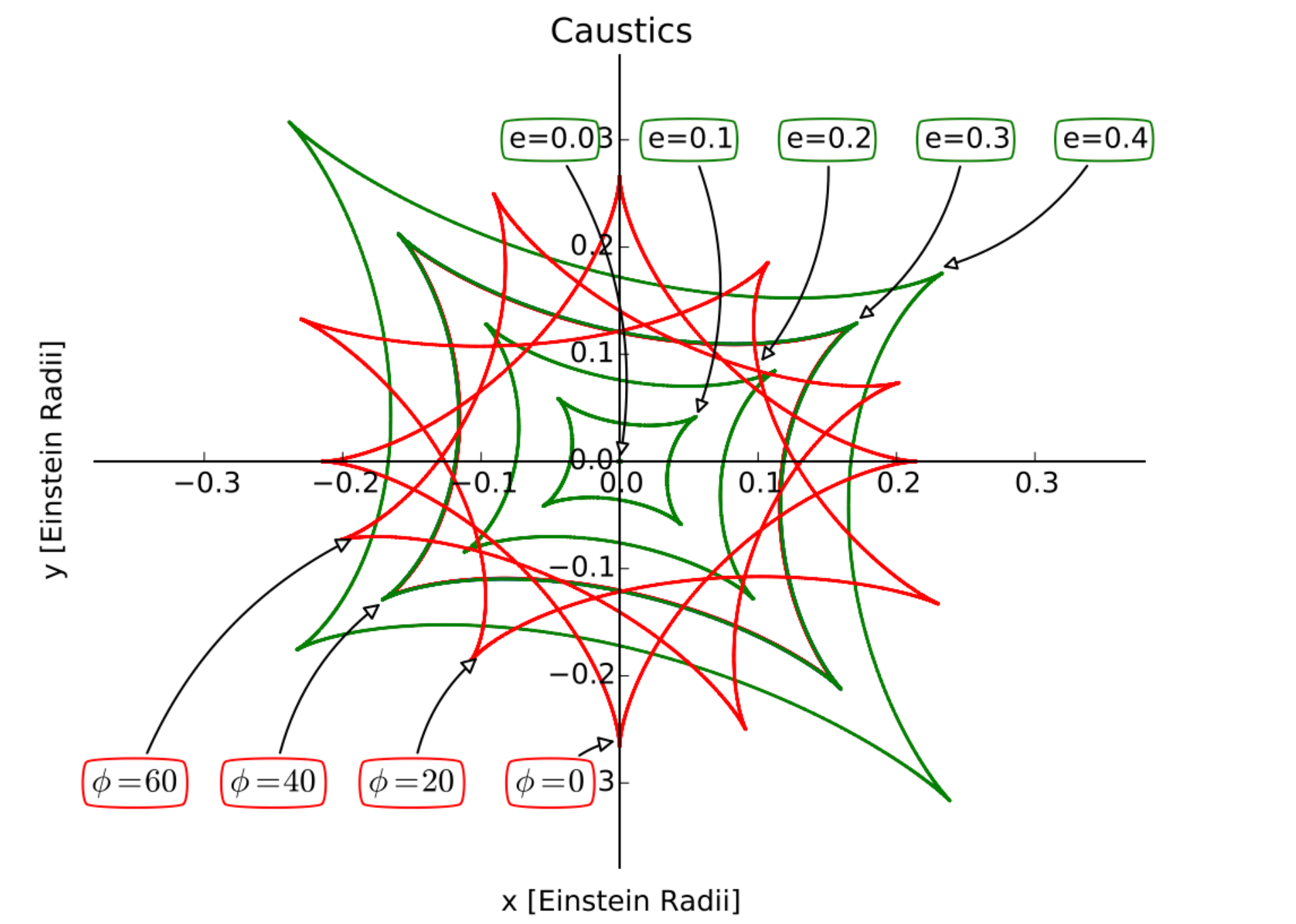}
\end{center}
\label{fig:Caustic}
\caption{ Caustics of an elliptical lens. 
                 {\bf Left:} Ray traced caustic.
			The caustic is a curve or surface to which each of the light rays is tangent, defining a boundary of an envelope of rays as a curve of concentrated light.
                         {\bf Right:} Caustics for a singular isothermal ellipsoid mass distribution normalized by the Einstein radius.  
                          Green lines represent caustics for a different ellipticities and a fixed angle $\phi=40^\circ$.
                          Red lines indicate caustics for a range of angles with a fixed ellipticity of e=0.3. Figure from (34). }
\end{figure}

%%%%%%%%%%%%%%%%%%%%%%%%%%%%%%%%%%%%%%
\subsection{Lens Model: Singular Isothermal Ellipsoid}
\label{sec:SIE}
%%%%%%%%%%%%%%%%%%%%%%%%%%%%%%%%%%%%%%

Another popular model used to describe a mass distribution of a lensing galaxy is the Singular Isothermal Ellipsoid (SIE).
The SIE model includes ellipticity in the lens potential. 
The lenses with an ellipticity in their mass distribution can form four lensed images,
as compared to the SIS model that produces only two observable lensed images.
The SIE has the three-dimensional radial profile of $\rho \propto r^{-2}$, and a convergence given by
\begin{equation}
\kappa = \frac{b_{SIE}(q)}{2\sqrt{\tilde{x}^2 +\tilde{y}^2/q^2}} \,,
\end{equation} 
where $e$ is an ellipticity, and the axis ratio is $q=1-e$, with $q=1$  for a spherical case. 
The normalization factor $b_{SIE}$ is related to the velocity dispersion, $\sigma$, as
\begin{equation}
b_{SIE}(1) = 4\pi \left( \frac{\sigma}{c} \right)^2 \frac{D_{LS}}{D_{OS}} \,,
\end{equation}
where the coordinates $\tilde{x}$ and $\tilde{y}$ are rotated by an angle $\phi$
 \begin{equation}
 \tilde{x}  = x \cos \phi + y \sin \phi \,, 
 \tilde{y} =  -x \sin \phi + y \cos \phi \,.
 \end{equation}

Different reflective or refractive surfaces can produce a variety of caustic curves.
Elliptical lenses produce a diamond-shaped inner caustic. 
Observationally, sources located inside the inner caustic are recognized by having four lensed images. 
Figure~\ref{fig:Caustic} shows caustics of elliptical lenses obtained using {\tt glafic} code \citep{2010PASJ...62.1017O}.
%resulting from deflection of rays in the spacetime curved by an elliptical galaxy located at redshift equals 1 and source at redshift 2. 
%The Caustics are obtained using {\tt glafic} code \citep{2010PASJ...62.1017O}.
The caustic size scales with the Einstein radius of the lens. 
The angle $\phi$ does change the orientation of the caustic, 
but has no influence on the caustic shape or size (see Figure~\ref{fig:Caustic}). 

For SIE with an ellipticity $e=0$, 
the caustic is represented by a point at the center of the lens. 
The lens ellipticity defines the caustic length. 
Lenses with greater ellipticity have larger caustics. 
Thus, the probability that a background source will be located close to the caustic increases with the lens ellipticity
and the Einstein ring radius.

If the source is located relatively close to the inner side of the caustic,
then even small differences  of the position of the source can produce large changes in magnifications and positions of the lensed images. 
This effect will be further used in this review to demonstrate the applications of caustics of elliptical galaxies as non-linear amplifiers.

%%%%%%%%%%%%%%%%%%%%%%%%%%%%%%%%%%%%%%%%%%%%%%%%
\subsection{Lensing Probability \label{sec:Probability}}
%%%%%%%%%%%%%%%%%%%%%%%%%%%%%%%%%%%%%%%%%%%%%%%%

Considering the emptiness of the universe and the vast separation between galaxies,
the probability that a source is gravitationally lensed can be expressed as optical depth. 
When the optical depth is smaller than 1, then it can be understood as a probability.
The optical depth for gravitational lensing was introduced by \cite{1983ApJ...267..488V}, 
and has become the standard way of determining the probability of lensing.

In the strong gravitational lensing regime, the source is located within one Einstein radius of the lens.
Thus, the optical depth, $\tau$, is an estimate of the number of lenses within the Einstein 
radius along the line-of-sight from the observer to a source \citep{1989ApJ...341..579N},
and is expressed as
\begin{equation}
\tau=\int{\rho_{lens}\pi r_E^2 dl} \, ,
\end{equation}
where $\rho_{lens}$ is the lens density along the line-of-sight.
 
The optical depth depends on the cosmological model  \citep{1964SvA.....8...13Z,1973ApJ...180L..31D,1992ApJ...393....3F}. 
Based on observations, the universe is homogeneous and isotropic on large scales and
is well described  by Friedmann-Lemaitre-Robertson-Walker (FLRW) geometry. 
The FLRW model is characterized by
the mean mass density $\Omega_M$ and the normalized cosmological constant $\Omega_{\Lambda}$.

The angular diameter distance for the FLRW model is 
\begin{equation}
d_S(z_1,z_2)=\frac{D_H}{1+z_2} \int_{z_1}^{z_2} \frac{dz}{\sqrt{\Omega_M (1+z)^3 + (1-\Omega_M)}}\, .
\label{eq:ds}
\end{equation}
where $D_H$ is the Hubble distance defined as $c/H_0$,  with H$_0=h\times$~100~km/s/Mpc, 
where $h$ is the reduced Hubble parameter.

The cross-section  for lensing by the SIS model described in Section~\ref{sec:SIS} is 
\begin{equation}
\Sigma=\pi r^2_{E}=16\pi^3\left(\frac{\sigma}{c}\right)^4\left(\frac{D_{OL}D_{LS}}{D_{OS}} \right)^2
\end{equation}
The differential probability of lensing is
\begin{equation}
\label{eq:tau}
d\tau =n_{0}(1+z_{L})^3\Sigma \frac{cdt}{dz_{L}}dz_{L} 
= F(1+z_{L})^3\left(\frac{D_{OL}D_{LS}}{D_{H}D_{OS}} \right)^2\frac{1}{D_H}\frac{cdt}{dz_L}dz_L \, ,
\end{equation}
where
\begin{equation}
\frac{cdt}{dz_{L}} = \frac{D_H}{1+z_L}\frac{1}{\sqrt{\Omega_M(1+z_L)^3+(1-\Omega_M-\Omega_\Lambda)(1+z_L)^2+\Omega_\Lambda)}} \, ,
\end{equation}
and $F$ is a quantity which measures the effectiveness of matter in producing double images \citep{1984ApJ...284....1T} and is expressed as
\begin{equation}
\label{eq:F}
F=16\pi^3 n_0 \left(\frac{\sigma}{c}\right)^4 D_H^3 \,.  
\end{equation}
The value of $F$ used by \citep{1992ApJ...393....3F} was 0.047. 
%The total probability of lensing is calculated as in equation (\ref{eq:TotalPro}).
%
Integrating Equation~\ref{eq:tau} from z=0 to z=2 results in the probability $ \tau\sim 0.001$.
Thus, one per thousand sources at redshift $\sim2$ is expected to be gravitationally lensed. 
Note, the third {\it Fermi} GBM gamma-ray burst catalog includes 1405 triggers identified as Gamma Ray Bursts (GRBs) \citep{2016ApJS..223...28N}. 
However, there is still a lack of convincing evidence of multiply-imaged GRB. 
Short nature of GRBs of the order of seconds would require continuous monitoring to detect delayed counterparts.
In the case of GBM, the lensing probability is limited by detector dead time.

In this review, the probability of lensing underlies estimation of expected number of lensed gamma-ray sources. 
The distances are calculated based on a homogenous Friedmann-Lema{\^i}tre-Robertson-Walker cosmology,  $h=0.673$, 
the mean mass density $\Omega_M=0.315$ and the normalized 
cosmological constant $\Omega_\Lambda=0.686$ \citep{2013arXiv1303.5076P}.

%%%%%%%%%%%%%%%%%%%%%%%%%%%%%%%%%%%%%%%%%%%%%%%%%%%%
\section{Sources \label{sec:Sources}}
%%%%%%%%%%%%%%%%%%%%%%%%%%%%%%%%%%%%%%%%%%%%%%%%%%%%

Maarten Schmidt's  discovery in 1956 that the 3C~273 hydrogen line Balmer series spectrum implied a redshift of 0.158
 was considered shockingly high for such bright object. 
Curiously, the 13 magnitude optical counterpart of 3C~273, one of the strongest extragalactic sources in the sky, was observed at least as early as 1887.
However, its point-like optical appearance associated it with a starlike object not interesting enough to follow up with spectroscopic observations.
A follow-up spectroscopic observation was conducted only after 1962, 
when a strong radio source was associated with this apparent starlike object.

The observed intraday variability implied small size  ($<$ light days), which led to invoking black holes, which were only theoretically speculated upon at that time.
Theories that accretion onto supermassive black holes (SMBHs) powers quasars and other less luminous active galactic nuclei (AGN) 
rapidly gained momentum \citep{1964ApJ...140..796S,1969Natur.223..690L,1971MNRAS.152..461L}. 

The first quasar discovery was quickly followed by many others.
Observations provided evidence that quasars were much more numerous at $z\sim2$ than they are now. 
After a period of vigorous accretion, a quiet SMBH remains in the center of a galaxy. 
Thus, many of the nearby galaxies host dead quasar engines in their centers, including our galaxy.
Now, it is well known that nearly every galaxy hosts a SMBH of millions to billions of solar masses in its center \citep{1995ARA&A..33..581K},
which may have co-evolved with its host \citep{2013ARA&A..51..511K}. 

The discovery of quasars with their large redshifts and corresponding unprecedented-large radio and optical luminosities 
generated a wide range of observational and theoretical investigations \citep{2013BASI...41....1K}.
Despite great progress in our understanding of the evolution and physical nature of these sources, 
there are numerous questions that cannot be answered directly using current and forthcoming facilities. 
Strong gravitational lensing provides a tool to improve angular resolution and sensitivity of current and future telescopes by orders of magnitude, 
and give a new path to resolve the inner regions of sources. 

This section is dedicated to presenting briefly properties of active galaxies (Section~\ref{sec:ActiveGalaxies}), 
supermassive black holes (Section~\ref{sec:SMBH}), and relativistic jets (Section~\ref{sec:jets}).

%%%%%%%%%%%%%%%%%%%%%%%%%%%%%%%%%%%%%%%%%%%%%%%%%%%%
\subsection{Active Galaxies \label{sec:ActiveGalaxies}}
%%%%%%%%%%%%%%%%%%%%%%%%%%%%%%%%%%%%%%%%%%%%%%%%%%%%

Galaxies hosting an active galactic nucleus (AGN) are stronger emitters than the nuclei of typical galaxies. 
Excess luminosity is not produced by stars but, rather, originates from a central engine powered 
 by ongoing accretion of magnetized plasma into a supermassive ($>10^6 M_{\odot}$) black hole.
 These SMBHs convert the gravitational energy of accreting matter into mechanical and electromagnetic energy making AGN  the brightest extragalactic sources accounting for a significant fraction of the electromagnetic energy output of the universe. 
A fraction of the energy released by the central engine is converted into heat and electromagnetic radiation inside the accretion disk and is radiated away by it. 
Some of the material processed through the accretion disk escapes the central engine as collimated jets and uncollimated wind outflows.

The unified structure of AGN demonstrates that they have similar internal structure \citep{1995PASP..107..803U,2000ApJ...545...63E}.
The observed characteristics of AGN depend on several properties including
 accretion rate \citep{2014ARA&A..52..589H}, 
 orientation \citep{1993ARA&A..31..473A,2015ARA&A..53..365N},
 the presence or absence of relativistic jets, % (e.g. Padovani, 2016), 
 and possibly the host galaxy and the environment.
Different components of the central engine produce emission dominating different energy ranges.
The accretion disc dominates the optical and ultraviolet (UV) bands. 
The obscuring material and dust surrounding the central engine is probed in the infrared (IR) band.
The corona created by non-collimated outflow can be traced by observing the X-ray emission. 
The collimated jet produces a significant excess of radio and gamma-ray emission.
The fraction of quasars with strong radio emission from the jet constitute 
about $\sim$10\% of the sample \citep{2002AJ....124.2364I,2014arXiv1401.1535K}.
The observational selection biases is crucial to account for to understand AGN physics and their role in galaxy evolution \citep{2012NewAR..56...93A}.
 
 Numerous subclasses of AGN have been defined based on their observed characteristics; 
 the most powerful AGN are classified as quasars.
AGN exhibiting a wide range of phenomena and different classes of AGN 
can be selected across the full range of the electromagnetic spectrum \citep{2017A&ARv..25....2P}
For recent results concerning the properties of AGN see \citep{2013FrPhy...8..609K}.

 The inner regions of active galaxies can host the most extreme and energetic phenomena in the universe including
relativistic jets, binary supermassive black holes, or recoiling supermassive black holes
 \citep{1980Natur.287..307B,1984RvMP...56..255B,2002ApJ...565..244H,2003ApJ...582..559V,2006AIPC..856....1M,2009ApJ...698..956C,2011MNRAS.412.2154B,2012Sci...338..355D,2017arXiv170306143B,2016ApJ...830...50M,2016arXiv161100554D,2016A&A...588A.125R,2017MNRAS.464.3131K,2017arXiv170604010P}.

%%%%%%%%%%%%%%%%%%%%%%%%%%%%%%%%%%%%%%%%%%%%%%%%%%%%
\subsection{Supermassive Black Holes \label{sec:SMBH}}
%%%%%%%%%%%%%%%%%%%%%%%%%%%%%%%%%%%%%%%%%%%%%%%%%%%%

Supermassive black holes are common objects in the universe. 
Nearly every galaxy contains a supermassive black hole at its center, 
with a mass ranging from millions to billions of solar masses.
Black holes are entirely specified by their mass, angular momentum, and electric charge (likely $\sim0$): the no-hair theorem.
For a compact  review on astrophysical black holes see \citep{2003Sci...300.1898B,2017arXiv171110256B}.

Discovery that quasars are distant sources led to a realization that extraordinarily efficient and exiguous engines power these objects. 
The only likely physical process able to produce observed energy is accretion onto SMBH. 
In accretion process, gas falling into the potential well converts a part of the released potential energy to thermal energy, which ultimately results in radiation. 
Any gravitating object has a characteristic luminosity called the Eddington limit (also referred to as the Eddington luminosity) 
at which outward radiative acceleration is balanced by the inward pull of gravity \citep{2013arXiv1312.6698N}.
The Eddington luminosity is defined as
\begin{equation}
L_E = \frac{4\pi G M m_p c}{\sigma_T} = 1.25\times 10^{38} \frac{M}{M_\odot} \mbox{erg\,s}^{-1} \,,
\end{equation}
where $M$ is the mass of the object, $m_p$ is the mass of the proton, 
and $\sigma_T$  is the Thomson cross-section for electron scattering. 
Bright quasars have luminosities $L\sim 10^{46}\,$erg$\,$s$^{-1}$,
which implies very massive objects, with mass $M>10^8\,M_\odot$. 

The gravitational radius of a black hole of mass  $M_{BH}\sim10^8\,M_\odot$ is
\begin{equation}
R_g = \frac{GM_{BH}}{c^2}  \approx  1.48 \frac{M_{BH}}{10^8M_\odot} 10^{13}\mbox{cm} \,. 
\end{equation}
Primary constraints on the physical scale of quasars come from the intrinsic variability \citep{2001sac..conf....3P} and microlensing \citep{2010ApJ...712.1129M}. 
Quasars show variability on a time scale of days, which implies size smaller than 100$\,R_g$.
The estimation of the quasar size using variability is based on an argument that an object cannot have large-amplitude variations on a time scale shorter than its light-crossing time.
The small sizes indicated by short variability time scales and the large masses inferred from the Eddington limit
point to SMBHs as the source of power in quasar's engines. 

The origin of seed SMBHs, their formation and evolution are still under debate.
The three most popular BH formation scenarios include the core-collapse of massive stars,  
dynamical evolution of dense nuclear star clusters, and collapse of a protogalactic metal-free gas cloud \citep{2016PASA...33...51L}.
%see a recent review on this topic by Latif \& Ferrara . 
Intriguingly, quasars are observed at redshifts even greater than $z>6$,
which implies the existence of SMBHs of a few billion solar masses within  $1\,$Gyr after the Big Bang. 
Understanding the growth of high redshift SMBHs is an essential problem in astrophysics. 
Frequently observed luminous high redshift quasars may represent only the tip of the iceberg. 
A large population of low luminosity AGN may remain undetected due to insufficient sensitivity of current facilities \citep{2018MNRAS.476.5016L}.

%%%%%%%%%%%%%%%%%%%%%%%%%%%%%%%%%%%%%%%%%%%%%%%%%%%%
\subsection{Relativistic Jets \label{sec:jets}}
%%%%%%%%%%%%%%%%%%%%%%%%%%%%%%%%%%%%%%%%%%%%%%%%%%%%

Relativistic jets are beams of plasma launched in the vicinity of accreting SMBHs.
There are two competing theories on the origin of the jet power. 
The first proposes that jets are powered by the gravitational energy of accreting matter that moves toward the black hole,  
 where jets may either be launched purely electromagnetically \citep{1976MNRAS.176..465B,1976Natur.262..649L}
 or as the result of magnetohydrodynamic processes at the inner regions of the accretion disk
 \citep{1984RvMP...56..255B,1982MNRAS.199..883B}. 
The second theory utilizes the rotational energy of a rotating black hole \citep{1977MNRAS.179..433B}. 
%The jets composition remain uncertain \cite{0004-637X-625-2-656}.

The relativistic motion of plasma in jets leads to multiple effects of the special theory of relativity 
including relativistic boosting, time dilution, and apparent superluminal motions. 
%When jets point toward the observer then strong relativistic effects are observed.
%
The radiation emitted by jets is Doppler boosted toward the observer by $\mathcal{D}^{n}$ 
\citep{1966Natur.211..468R,2003A&A...406..855M,2007ApJ...658..232C}. 
The Doppler factor is defined as 
\begin{equation}
\mathcal{D}=[\Gamma(1-\beta \cos{\theta_{obs}})]^{-1} \,,  
\end{equation}
with the Lorenz factor 
\begin{equation}
\Gamma=\frac{1}{\sqrt{1-\beta^2}} \,,
\end{equation}
where $\beta=v/c$ is the velocity of moving plasma, $v$, in units of the speed of light $c$,  
and  $\theta_{obs}$ is the angle to the line-of-sight with the observer. 
The exponent $n$ combines effects due to the K correction \citep{2002astro.ph.10394H} 
and the Doppler boosting caused by relativistic aberration, 
time dilation, and the solid angle transformation \citep{1995PASP..107..803U}.
In the calculations presented in this review, the index $n=4$ is assumed.

%The differences in quasar appearance can be explained largely by the jet orientation.
The emission experiences strong relativistic boosting when the jet is pointed close to the line-of-sight ($<20^\circ$).
Relativistic beaming changes the apparent beam brightness, as a result, only the side of the jet pointed toward the observer is visible, and the resulting extremely luminous object is called a blazar. 
Features used in blazar classification include the presence of a compact radio core, 
with flat or even inverted spectrum, extreme variability (both in timescale and in amplitude) at all frequencies, and a high degree of optical and radio polarization \citep{2009A&A...495..691M}.

Non-thermal emissions produced by a relativistic jet dominates the broadband spectrum of blazars. 
The spectral energy distribution (SED) of blazars is characterized by two broad spectral components. 
A low-energy component extends from the radio up to optical/UV/X-rays is produced by the synchrotron radiation 
of relativistic electrons. 
The high-energy component extending from X-rays to gamma-rays, according to recent interpretations, 
is produced by inverse-Compton (IC) radiation with a possible source of seed photons, being either the synchrotron radiation, 
the broad line region (BLR), or the dusty torus (DT).

Blazars are divided into two classes: flat spectrum radio quasars (FSRQs) and BL Lac objects; 
FSRQs are distinguished by the presence of broad emission lines, which are absent or very weak in BL Lac objects.
The high-energy component of FSRQs is usually much more luminous than the low-energy one.
The high-energy component of BL Lac objects results from the Comptonization of synchrotron photons. 
The luminosity at the peak of the high-energy component is comparable or lower than the synchrotron peak luminosity \citep{1978PhyS...17..265B}.

Relativistic jets of blazars provide environments to accelerate particles to velocities close to the speed of light.
The two most popular processes used to explain particle acceleration in relativistic jets are internal shock scenario \citep{1994ApJ...421..153S,2001MNRAS.325.1559S},
and reconnection of magnetic field \citep{1992A&A...262...26R,2002A&A...391.1141D}
The internal shock scenario assumes an instability in the central engine, which results in ejection of shells of plasma \citep{1978PhyS...17..193R}.
The shells with inhomogeneous velocity or mass distribution ``catch up",  a nonelastic collision occurs,
and particles are accelerated through the first-order Fermi mechanism 
- a process in which particles scatter between the upstream and downstream regions of shocks to gain energy \citep{1987PhR...154....1B,1538-4357-682-1-L5,2002A&A...394.1141O}.
Acceleration of particles in shocks is commonly used to model non-thermal phenomena in the universe  
using Monte Carlo test particle simulations  \citep[e.g.,][]{1991MNRAS.249..551O}
and semianalytic kinetic theory methods \citep{2000ApJ...542..235K,2001MNRAS.328..393A,2005PhRvL..94k1102K}.

The observed spectral energy distribution (SED) of blazars can be well reproduced with shock scenario 
\citep{2009MNRAS.397..985G,2011A&A...530A...4A,2012ApJ...760...69N,2014A&A...567A.113B}.
However, the first-order Fermi mechanism requires relatively long timescales of the order of days to sufficiently accelerate particles. 
Observation of blazars show variability down to (sub-)hour time scales \citep{2007ApJ...664L..71A,2011A&A...530A..77F}, challenging the shock scenario.  

Magnetic reconnection was proposed as a more likely candidate that shocks for explaining short variability timescales observed in the jet emission.
During an event of magnetic reconnection, the annihilation of field lines of opposite polarity transfers the field energy to the particles.
It is still under debate if shocks or magnetic reconnection accelerate particles in relativistic jets \citep{2015MNRAS.450..183S}.
Both mechanisms are based on assumptions that the energy dissipation is happening at small distances, $\sim\,$parsecs from the central engine. 
The recent observations show evidence that variable emission can be produced more than a dozen of parsecs from the central engine, 
which challenges both scenarios of particle acceleration.

Jets transport energy and momentum over even megaparsec distances \citep{1974MNRAS.169..395B}.
Radio interferometry resolves the details of complex jet structure that  includes
hotspots and blobs \citep{2008Natur.452..966M,2011ApJS..197...24M,2012ApJ...755..174G,2012ApJS..203...31M}. 
Improved angular resolution of current X-ray satellites demonstrates 
that the high energy emission from jets also form structures as large as 
hundreds of kpcs \citep{2006ARA&A..44..463H,2007ApJ...662..900T,2002ApJ...570..543S}.  
At gamma rays, the technology is inadequate to resolve the sources. 
However, the short variability timescales, $<1\,$day, suggest that the sources of the gamma-ray radiation 
during a flare is of the order of $10^{-3}$~parsec \citep{2011AdSpR..48..998S}. 
To explain the observed rapid variability and to avoid catastrophic pair production in blazars, 
models assume that the $\gamma$-rays are produced in compact emission regions moving 
with relativistic bulk velocities in or near the parsec scale core \citep{1995MNRAS.273..583D}. 
However, recent detection of sub-TeV emission from FSRQs suggests 
that  the blazar zone can be located several parsecs away 
from a SMBH \citep{2012MNRAS.425.2519N}.
It remains unclear whether the radiation source is the same at all energies. 
The source of radiation may be close to the base of the jet or it may originate from blobs of plasma 
moving along the jet at relativistic speeds.

%%%%%%%%%%%%%%%%%%%%%%%%%%%%%%%%%%%%%%%%%%%%%%%%%%%%
\subsection{Case Study:  M87}
%%%%%%%%%%%%%%%%%%%%%%%%%%%%%%%%%%%%%%%%%%%%%%%%%%%%

The giant elliptical galaxy M87 in the core of the Virgo cluster, located 16.7 Mpc \citep{2005ApJ...634.1002J}  away, hosts the first discovered jet: 
``A curious straight ray lies in a gap in the nebulosity in p.a. 20deg, apparently connected with the nucleus by a thin line of matter`` (Curtis 1918). 
At that time, the nature of this extended feature was not understood.
Today, M87 serves as one of the best and the largest particle acceleration laboratories.
The M87 jet is among the most studied sources with observations from its event horizon of the SMBH residing in the center of galaxy \citep{2015ApJ...805..179B} 
to 20-30 kpc long jet cavity \citep{2008ApJ...676..880M}. 

M87 has a blazar-like core and a relativistic jet, but it has lower luminosity than the typical high redshift blazar. 
The M87 jet consists of bright knots of radio, optical, and X-ray emission spread throughout a projected distance of 1.6 kpc  (81 pc/").
The optical and X-ray emission correspond to radio features. 
The bright knots are observed with apparent velocities, $\beta_{app}$, within a range of 4c-6c \citep{1999ApJ...520..621B}.
Based on the standard picture of relativistic boosting \citep{1966Natur.211..468R}, 
the observed apparent velocities require that the jet must be less than 19$^\circ$ from our line-of-sight~\citep{1999ApJ...520..621B}.

The M87 jet produces variable emission. 
It has been believed that such variable emission is produced within 1~pc from the SMBH. 
Observations with the {\it Chandra X-Ray
Observatory} \citep{2007Ap&SS.311..329M} show that substantially  increased X-ray emission originates from at least two regions:
the core of M87 and the  HST-1 knot \citep{1999ApJ...520..621B}. 
 The HST-1 knot located at a projected distance of $\sim$60~pc  from the core is one of the most interesting features along the M87 jet.
The {\it Chandra X-Ray Observatory} monitoring program revealed increased intensity by more than a factor of 50 from the knot HST-1 \citep{2006ApJ...640..211H,2009ApJ...699..305H}.
VERITAS, H.E.S.S. and MAGIC all detected VHE emission simultaneous with the increased X-ray luminosity from the HST-1. 
The exact location of the  source of very high-energy gamma rays remains unclear due to limited resolution of gamma-ray instruments. 

Models where the emission originates from spatially distinct
knots, as in M87, differ in the underlying fundamental physics 
\citep{1992A&A...256L..27D,1996ApJ...461..657B,1994ApJ...421..153S,2013ApJ...779...68S,2009ApJ...704...38S,2009MNRAS.395L..29G,2003APh....18..593M,2013ApJ...768...54B,2004A&A...419...89R,2005ApJ...626..120S,2006MNRAS.370..981S,2014ApJ...780L..27M}.
Thus, the location of gamma-ray flares is crucial  for understanding  particle acceleration  and magnetic fields at both small and  large distances from  SMBHs.

In this review, M87 is used as a toy model to demonstrate scales and illustrate ideas.
If an M87-analog were located at redshift equals~1 and there would a lensing galaxy of $\sim 10^{11}\,\mbox{M}_{\odot}$,
the  Einstein radius of the lens would be $0.45"$.
If we project a M87-like  jet into the source plane, 
then the projected distance between the core and HST-1 would
correspond to $\sim2\%\,r_E$. 
As will be reviewed in the following sections, 
a difference of  $\sim2\%\,r_E$ will produce significance change in expected time delay and magnification ratio.

%%%%%%%%%%%%%%%%%%%%%%%%%%%%%%%%%%%%%%%%%%%%%%%%%%%%
\section{Resolving Sources Using Strong Gravitational Lensing \label{sec:Resolving}}
%%%%%%%%%%%%%%%%%%%%%%%%%%%%%%%%%%%%%%%%%%%%%%%%%%%%

Discovering the nature of complex sources residing in the inner regions of active galaxies requires an ability to zoom within a few parsecs. 
Such resolution is currently available only with radio telescopes. 
Radio observatories can resolve sources of radiation with an angular resolution and absolute astrometry below 1~milliarcsecond. 

Resolution of our telescopes decreases with photon frequencies. 
At gamma rays, angular resolution degrades by seven orders of magnitude to less than 0.1~deg. 
Gamma-ray emission is captured by particle detectors launched into space. 
The angular resolution of gamma-ray satellites is limited 
by our ability to reconstruct the direction of photons from the trajectories of secondary particles, which is intrinsically limited by physical effects, such as nuclear recoil. 
%The angular resolution of gamma-ray telescopes is energy dependent and is of the order of one degree. 
The advantages of gamma-ray observations are an excellent temporal resolution and extreme source variability.
These temporal characteristics of gamma-ray sources allow us to detect gravitationally induced time delays. 

It has been commonly assumed that the multi-wavelength emission of active galaxies originates from a single compact region. 
Current and future facilities do not have sufficient angular resolution or astrometry to test this assumption. 
Strong gravitational lensing produces multiple images of the source with gravitationally induced time delays between these lensed images. 
Positions of lensed images and time delays depend on the location of the sources. 
Positions of the lensed images and their time delays probed at different energies and times give us a unique insight into the origins of the emission. 
The model of the lens combined with cosmological parameters allow us to convert the offsets in time delays and position of lensed images into physical separation of the sources. 

This review focuses on three approaches of using strong gravitational lensing to resolve origins of the emission spatially;
\begin{enumerate}[I.]
  \item {\bf The time delay approach} - turns the difference in time delays measured at different energies or times into distance between emitting regions 
  \item {\bf The Hubble parameter tuning approach} - combines time delay with  positions of the lensed images into the spatial separation of sources, 
  \item {\bf The caustic of galaxies as non-linear amplifiers approach} - use positions of lensed images to find multiple sources. 
\end{enumerate}
The methods of using strong gravitational lensing to infer the spatial origins of the emission are described in the following sessions, with a focus on study cases and future applications. 

%%%%%%%%%%%%%%%%%%%%%%%%%%%%%%%%%%%%%%%%%%%%%%%%%%%%
\section{The Time Delay Approach \label{sec:TDA}}
%%%%%%%%%%%%%%%%%%%%%%%%%%%%%%%%%%%%%%%%%%%%%%%%%%%%

The high-energy sky is dominated by the
most extreme and puzzling objects in the universe.
These sources harbor powerful jets,
the largest particle accelerators in the universe,
producing radiation ranging from radio wavelengths
up to very high-energy gamma rays.
Our ability to study the gamma-ray radiation
from these jets is observationally limited
by the poor ($>0.1^\circ$) angular resolution of the
detectors. This angular resolution is strongly
limited by physical effects such as nuclear recoil
and is unlikely to improve significantly in future
instruments.
Our inability to validate the assumption of the origin of the emission due to the limited resolution of instruments
strengthen the assumption that the  regions close to supermassive black holes were associated with producing gamma-ray emission. 
This assumption of the origin of the emission impacts our understanding of mechanisms of particle acceleration.

Gravitationally-induced time delays %measured using gamma-ray light curves 
provide a new route to resolve the origin of gamma-ray emission down
to tens of parsecs.
The gravitationally-induced time delay analysis of PKS~1830-211 improved angular resolution at gamma-ray 10,000 times 
and shows that gamma-ray flares do not always originate 
from regions close to the supermassive black hole,  
indicating existence of a particle acceleration mechanism 
capable of producing variable gamma-ray emission even kpcs from the central engine.  
Detection of multiple gravitationally-induced time delays originating from various emitting regions along the relativistic jet challenges our understanding of 
the particle acceleration mechanism, the jet environment at large distances from SMBHs, 
and confront the use of gravitationally-induced time delays as a cosmological probe. 

Section~\ref{sec:TDidea} outlines the idea of using gravitationally-induced time delays to resolve the origin of emission. 
Section~\ref{sec:TDGamma} reviews strong gravitational lensing at gamma rays. 
The application of the method relies on time delay measurement from poorly resolved or unresolved sources.
Thus, in Section~\ref{sec:method}, the Double Power Spectrum (DPS) 
method optimized for measuring time delays from unresolved sources with long and well sample data is reviewed. 
Section~\ref{sec:processing} briefly presents the signal processing procedure, which was essential in detecting time delays. 
Section~\ref{sec:MPM} describes the Maximum Peak Method (MPM) introduced to constrain time delays  
for gamma-ray flares with insufficient photon statistics to apply Fourier transform based methods.
Monte Carlo (MC) simulations to validate these methods are discussed in Section~\ref{sec:MC}. 
Finally, the methods are demonstrated on the gravitationally lensed blazar PKS~1830-211,
and its case study is reviewed in Section~\ref{sec:casePKS1830}. 

%%%%%%%%%%%%%%%%%%%%%%%%%%%%%%%%%%%%%%%%%%%%%%%%%%%%
 \begin{figure}
 \centering
 \includegraphics[width=9.3cm,angle=0]{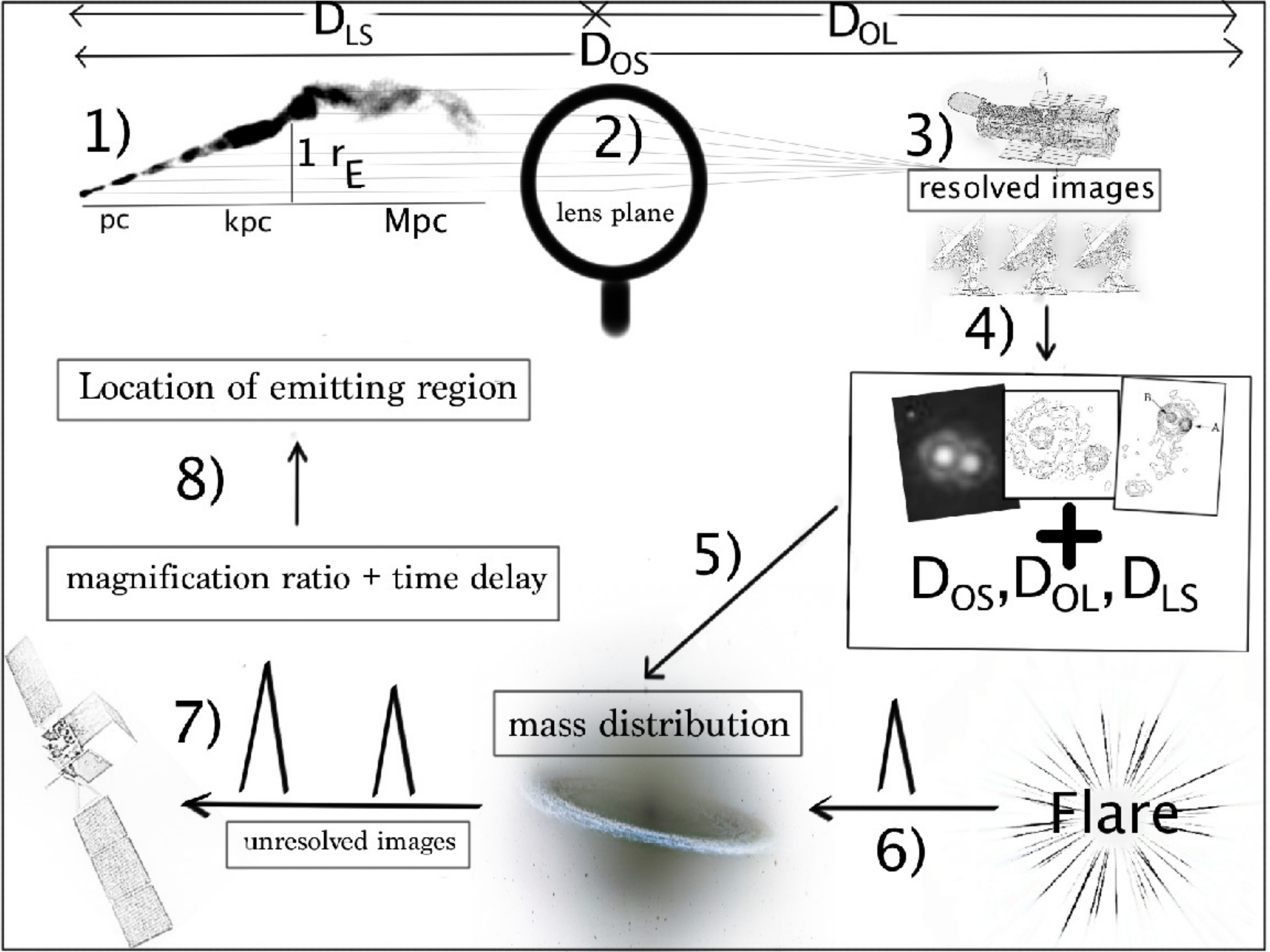}%
 \caption{\label{fig:sketch}  {Steps in the application of strong gravitational lensing 
                        to resolve jet structures using gravitationally-induced time delays. 
		     Figure from \cite{2014arXiv1403.5316B}. }
		 }  
 \end{figure}
%%%%%%%%%%%%%%%%%%%%%%%%%%%%%%%%%%%%%%%%%%%%%%%%%%%%
\subsection{Approach \label{sec:TDidea}}
%%%%%%%%%%%%%%%%%%%%%%%%%%%%%%%%%%%%%%%%%%%%%%%%%%%%

%{\bf Motivation: Turning temporal resolution into spatial resolution. }
High energy detectors have a limited angular resolution, 
however they have an excellent temporal resolution to monitor variable emission. 
Strongly lensed sources provide a unique ability to turn temporal resolution into spatial resolution 
by combining precise measurements of time delays with the lens model to infer the origin of the variable emission.  

Figure~\ref{fig:sketch}  shows {\bf Steps} in applying time delay approach to resolving the origin of variable emission proposed by Barnacka et al. \citep{2014arXiv1403.5316B}. 
%The most variable emission is produced in relativistic jets.
%
The brightest and the most variable quasars are those with jets pointed toward the observer. 
The jets pointed toward the observer exhibit strong relativistic effects including Doppler boosting and time dilation. 
These extragalactic jets extend from sub-parsec up to even megaparsec scales.
When a jet is pointed close to the line-of-sight to the observer 
then the projected size of such jet is of the order of a few kiloparsecs (see {\bf Step~1}). 

A galaxy located close to the line-of-sight between the source and observer acts as a lens with an Einstein radius of a few kpcs.
Thus, the projected size of the relativistic jet is comparable to the size of the lens.
Radiation emitted in different regions of the jet approaches  the lens plane at different distances from the center of the lens.
Differences in path length and traversed gravitational potential result in a different  magnification ratio  
and time delay between lensed images of the source (see {\bf Step~2}).

Observations at radio and optical wavelengths provide resolved images (see {\bf Step~3}). 
They also provide redshifts of the lens and source. 
{\bf Step~4} shows example images of the lensed system B2~0218+35 observed using optical and radio telescopes  
\citep{2000MNRAS.311..389J,1995MNRAS.274L...5P,1999MNRAS.304..349B}.
The resolved images and the known distances allow us to reconstruct the mass distribution of the lens (see {\bf Step~5}).

The location of the high-energy emitting region cannot be directly resolved with current instruments.
However, variability timescales observed for distant sources of high-energy emission can be as short as a few hours implying  a compact emission region ($\ll\,$pc).
These variability timescales are short compared to the time delay. 
Gravitationally-induced time delays are of the order of weeks to months. 
Thus the time delay and corresponding magnification ratio for the delayed counterparts of the radiation source can be estimated even when lensed images of the source are unresolved (see {\bf Step~6} and~{\bf7}).
Measurement of the time delay and magnification ratio for  the lensed flare 
can be used to limit the location of the high energy emitting region along the relativistic jet (see {\bf Step~8}).

The potential of the time delay approach can be demonstrated using a toy model based on M87 following  \citep{2014arXiv1403.5316B}. 
At the distance of M87,  the angular resolution of gamma-ray instruments corresponds to a projected size of 30~kpc, 
500~times greater than the projected distance between the core and HST-1 of $60\,$pc.
Thus, even for M87 it is impossible to distinguish if the variable gamma-ray emission originates from a region close to the base of the jet or the HST-1 knot. 
An improvement in the angular resolution of the order of 1000 is required to resolve even nearby sources.
Future instruments like the Cherenkov Telescope Array (CTA) will provide improvement in angular resolution by a factor of a few \citep{2011ExA....32..193A}. 

Now, let us investigate an M87 analog placed at redshift $z \sim 1$ 
with a lens close to the line-of-sight at $z \sim 0.6$,
and assuming that the orientation of the M87 jet relative to the line-of-sight is the same as observed at its true redshift. 
For the M87 toy model, a 1\% $r_E$ difference in position of the source (1\% $r_E$ = 22~pc) 
results in the time delay difference of $\sim0.6$~days. 
The displacement of $\sim$0.02$r_E$ in the source plane between the core  of M87 and HST-1
changes the time delay by almost 2 days
and  the magnification ratio by $\sim 0.2$.
Such a large differences in time delay and magnification ratio can be measured using existing facilities. 
Thus, the time delay approach can be used to elucidate complex structure of distant quasars.

M87 is not gravitationally lensed, at least not when observed from our galaxy, 
but at least $20$ gravitationally-lensed quasars are available in the  
CLASS (Cosmic Lens All-Sky Survey) and JVAS (Jodrell/VLA Astrometric Survey)
\footnote{http://www.jb.man.ac.uk/research/gravlens/lensarch/lens.html} surveys. 
These samples consist of radio-loud quasars - excellent candidates to apply the time delay approach.
 
%%%%%%%%%%%%%%%%%%%%%%%%%%%%%%%%%%%%%%%%%%%%%%%%%%%%
\subsection{Time Delays and Gamma Rays  \label{sec:TDGamma}}
%%%%%%%%%%%%%%%%%%%%%%%%%%%%%%%%%%%%%%%%%%%%%%%%%%%%

The {\it Fermi} Gamma-ray Space Telescope has continuously monitored the entire sky since 2008, 
detecting photons from the most luminous and variable objects in the universe (e.g. blazars and gamma-ray bursts).  
The {\it Fermi} satellite carries two instruments: the Large Area Telescope (LAT; 25~MeV to $>$ 300~GeV) 
and the Gamma-ray Burst Monitor (GBM; 8~keV to 40~MeV).
%The majority of the sources detected by {\it Fermi}/LAT are blazers.

The {\it Fermi} satellite has provided detection of more than $1000$ blazars, but their redshifts are generally unknown. 
Using the formalism described in Section~\ref{sec:Probability}, and a sample of 370 FSRQs with known redshift  listed in the 2nd Fermi catalogue \citep{2012ApJS..199...31N},
the number of expected gravitationally-lensed systems observed by the {\it Fermi} satellite  
is $\sim$10 \citep{2013arXiv1307.4050B}.
However,  including the effectiveness of matter in producing double images defined by Equation~\ref{eq:F},
only about $\sim$5\% of these systems will produce multiple images of the source. % \citep{1992ApJ...393....3F}.
%Where F  measures the effectiveness of matter in producing double images. 
%One estimate of F is 0.047 \citep{1992ApJ...393....3F}. 
The lensing probability may be  increased by  magnification  bias \citep{1984ApJ...284....1T,1993LIACo..31..217N}.
Thus, the number of lensed gamma-ray sources may exceed the estimate based only on the FSRQs. 
 
Among the FSRQs detected at gamma rays, PKS~1830-211  and B2~0218+35
are well known gravitationally lensed systems. 
The detection of a gravitational-lens induced time delay of $27\pm 0.5$~days in the light curve of PKS~1830-211  during its low state
by \cite{2011A&A...528L...3B} provided 
the first evidence for strong gravitational lensing at gamma rays.

The launch of the {\it Fermi} satellite in 2008 opened 
a unique opportunity to characterize the long-term variability of distant gamma-ray blazars.    
{\it Fermi}/LAT surveys the entire sky in 190~min,  regularly sampling blazar light curves with a period of a few hours. 
{\it Fermi}/LAT provides very long and evenly sampled data. 
The data include the time of detection of each gamma-ray photon,
along with an estimation for its energy and direction on the sky. 

%
%relativistic jets powered by supermassive black holes 
%which happened to be pointed toward Earth. 
%
%Figure~\ref{fig:lc_whole} presents the time series showing flux of gravitationally lensed blazar B2~0218+35 measured by {\it Fermi}/LAT.
{\it Fermi}/LAT detects one photon per day from a typical blazar  in its quiescent state. 
However, when blazars flare,  the flux can increase by orders of magnitude. % (see Figure~\ref{fig:lc_whole}). 
During a flare,  the observed variability time scales intrinsic to the source  range from minutes to days. 
In contrast, the gravitationally induced time delays are in the range from weeks to months. 
This separation of timescales facilitates the measurement of gravitationally-induced time delays. 
%
%Data provided by {\it Fermi}/LAT are well suited to study temporal behavior of blazars. 

Blazar emission appears  to be stochastic  \citep{2014ApJ...791...21F,2013ApJ...773..177N,2014ApJ...786..143S},
with a power spectral density (PSD) given by
%The power spectral density (PSD) is inversely proportional to the frequency, $f$,  of the signal to the power $\alpha$:
%
\begin{equation}
\label{eq:PSD}
S(f) \propto 1/f^{\alpha} \,,
\end{equation}
where $f$ is the signal frequency (here, $f$ is given by the Fourier transform of the brightness;
not the frequency of the photon, which is denoted as $\nu$). 
Typically, blazers  have $\alpha \sim 1\, - \,2$.

This random variability is often referred to as {\it noise} intrinsic  to the source (not a result of the measurement error),  
which is a result of stochastic processes  \citep{2003MNRAS.345.1271V}.
Astronomers refer to these stochastic fluctuations as {\it signal} \citep{1978ComAp...7..103P}. 
In general, the temporal behavior of quasars is simply represented by stochastic fluctuations \citep{2013ApJ...773..177N,2014ApJ...786..143S}. 
The time series of a strongly lensed source has the same temporal evolution, but it additionally contains a time delay.
%The lens itself is not a gamma-ray emitter at a detectable level.
Gamma-ray detectors do not spatially resolve emission of lensed images.  
Thus, the time series can be constructed as a sum of the components of a gravitationally lensed gamma-ray blazar
\begin{equation}
\label{eq:lc}
S(t) = \sum_{i}^{N}\frac{s(t+a_i)}{b_i}\,,
%S(t) = s(t)+s(t+a)/b\,,
\end{equation}
where $S(t)$ is the  unresolved light curve of the lensed blazar, 
composed of the sum of N lensed images of the source with intrinsic variability $s(t)$. 
The temporal behavior of individual images is determined  by the source,
but the images are shifted in time 
by the time delay, $a_i$, 
and with the magnification ratio between lensed images, $b_i$.

One of the issues of very high energy observations is  that gamma-ray photons emitted from sources at cosmological distances may 
be absorbed by the gamma-gamma interaction as they travel through various photon fields.
\citep{2014arXiv1404.4422B} shows that gamma rays can avoid absorption by being deflected 
by the gravitational potential of the luminous massive galaxy located close to the line-of-sight.
The collective photon fields from lensing galaxies typically do not produce any measurable excess gamma-gamma opacity 
beyond that of the extragalactic background light (EBL; \citep{2010ApJ...723.1082A,2012Sci...338.1190A}. 
The EBL can reduce the observed gamma-ray flux of lensed blazars, but the images of a given blazar will be changed by the same fractional amount. 
Thus, the magnification ratios and the time delays at gamma rays between the images remain unchanged by the EBL.

%The following sections present methods developed to measure time delays from unresolved sources. 

%%%%%%%%%%%%%%%%%%%%%%%%%%%%%%%%%%%%%%%%%%%%%%%%%%%%
\subsection{Double Power Spectrum Method \label{sec:method}}
%%%%%%%%%%%%%%%%%%%%%%%%%%%%%%%%%%%%%%%%%%%%%%%%%%%%

Measuring gravitationally-induced time delays is challenging.
Such measurement requires long monitoring campaigns. 
In addition, the gravitationally lensed source has to be variable. 
Optical and radio monitoring of gravitationally lensed sources have resolved multiple images of lensed sources for 
a number of measured time delays 
\citep{2002ApJ...581..823F,2011A&A...536A..44E,2013A&A...557A..44R,2013A&A...556A..22T,2013A&A...553A.121E}. 
Unevenly spaced data resulting from, for example, weather and/or observing time allocation,
are a challenge for light-curve analysis. 
A number of techniques have been specially developed to utilize 
these resolved multiple light curves of lensed images  with unevenly sampled data 
\citep{1988ApJ...333..646E,1992ApJ...385..404P,1992ApJ...398..169R,2001A&A...380..805B,1998A&A...337..681P,2005ApJ...626..649P,1982ApJ...263..835S,1987AJ.....93..968R,1996MNRAS.282..530G,2014MNRAS.441..127G,2011arXiv1105.5991H}.

The observed gamma-ray light curve is a sum of lensed images. 
The Double Power Spectrum (DPS) method was developed 
to measure time delay from unresolved light curves that include delay components of intrinsic source variability,
and was optimized for low photon statistics and long evenly sampled data. 
Here, I briefly introduce the DPS method. 
The individual steps are given in  \citep[Appendix~A]{2015ApJ...809..100B}. 
For simplicity, the case with two lensed images;
 $a_0=0$ and $b_0=1$, is considered.
The signal in the time domain is $S(t) = s(t)+s(t+a)/b$.
The  Fourier transform of the first component, $s(t)$, results in $\tilde{s}(f)$.
The second component transforms to the frequency domain as $\tilde{s}(f) e^{-2\pi if a}/b$.
The observed time series $S(t)$ transforms into 
\begin{equation}
\label{eq:fft}
\tilde{S}(f) = \tilde{s}(f) (1 + b^{-1} e^{-2\pi i f a}) \,,
\end{equation}
in Fourier space.

The first power spectrum of the source is the square modulus of $\tilde{S}(f)$:
\begin{equation}
\label{eq:fps}
|\tilde{S}(f)|^{2} = |\tilde{s}(f)|^{2}(1 + b^{-2} + 2b^{-1} cos(2\pi f a)) \,.
\end{equation}
The first power spectrum is the product of the intrinsic power spectrum of the source, $|\tilde{s}(f)|^2$,
 times a periodic component with a period (in the frequency domain) equal to the inverse
of the relative time delay $a$.
Therefore, the period of the pattern in the first power spectrum need to be obtained  to find  time delay $a$. 
The transformation of the first power spectrum, which is in the frequency domain, to Fourier space 
brings the signal back to the time domain.
After this 2$^{nd}$ transformation,  
peaks in the second power spectrum correspond to time delays  present in the original time series.
The method is similar to the Cepstrum method widely
 used in speech processing and seismology \citep{Bogert1963}. 
 
 %%%%%%%%%%%%%%%%%%%%%%%%%%%%%%%%%%%%%%%%%%%%%%%%%%%%
\subsection{Signal Processing \label{sec:processing}}
%%%%%%%%%%%%%%%%%%%%%%%%%%%%%%%%%%%%%%%%%%%%%%%%%%%%

Detection of time delay in data with sparse statistics is possible thanks to a signal processing procedure optimized using MC simulations. 
As \cite[Appendix~A]{2015ApJ...809..100B} demonstrates, 
without accurate signal processing, time delays would be difficult or even impossible to detect.
Thus, the signal processing distinguishes between significant detections and no detections.
Here, I briefly review steps of signal processing and their purpose. 

The DPS method divides the analysis into three stages.
The first stage is a preparation of the input time series, the light curve.
The second stage is a calculation of the first power spectrum (FPS),
and the last stage is a calculation of the second power spectrum (SPS). 
An optimal signal processing procedure is applied at each stage of the analysis. 

The steps in this signal processing are based on widely used methods \citep{1975dsp..book.....O,1971A&A....13..169B},
and are outlined  in detail  in \cite{2011A&A...528L...3B,2015ApJ...809..100B},
and each step is optimized using Monte Carlo Simulations. 
%Time series of a power law noise were generated with introduced artificial time delay. 
The optimized signal processing applied on all of these stages of analysis  distinguishes the DPS method 
from the Autocorrelation function for which signal processing is applied only in the time domain.  

The signal processing starts with preparing the input by removing the mean from the time series and windowing the data.
Windowing can improve frequency resolution and reduce spectral leakage 
caused by discontinuities in the original noninteger number of periods in a signal. 
Windowing reduces the amplitude of the discontinuities at the boundaries of each finite time series. 
In each step, the window function is optimized to
balance the sharpness of the peak of a periodic signal with the spectral resolution. 

If a time delay is present in the time domain, 
it manifests in the frequency domain of the First Power Spectrum
as a periodic pattern with a period inversely proportional to the time delay revealed by Equation~\ref{eq:fps}. 
Thus, to preserve the maximum resolution of the the FPS, a rectangular window is used. 
Zero padding is applied to avoid the large power at low frequency caused by the discontinuity 
at the beginning and the end of the time series.
Also, to eliminate effects of aliasing points are upsampled without interpolation.
Upsampling the points does not change intrinsic variability of the source, 
but shifts the Nyquist frequency allowing the power go to zero when the frequency approaches the  Nyquist frequency. 

The FPS calculated using the time series prepared using the above procedure serves as an input for the SPS. 
Again, the input is processed before applying Fourier transform. 
At this stage, the goal is to obtain the period of the periodic pattern.
The pattern is convolved with the intrinsic variability of the source represented in the FPS by a power law.
Intrinsic variability of the source introduces the trend into the signal.
To make signal stationary, the signal is ``flattened"  by taking the logarithm of the power spectrum. 
The part of the spectrum at low frequency with large amplitude resulting from power law noise is removed. 
Next, the mean from the series is extracted, windowing and  zero padding is applied. 
At this stage, the signal of interest  is characterized as a peak around the true value of the time delay.
Thus,  the Bingham window\footnote{http://www.vibrationdata.com/tutorials/Bingham\_compensation.pdf} is applied to sharpen the peak around value of the time delay.
The signal processing procedure allows us to maximize detection of the time delay 
by reducing negative impact of the intrinsic variability of the source and minimizing time series effects. 
%I apply such prepared procedure to the data. 
 
 %%%%%%%%%%%%%%%%%%%%%%%%%%%%%%%%%%%%%%%%%%%%%%%%%%%%
\begin{figure*}
%\vskip 1cm
\begin{center}
\includegraphics[width=15.4cm,angle=0]{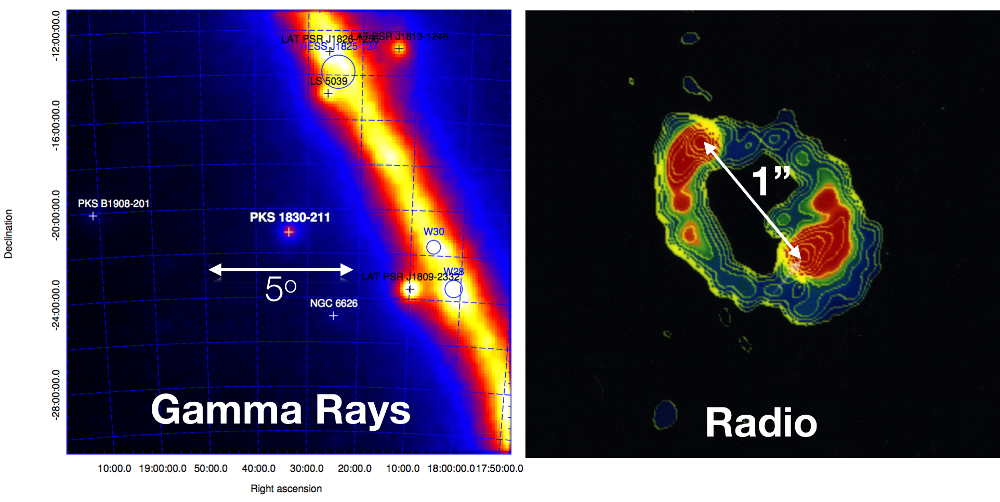}
\end{center}
\caption{\label{fig:pks1830}
                           {\bf Left:} {\it Fermi}/LAT count map around the position of PKS 1830-211.
                            The map contains photons in the energy range from 200 MeV to 300 GeV \citep{2015ApJ...809..100B}.
                            {\bf Right:} MERLIN image of PKS 1830-211 at 5 GHz \citep{1991Natur.352..132J}. }
\end{figure*}
 
 %%%%%%%%%%%%%%%%%%%%%%%%%%%%%%%%%%%%%%
\subsection{Maximum Peak Method \label{sec:MPM}}
%%%%%%%%%%%%%%%%%%%%%%%%%%%%%%%%%%%%%%

The Maximum Peak Method (MPM) was developed to deal with gamma-ray flares that occur as short and isolated events.  
During a flare, even dozens or hundreds  of gamma-ray photons are detected, 
contrary to quiescent state when on average one gamma-ray photon is detected. 
In such cases, the photon statistics are insufficient to extract the time series around isolated flares, 
which precludes use of Fourier transform based methods to extract time delays. 
However, isolated flares are ideal for the direct search of gravitationally delayed counterparts. 
Here, I briefly introduce the MPM  developed to constrain gravitationally induced time delay for short flares 
using predictions of the lens model  \citep{2015ApJ...809..100B}.  

The gamma-ray flux before and after isolated flares corresponds to a quiescent state.
Thus, bins with a more prolonged integration time are necessary to have enough photon statistics to build a light curve including time before and after the flare.
The short duration of gamma-ray flares relative to the expected time delays are an essential element in the analysis of unresolved gamma-ray light curves. 
The echo flares can be searched for in successive bins directly because gamma-ray flares can be identified as distinct events in the time series, 
and the range of expected time delays and corresponding magnification ratios can be predicted using a model of the lens.  

In lensing systems with two lensed images, the brighter lensed image is always followed by less luminous echo flare.
Thus, the method starts by identifying the first brightest flare.
Next, the flux ratio between the bin with the largest flux (the flare) and flux in successive bins is calculated.
The flux ratio is a proxy for magnification ratio between lensed images. 
Thus, the flux ratio can be used to compare with the lens model predictions. 

The MPM  enables us to extract additional physical constraints from the time series.
As mentioned, the flux ratios constrain the magnification ratios which are not accessible by the DPS method. 
Thus, the calculated flux ratios at different times from the flare (the bin with the largest flux) are compared to the magnification ratios  predicted by the lens model. 
This approach allows us to identify  time delays where the ratio of fluxes is consistent with the predicted magnification ratio. 
If there are bins consistent with the expected magnification ratio at given time from the brightest flare, the bins are identified as echo flares,
and their time since the brightest flare is used as a constraint on time delay. 
The details of the MPM with Monte Carlo evaluation are presented in \citep[ Appendix B]{2015ApJ...809..100B}.

Methods like the ACF or the DPS are very well suited for analyzing long periods of gamma-ray 
activity when light curves can be extracted with short binning ranging from 12 hours to 1 day. 
The MPM  complements the DPS method  by providing a method to constrain time delays for isolated flares.

%%%%%%%%%%%%%%%%%%%%%%%%%%%%%%%%%%%%%%%%%%%%%%%%%%%%
\subsection{ Monte Carlo Simulations \label{sec:MC}}
%%%%%%%%%%%%%%%%%%%%%%%%%%%%%%%%%%%%%%%%%%%%%%%%%%%%

Monte Carlo simulations are a traditional and powerful tool for calibrating the analysis of time series. 
They are important in the case of sparsely sampled data and they 
are necessary for evaluating the significance of an apparent time delay detection (Vaughan 2005).

A crucial element of the analysis is evaluation of the statistical significance of the detection of a time delay. 
An observed modulation of the signal could be a real time delay or it could arise purely by chance. 
Monte Carlo simulations provide a way to compute the probability of detections as opposed to false positives.

 Monte Carlo simulations should be used to assess the chance of detecting a real signal at a given significance level. 
 For such purposes,  the simulated light curves contain artificial time delays with the appropriate magnification ratio. 
 For each combination of noise spectrum, time delay, and magnification ratio, the detectability depends on the analysis method.
 In some cases where, for example, the magnification range is large and the time series is too short, 
 some methods may not detect the time delay at all.

%%%%%%%%%%%%%%%%%%%%%%%%%%%%%%%%%%%%%%%%%%%%%%%%%%%%
\subsection{Case Study: PKS~1830-211 \label{sec:casePKS1830}}
%%%%%%%%%%%%%%%%%%%%%%%%%%%%%%%%%%%%%%%%%%%%%%%%%%%%

The first gravitationally-induced time delay at gamma rays was detected using the first two years of {\it Fermi}/LAT data of PKS~1830-211 during its quiescent state. 
The detected time delay at gamma rays is consistent with radio time delay indicating consistent spatial origin of the gamma ray and radio emission in the quiescent state. 
After a few years of quiescent state of emission observed by  {\it Fermi}/LAT, 
PKS~1830-211 underwent series of gamma-ray flares. 
Surprisingly, the time delays measured during gamma-ray flares do not match radio time delay. 
Multiple time delays from the source cannot be explained by any lens complexity
and point  to the complex source structure. 

Here, I review how the time delay approach was used to elucidate the spatial origin of gamma-ray flares of PKS~1830-211.
In Section~\ref{sec:PKS1830_Properties}, I describe properties of the PKS~1830-211 system, 
one of the brightest gravitationally lensed radio sources with complex radio jet structures. 
In Section~\ref{sec:PKS1830_Flares}, I summarize results of the time delay analysis of four gamma-ray flares.
In Section~\ref{sec:PKS1830_Origin}, I review how these time delays were used to find that  the origin of the first two  flare is consistent with the position of the radio core, 
and  the other two gamma-ray flares originated  along the relativistic jet at very large distance ($>\,$kpcs) from the core. 
The discussion is in Section~\ref{sec:Discussion}.

\begin{figure*}
%\vskip 1cm
\begin{center}
\includegraphics[width=15.4cm,angle=0]{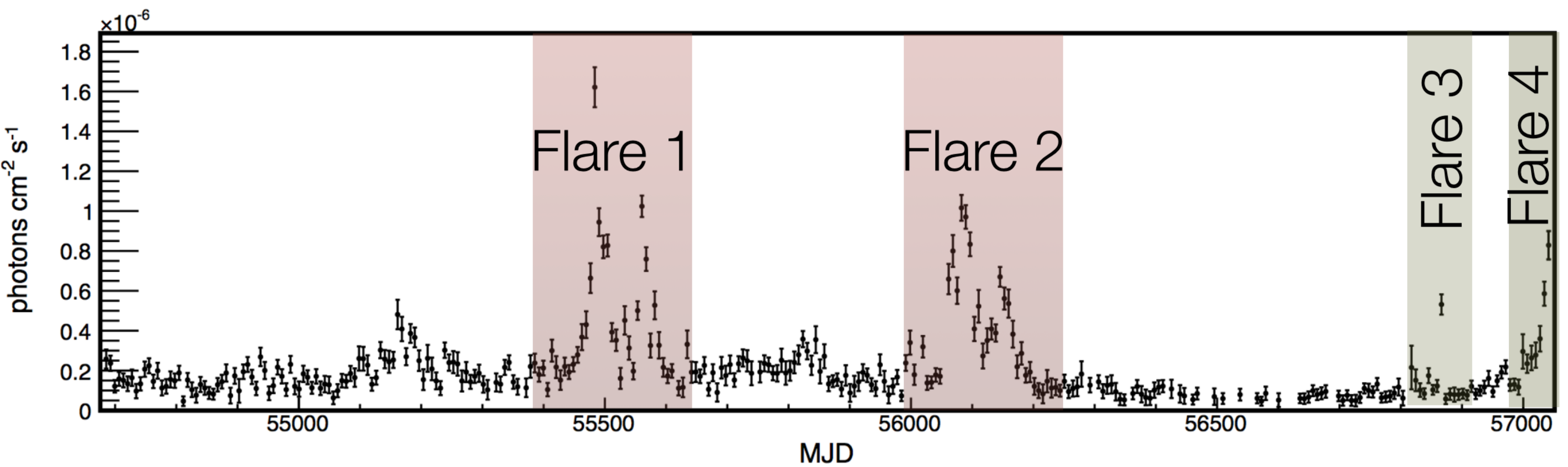}
\end{center}
\caption{\label{fig:lc_whole_PKS1830} 
                           {\it Fermi}/LAT light curve of PKS 1830-211 from August 2008 through February 2015.
                           The fluxes are shown in seven-day bins.
                           The energy range of selected gamma-ray photons is 200 MeV to 300 GeV \citep{2015ApJ...809..100B}. 
                           Time is represented in days since November 17, 1858 (Modified Julian Date: MJD).
                           Figure from \citep{2015ApJ...809..100B}.
                           }
\end{figure*}

%%%%%%%%%%%%%%%%%%%%%%%%%%%%%%%%%%%%%%%%%%%%%%%%%%%%
\subsubsection{Properties of the System \label{sec:PKS1830_Properties}}
%%%%%%%%%%%%%%%%%%%%%%%%%%%%%%%%%%%%%%%%%%%%%%%%%%%%

In 1991 the source PKS 1830-211 was identified as an unusually strong, $\sim10\,$Jy, 
gravitationally lensed system with extended emission forming a ring-like structure \citep{1991Natur.352..132J}. 
Figure~\ref{fig:pks1830} shows a radio image of PKS1830-211 with two bright lensed images of the radio core separated by roughly one arcsecond, 
and fainter lensed images of knots along the jet forming the ring-like structure. 
This peculiar ring-like structure reveals alignment of the jet to be roughly perpendicular to the line of separation of the two lensed images of the core. 
The lens responsible for bending light from this radio jet is a face-on spiral galaxy  located at redshift $z = 0.886$  \citep{1996Natur.379..139W,2001ASPC..237..155W, 2002ApJ...575..103W}. 
The radio jet is a part of a quasar located at redshift $z=2.507$ \citep{1999ApJ...514L..57L}. 
 
The lens splits emission of the source into two lensed images that travel through two paths of different length, traversing different gravitational potential. 
Emission from the fainter image arrives delayed in respect to the brighter image.
The radio monitoring program lasting for 18 months resulted in the measurement of this gravitationally induced time delay of $26^{+4}_{-5}\,$days,
and a magnification ratio between these two lensed images of $1.52\pm0.05$ \citep{1998ApJ...508L..51L}. 

Our understanding of bending properties of the gravitational lens relies on a reconstruction of the mass distribution of the lensing galaxy.
The mass distribution of the lens in the PKS~1830-211 system has been modeled \citep{2002ApJ...575..103W}.
The best model to date yields a singular isothermal ellipsoid (SIE)  with ellipticity $e=0.091$ and a lens oriented at 
$86.1^\circ$ \citep{Sridhar2013}.

The mass distribution of the lens is used to predict a time delay and magnification ratio for a given position of a source. 
Such constructed time delay map can be used to constrain  the position of the source once the time delay is measured.
The time delay map will allow us to constrain distances between emitting regions when multiple time delays are measured. 
%
%In the following sections, I show gravitationally induced time delays measured at gamma rays 
%and the potential of gravitational lensing to turn temporal information to constrain the spatial origin of gamma-ray emission. 

\begin{figure*}
%\vskip 1cm
\begin{center}
%-------------------------------- Flare 1--------------------------------%
\includegraphics[width=7.9cm,angle=0]{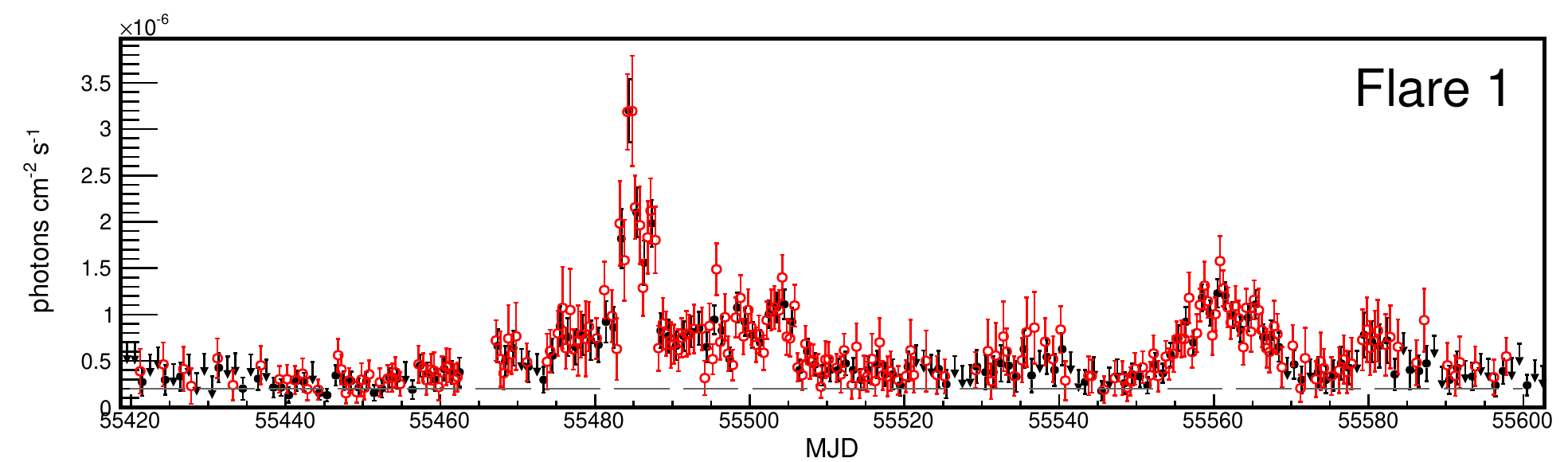} 
\includegraphics[width=7.9cm,angle=0]{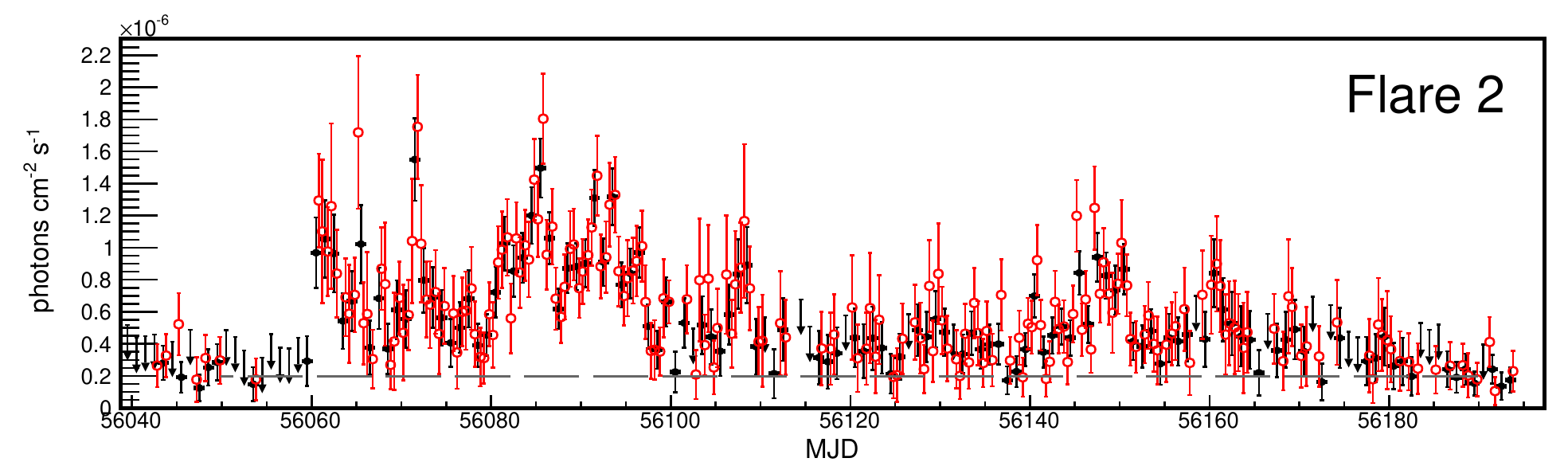} \\
\includegraphics[width=7.9cm,angle=0]{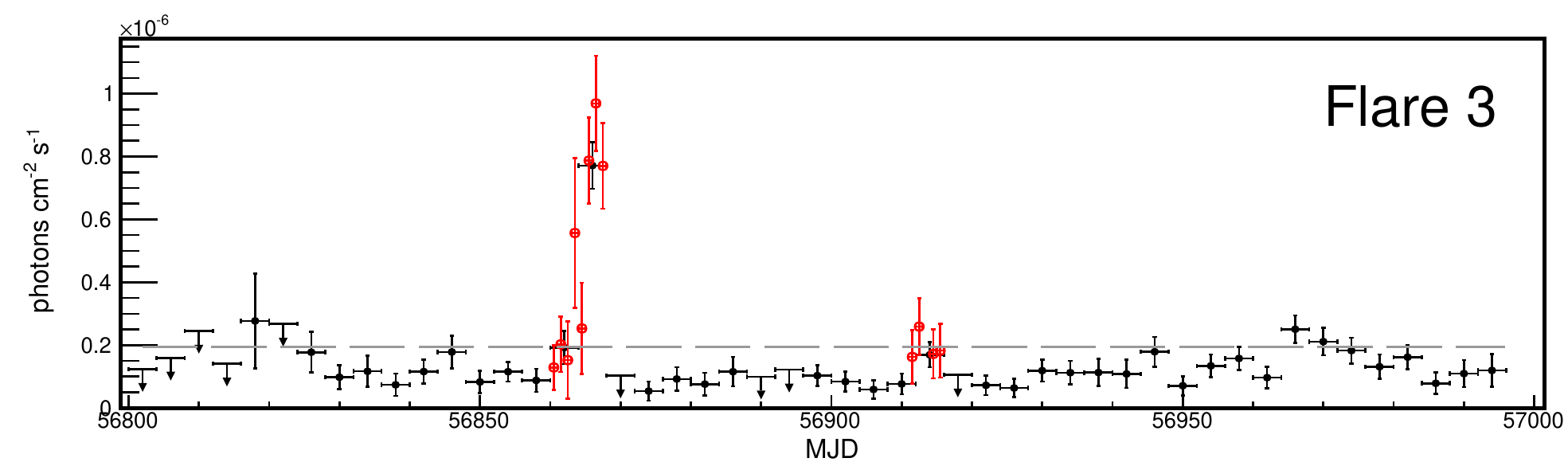} 
\includegraphics[width=7.9cm,angle=0]{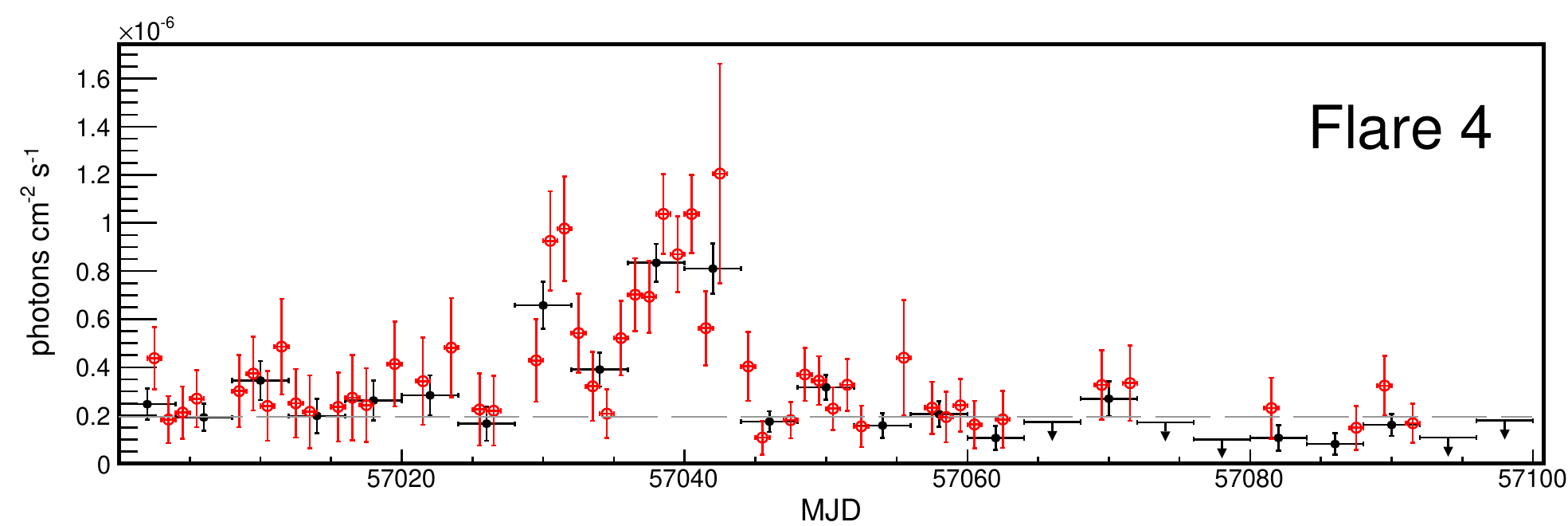} \\
\end{center}
\caption{\label{fig:lc_flares} 
                           {\it Fermi}/LAT light curves of flaring  activity of PKS 1830-211. 
                           Flare~1 and Flare~2 are shown using one-day binning  (black-filled circles),
                           and with 12-hour binning (red open circles). 
                           Flare~3 and Flare~4  are shown with  four-day binning  (black-filled circles),
                           and  one-day binning (red open circles). 
                           Figures from \cite{2015ApJ...809..100B}.
                           }
\end{figure*}

%-------- Gamma-ray flares ----------%
%%%%%%%%%%%%%%%%%%%%%%%%%%%%%%%%%%%%%%%%%%%%%%%%%%%%
\subsubsection{Time Delays at Gamma Rays}
\label{sec:PKS1830_Flares}
%%%%%%%%%%%%%%%%%%%%%%%%%%%%%%%%%%%%%%%%%%%%%%%%%%%%

Figure~\ref{fig:pks1830} shows the gamma-ray sky around the position of PKS~1830-211. 
Poor angular resolution at gamma rays forces integration of the signal from a significant fraction of the sky ($\sim1^\circ$).
The gamma-ray map demonstrates that PKS~1830-211 is an isolated source. 
As such, the gamma-ray photons extracted around the position of PKS~1830-211 are not contaminated by other sources.  
The lensed images of PKS~1830-211 are not resolved by {\it Fermi}/LAT. 
The observed gamma-ray flux is a sum of lensed images. 

The gamma-ray light curve of PKS~1830-211 is shown in Figure~\ref{fig:lc_whole_PKS1830}
with two $\sim100\,$days active periods (red area; Flares 1 and 2)
and two isolated individual flares (green area; Flares 3 and 4) 
These active periods are presented individually in Figure~\ref{fig:lc_flares}. 

The temporal behavior of PKS~1830-211 can be divided into extended quiescent states and flaring periods. 
The quiescent state is characterized by stochastic variability with power-law noise of $1/f^{1.24\pm0.12}$ \citep{2015ApJ...799..143A,2015ApJ...809..100B}. 
Such variability can be described with a pink noise model.  
The flaring state is a period of rapid variability with time scales ranging from days to minutes. 
Such rapid variability is typical of blazars.

\begin{figure}
\begin{center}
\includegraphics[width=5.cm,angle=-90]{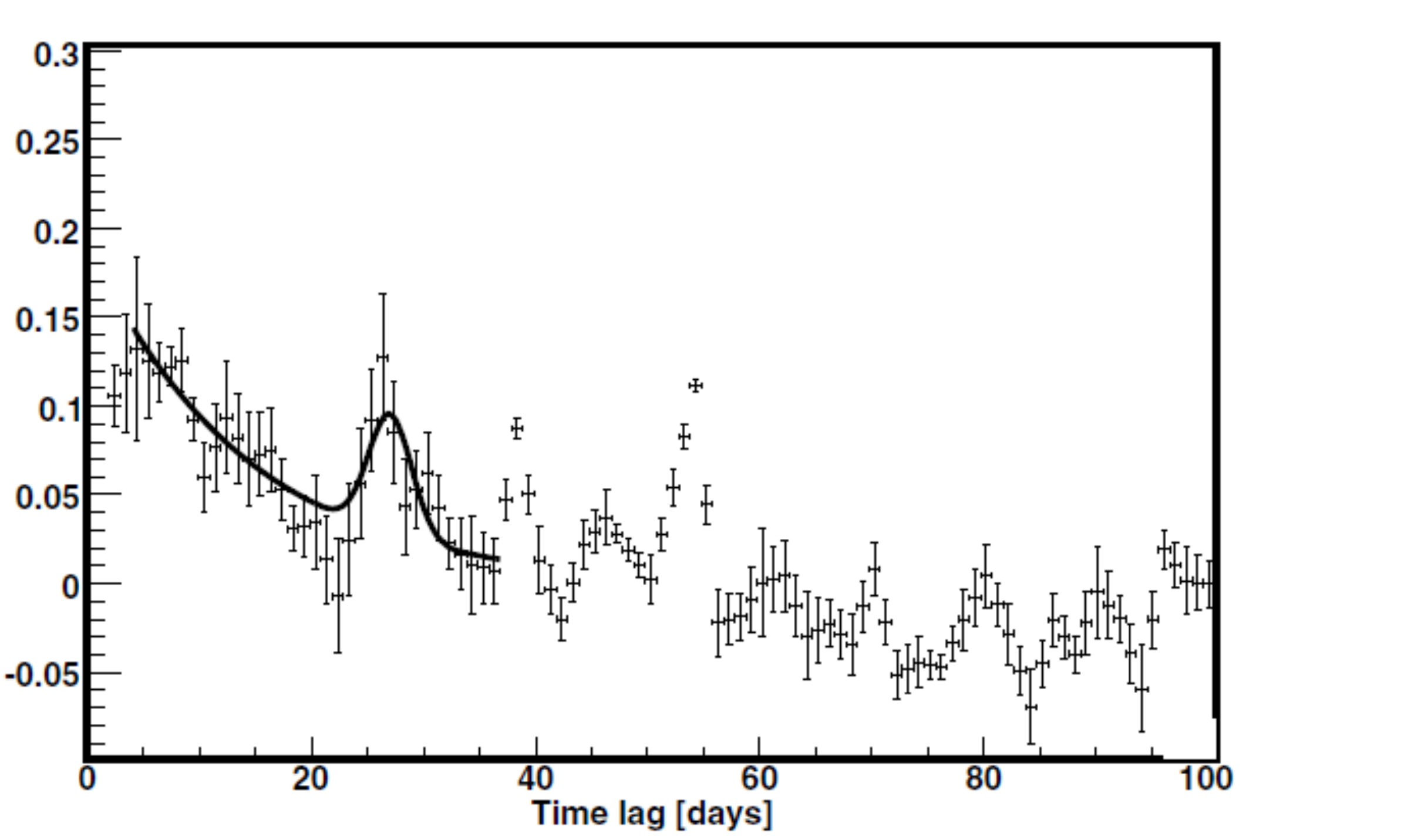}
\end{center}
\caption{\label{fig:FirstTD} 
			Autocorrelation function of PKS~1830-211 during quiescent state observed by {\it Fermi}/LAT from 2008 until 2010.
                           The solid line is a fit of an exponential function representing intrinsic variability of PKS~1830-211 with added  Gaussian profile representing time delay.
                           Figure from \cite{2011A&A...528L...3B}.
                           }
\end{figure}

Long and evenly sampled light curves provided by {\it Fermi}/LAT are ideal for applying Fourier transform based techniques to measure time delays. 
The gamma-ray light curve of PKS~1830-211  in its quiescent state monitored by the {\it Fermi} satellite from
its launch in August 2008 until November 2010 was used to detect the first gravitationally-induced time delay at gamma rays \citep{2011A&A...528L...3B}. 
This time delay was found using the autocorrelation function and the double power spectrum method reviewed in Section~\ref{sec:method},
and using the signal processing procedure outline in Section~\ref{sec:processing}.
The DPS method resulted in a time delay of $27.1\pm 0.6\,$days at a $4.2\sigma$ level \citep{2011A&A...528L...3B}.  
The results obtained using the autocorrelation function are in a perfect agreement with the DPS method,
and are shown in Figure~\ref{fig:FirstTD}.
However, the autocorrelation method results in much lower significance of the detection due to contamination from the intrinsic variability of the source 
present as an exponential component in Figure~\ref{fig:FirstTD}.

Starting at the end of 2010, PKS~1830-211 went through series of vigorous outbursts highlighted in Figure~\ref{fig:lc_flares}. 
Flares~1 and~2 have temporal behavior characterized by a set of very bright flares. 
The autocorrelation function of Flare~1 reported in \cite{2015ApJ...809..100B} shows a broad feature at a time delay of $17.9\pm7.1\,$days at $\sim2\sigma\,$level, 
which agrees with the time delay of $19\pm1\,$days obtained by \cite{2015ApJ...799..143A} also using  the autocorrelation function. 
The time delay detected during Flare~1 using the DPS method is $23\pm0.5\,$days above $2\, \sigma$ level \citep{2015ApJ...809..100B}.
To further investigate whether the time delay detected during Flare~1 using the DPS method is indeed gravitationally induced 
\cite{2015ApJ...809..100B} used the MPM method (see Section~\ref{sec:MPM}), which combines the observations with predictions of the lens model. 
If the time delay is  gravitationally induced then the ratio of fluxes between 
the brightest peaks and fluxes  at  periods after the peak corresponding to the time delay should be consistent with the magnification ratio 
expected from the model of the lens. 
The magnification ratio of $\sim 2$ is expected for the time delay of $\sim20\,$days if emission originates along the radio jet. 
Following the two largest outbursts in Flare~1, 
it is found that successive periods corresponding to the time delay do have consistent flux ratios with magnification ratio expected from the model of the lens. 
This strengthens evidence that the time delay of  $23\pm0.5\,$days detected during Flare~1 is induced by gravitational lensing. 

The significance of the time delay detection is an essential tool to distinguish a gravitationally induced time delay from a spurious fluctuation. 
Sampling and photon statistics during gamma-ray flares are limited but at the same time very well defined. 
The well defined time series allows us to reconstruct the signal using Monte Carlo simulations to investigate expected performance of different methods of time delay estimation. 
Including properties of the time series such as duration, binning, and spectral index, \cite{2015ApJ...809..100B} performed a million MC realizations of flare-like events. 
This approach allows us to distinguish real signal from spurious fluctuations. 

Simulated time series with properties of the real light curves and artificially  induced time delays with magnification ratios consistent with the lens model
 allows us also to predict significance of the time delay detection. 
Taking into consideration limitations of gamma-ray  observations of flares of PKS~1830-211 and using the Monte Carlo simulations, 
\cite{2015ApJ...809..100B}  estimated 90\% probability to detect time delay of $\sim 20\,$days at $2\sigma\,$level, 
 75\% probability to detect it at $3\sigma\,$level, and below 40\% chance to detect it above $4\sigma\,$level. 
 Thus, the significance of the detection of the time delay of  $23\pm0.5\,$days using the DPS method is consistent with the expectations. 
 Note that such approach can be used as an auxiliary approach to eliminate time lags that are not due to gravitational lensing.

Now reviewing Flare~2, the time delay reported by \cite{2015ApJ...809..100B}  using the DPS method was $19.7\pm1.2\,$days,
and is detected at significance level greater than $3\, \sigma$. 
The ACF yielded consistent values, but at lower significance which is consistent with the predictions of the Monte Carlo simulations. 
The MPM also provide a confirmation that this time delay is consistent with the lens model. 

Flares~3 and~4 were very short and  lasted for only  a few days. 
The light curve extracted around these periods do not have enough photons to effectively apply the Fourier transform based methods,
such as ACF or DPS. 
However, such isolated flares are perfect for methods searching directly for delayed counterparts, like the MPM. 

During Flare~3, the emission increased by a factor of 5 relative to the average flux.  
The maximum peak method shows  the time delay range consistent  with the expected magnification ratio in time delay range of 46-50 days. 
At such time delays, the predicted magnification ratio is $\sim5$. 
Longer time delays would result with greater magnification ratio between images, 
and flux of the echo flare would be below quiescent emission. 
Thus, echo flares with time delay greater than 48 days would not be detectable for Flare~3. 
As a result, Flare 3 must have a time delay equal to or larger than 48~days. 

\begin{figure*}
\begin{center}
\includegraphics[width=14.cm,angle=0]{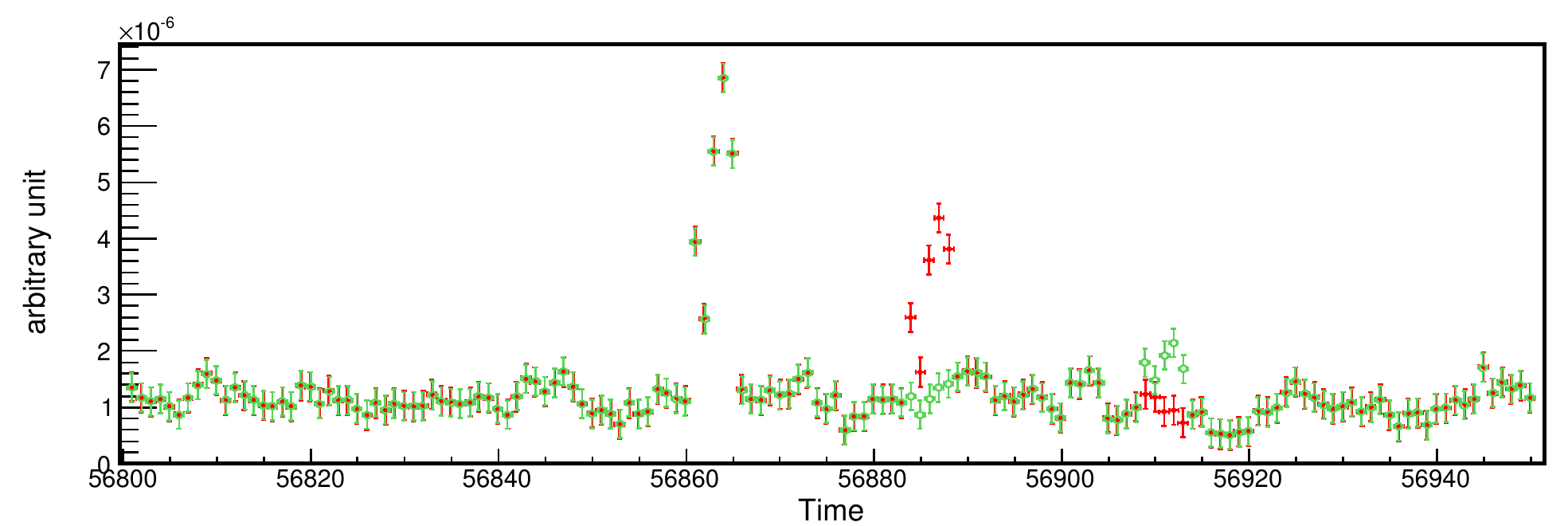}
\end{center}
\caption{\label{fig:MCSingle} 
			Monte Carlo simulation of Flare-3-like event.
			Green points  indicate artificial light curve with a time delays of 48~days and magnification ratio of 4.5.
			Red points is the same artificial light curve but with 23~days lag and magnification ratio of 1.8. 
			Arrows set boundary on the projection of relativistic jet. 
                           Figure from \cite{2015ApJ...809..100B}.
                           }
\end{figure*}

The time delay expected from the radio core is in the range  20-30 days with magnification ratio $\sim1.5$. 
If Flare~3 originated from the region close to the radio core, 
the echo flare would appear with a flux at least twice the average gamma-ray emission of PKS~1830-211. 
Figure~\ref{fig:MCSingle} shows a simulation how gamma-ray light curve would look like if there was $23\,$day time lag.
The {\it Fermi}/LAT continuously observes the gamma-ray sky. 
Thus, it would be impossible to not detect the gamma-ray echo of Flare~3 if the time delay was shorter than at least 30 days. 
The absence of detection of this echo flare 20 to 30 days after  Flare~3 provides very strong evidence 
that the Flare~3 does not originate from the radio core region. 

Similarly, Flare~4 shows lack of expected time delays, which indicates that Flare~4, like Flare~3, 
does not originate from the radio core. 
 \cite{2015ApJ...809..100B} show that  the time delay must be greater than $\sim50\,$days. 
 
%%%%%%%%%%%%%%%%%%%%%%%%%%%%%%%%%%%%%%%%%%%%%%%%%%%%
\subsubsection{Spatial Origin of Gamma-ray Flares}
\label{sec:PKS1830_Origin}
%%%%%%%%%%%%%%%%%%%%%%%%%%%%%%%%%%%%%%%%%%%%%%%%%%%%

\begin{figure}
%\vskip 1cm
\begin{center}
\includegraphics[width=6.9cm,angle=0]{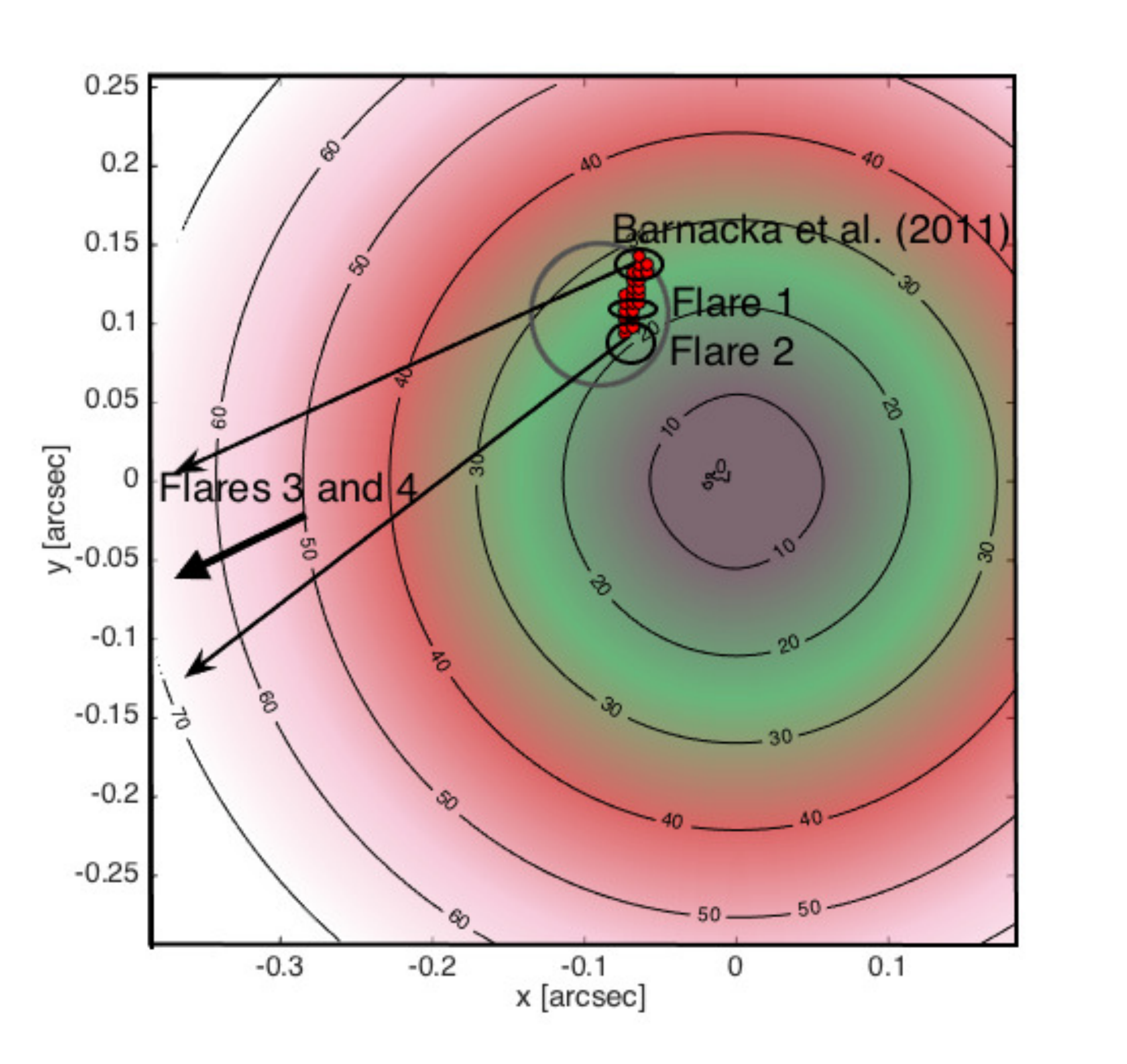}
\end{center}
\caption{\label{fig:jet_resolved} Time delay map with resolved positions of gamma-ray emission. 
						The zero time delay corresponds to the lens center. 
						 The color pallet and contours indicate the time delays are in days.
						  Figure from \cite{2015ApJ...809..100B}.}
\end{figure}

Equipped with time delays measured at gamma rays and a time delay map, 
we can map the origin of gamma-ray emission. 
%
%Here we review a determination of the spatial origin of emission using time delay map. 
%The time delay as a function of the position of the source in the lens plane can be determined using a model of the lens. 
Figure~\ref{fig:jet_resolved} shows the time delay map obtained using the model of the lens reconstructed by \citep{Sridhar2013}. 
The region of the time delay map consistent with the radio time delay of 20-30 days is shown in green. 
The time delay alone does not provide unequivocal localization of emitting region. 
The position of the core and the alignment of the jet in the lens plane is required to obtain robust constraints on the spatial origin of the emission. 
The position of the lensed images can be used to put additional constraints on the location of the radio core. 
The positions of the lensed images of the radio core reduce constraints on the position of the radio core to the gray circle in Figure~\ref{fig:jet_resolved}. 
The red filled points  are further constraints  on the position of the radio core using the lens model and 
the time delay and magnification ratio measurements by \cite{1998ApJ...508L..51L}.

The jet alignment is limited by the well-resolved radio ring-like structure of PKS~1830-211,
and long arrows indicate its boundaries. 
Once the alignment of the jet in the lens plane is reconstructed,
the time delay map can be used to locate regions along the jet using measured time delays.
The black ellipses show predicted origin of gamma-ray emission along the jet.
The time delay of $27.1\pm0.5\,$days measured by \cite{2011A&A...528L...3B}
using the gamma-ray light curve in the quiescent state is indicated by the top ellipse.
The following two ellipses indicate positions of  Flare~1 and Flare~2 based on the measured gamma-ray time delays. 
The spatial origin of Flare~1 and Flare~2  is consistent with the position of the radio core. 
The short arrow indicates constraints from Flares~3 and 4. 
The time delays $\gtrsim 50$~days imply that  
 the emitting region must be located at a projected distance of $\gtrsim$ 1.5~kpc from the radio core. 

%%%%%%%%%%%%%%%%%%%%%%%%%%%%%%%%%%%%%%%%%%%%%%%%%%%%
\subsection{Discussion \label{sec:Discussion}}
%%%%%%%%%%%%%%%%%%%%%%%%%%%%%%%%%%%%%%%%%%%%%%%%%%%%

The time delay approach allows us to resolve the origin of emission using temporal resolution spatially. 
This approach is especially beneficial for high energy observations, 
where angular resolution is intrinsically limited by challenges of building X-ray and gamma-ray satellites,
including small effective area and physical processes like nuclear recoil. 

The time delay approach applied to the case study of gamma-ray observations of PKS~1830-211 revealed the spatial origin of four flares. 
Flares~1 and~2 originate from a region of $\sim$100~pc around the core.
At the redshift of $z=2.507$, where PKS~1830-211 is located, a projected distance of 100~pc corresponds to $\sim0.02\,$arcsecond,
which improves the angular resolution times at gamma rays $\sim10000$.

Intriguingly, the time delay analysis shows that Flare~3 and~4 originated at least $1.5\,$kpc from the central engine, 
indicating that gamma-ray flares can be produced at multiple emitting regions at vast distances from the central engine.
The origin of gamma-ray flares is a subject of intense debate  \citep{2014ApJ...789..161N,2010MNRAS.405L..94T}.
In  theoretical modeling, it is generally assumed that  gamma-ray flares originate from 
 regions close to the central engine, typically on  parsec scales 
 \citep{2014ApJ...796L...5N,2014A&A...567A.113B,2014ApJ...789..161N,2015MNRAS.448.3121H,2011ApJ...733...19T}.  
Thus, the spatial origin of  Flare~3 and~4 challenges our understanding of physical mechanisms responsible for producing variable emission in relativistic jets. 

Multiple variable emitting regions place limitations on the use of the most variable quasars for measurement of the Hubble parameter based on time delays.
 \cite{2015ApJ...799...48B} point out that even a small spatial offset, for example $5\%$ of Einstein radius,  
 between the resolved position of the core and  site of variable emission may result in
 a bimodal distribution of values of Hubble parameters characterized by an RMS of  $\sim12\, [\mbox{km}\, \mbox{s}^{-1}\,\mbox{Mpc}^{-1}]$. 
The complex structure can be an essential source of systematics in the measurement of the Hubble parameter from gravitationally induced time delays. 

A limiting factor in any lensing analysis is the precise model of the lens and alignment of the jet. 
In the case of PKS~1830-211 very conservative position of the core and jet alignment was used. 
More detail analysis of radio observations can yield better constraints on the positions of the radio core and knots along the jet. 
Improvement in the accuracy of the measurement of the radio time delay from $5\,$days down to $0.5\,$days 
could provide localization of gamma-ray emission in respect to radio core to within 10$\,$ pc. 

The case study presented in this section focused on applying the time delay approach to gamma-ray observations. 
However, the technique can be extended to other wavelengths, for example, optical and X-ray observations, 
where variable gravitationally lensed sources can be monitored and time delays with high precisions can be measured. 
The time delay accuracy defines the spatial resolution.
For example, for a source observed at a redshift of 1, accuracy in the time delay of 3~hours allows resolving the emission down to  1~mas. 
The time delay approach can be applied to long-term monitoring data at any wavelength. 
Thus, the time delay method has a potential of providing resolution of radio telescopes at any frequency, 
for variable gravitationally lensed sources. 
Moreover, Euclid, LSST, or SKA will monitor many variable gravitationally lensed sources,
thus enabling a probe of the nature and evolution of radiation from these sources.

%%%%%%%%%%%%%%%%%%%%%%%%%%%%%%%%%%%%%%%%%%%%%%%%%%%%
\section{The Hubble Parameter Tuning Approach \label{sec:HPT}}
%%%%%%%%%%%%%%%%%%%%%%%%%%%%%%%%%%%%%%%%%%%%%%%%%%%%

Challenges of resolving multi-wavelength emission of the inner regions of active galaxies led us to an assumption 
that a well-resolved position of the compact radio core traces the location of SMBH. 
Here, I review the Hubble Parameter Tuning (HPT) approach, 
which combines well-resolved positions of the lensed images, gravitationally-induced time delays, 
and precise cosmology to pinpoint the spatial origin of emission. 
The HPT method applied to gravitationally lensed blazar B2~0218+35 provided evidence
that the radio core for some of the distant sources can be more than dozens of parsecs from the central engine. 
Such an offset between the radio core and the SMBH questions our understanding of the physical origin of the radio core and its connection to gamma-ray flares. 

Gravitational lensing combined with a new generation of scientific instruments gives us new paths to explore physical phenomena present in inner regions of active galaxies.   
Radio telescopes with their excellent angular resolution ($<\,$mas) image detailed structures of relativistic jets.
However, these very well-resolved radio observations cannot be directly compared with observations at higher frequencies 
due to the insufficient astrometric accuracy or angular resolution of the latter. 
Thus, it has been assumed that the well-resolved radio structures must be located close to the SMBH. 

The time delay approach allows us to find the spatial origin of emission of sources poorly resolved or even unresolved 
but variable with a long sample of data available. 
However, there might be no variability or not enough data to measure time delays. 
Measuring time delays is observationally demanding and challenging. 
Thus, in many circumstances, precise measurement of the time delay might be difficult or even impossible. 
For example, at radio frequencies sources are less variable in comparison to higher frequencies. 
%The processes responsible for emission in active galaxies result in less rapid variability 
The smaller variability impedes precise measurements of radio time delays and limits the application of the time delay approach.  
%In addition, observational gaps in the data due to weather may introduce systematics to these time delay estimations. 
However, radio telescopes have excellent angular resolution.
The positions of lensed images also depend on the source location, as well as, gravitationally-induced time delays. 
%which in principle could be used instead of radio time delays to investigate the origin of emission.  
The HPT method allows us to combine well-resolved position of lensed images and gravitationally-induced time delays 
to investigate the spatial relation of multi-wavelength emission.  

%%%%%%%%%%%%%%%%%%%%%%%%%%%%%%%%%%%%%%%%%%%%%%%%%%%%
\subsection{Approach}
%%%%%%%%%%%%%%%%%%%%%%%%%%%%%%%%%%%%%%%%%%%%%%%%%%%%

The combination of gravitationally-induced time delays 
and the model of the lens
has traditionally been used to measure the Hubble parameter.
Gravitationally-induced time delays are fundamental measurements in cosmology. 
In principle, they provide a measurement of  the Hubble parameter 
independent of the distance ladder 
\citep{1964MNRAS.128..307R,1997ApJ...475L..85S,2002MNRAS.337L...6T,2002ApJ...578...25K,2003ApJ...599...70K,2007ApJ...660....1O,2013ApJ...766...70S,2014MNRAS.437..600S}. 

The positions and magnifications of lensed images are dimensionless.  
While, time delays are proportional to the ratio of distances (see Equation~\ref{eq:D}), 
where the distances scale with the Hubble distance, $D_H$ (see Section~\ref{sec:Probability}). 
The Hubble distance is inversely proportional to H$_0$. 
As a result, the Hubble parameter enters into the time delay calculation (Section~\ref{sec:TimeDelay}).

Time delay depends on the source position and the model of the lens. 
For the SIS mass distribution, the model of the lens reduces to the positions of the lensed images. 
The positions of the lensed images also depend on the source location. 
It has been traditionally assumed that the positions of the lensed images and time delay corresponding to the same source position. 
If the time delay and resolved position of the lensed images indeed correspond to the same source position 
then the reduced Hubble parameter $h$, introduced in Section~\ref{sec:Probability}, 
can be obtained using Equation~\ref{eq:tdSIS} with the distance ratio defined as $D=d/h$
\begin{equation}
\label{eq:h}
h = \frac{d(1+z_L)(\theta_B^2-\theta_A^2)}{2c\,\Delta t}\,.
\end{equation}

The three essential ingredients of Equation~(\ref{eq:h}) are:
the reduced Hubble parameter ($h$), the positions of the lensed images ($\theta_A$ and $\theta_B$), and the time delay between the images ($\Delta t$).
Monitoring of lensed sources is observationally demanding and is usually performed with instruments with limited angular resolution. 
As a result, it is unknown if the observed variability corresponds to well-resolved radio structures of the source. 
Due to variability timescales, it has been commonly assumed that the variable emission must be produced at small, $<\,$pc, 
from the radio core. 
However, if variability used to measure time delay originates from a different location than the well-resolved lensed images, 
then Equation~(\ref{eq:h}) will return incorrect value of the Hubble parameter. 
The systematic offset between the ``true'' and estimated value of the Hubble parameter will correspond to 
the offset between the source of variable emission and well-resolved emission. 

As has been demonstrated by \cite{2015ApJ...799...48B}, the value of the Hubble parameter measured based 
on gravitationally-induced time delays is very sensitive to the spatial offset between 
the position of the core and the position of the variable-emitting region where the time delay originates. 
However, the problem can be inverted and the value of H$_0$ measured by  other techniques
can be used to find the spatial offset between the position of well-resolved images of lensed jets
and the position of the variable-emitting region where the time delay originates.

Here, the steps in applying the HPT approach are briefly reviewed. 
If  the variable emission originates from the same region as resolved lensed images then Equation~(\ref{eq:h})  will result in ``true'' value of H$_0$.
However, if there is an offset between the radio core and the variable emitting region, 
the Hubble parameter derived from the time delay will differ from the independently measured ``true'' value.
This difference between the  ``true'' and estimated value of H$_0$ depends on the distance between the radio core and the spatial location of the variable emission. 
The offset in  the  ``true'' and estimated value of the Hubble parameter corresponds to the spatial offset between resolved radio core and site of variable emission.

The Hubble parameter, measured with a variety of independent methods, provides a route to exploring 
the spatial origin of the emission of quasars. 
The precisely measured Hubble parameter can be used to evaluate the offset 
between the radio core and for example the site of the gamma-ray emission.
The significance of the separation of emitting regions is limited by the uncertainty in the H$_0$ measurement.
This method  called the Hubble Parameter Tuning (HPT) approach was proposed in \cite{2015ApJ...799...48B}. 
In the following Section~\ref{sec:CaseB20218}, I review the application of the HPT approach
based on observations of  B2~0218+35.

%%%%%%%%%%%%%%%%%%%%%%%%%%%%%%%%%%%%%%%%%%%%%%%%%%%%
\begin{figure}
%\vskip 1cm
\begin{center}
%Flare1_ACF.eps  UL not included in the CL
%/dane/Lensing/data/PKS1830/whole/mag_vs_distance_SIE.gp
\includegraphics[width=9.5cm,angle=0]{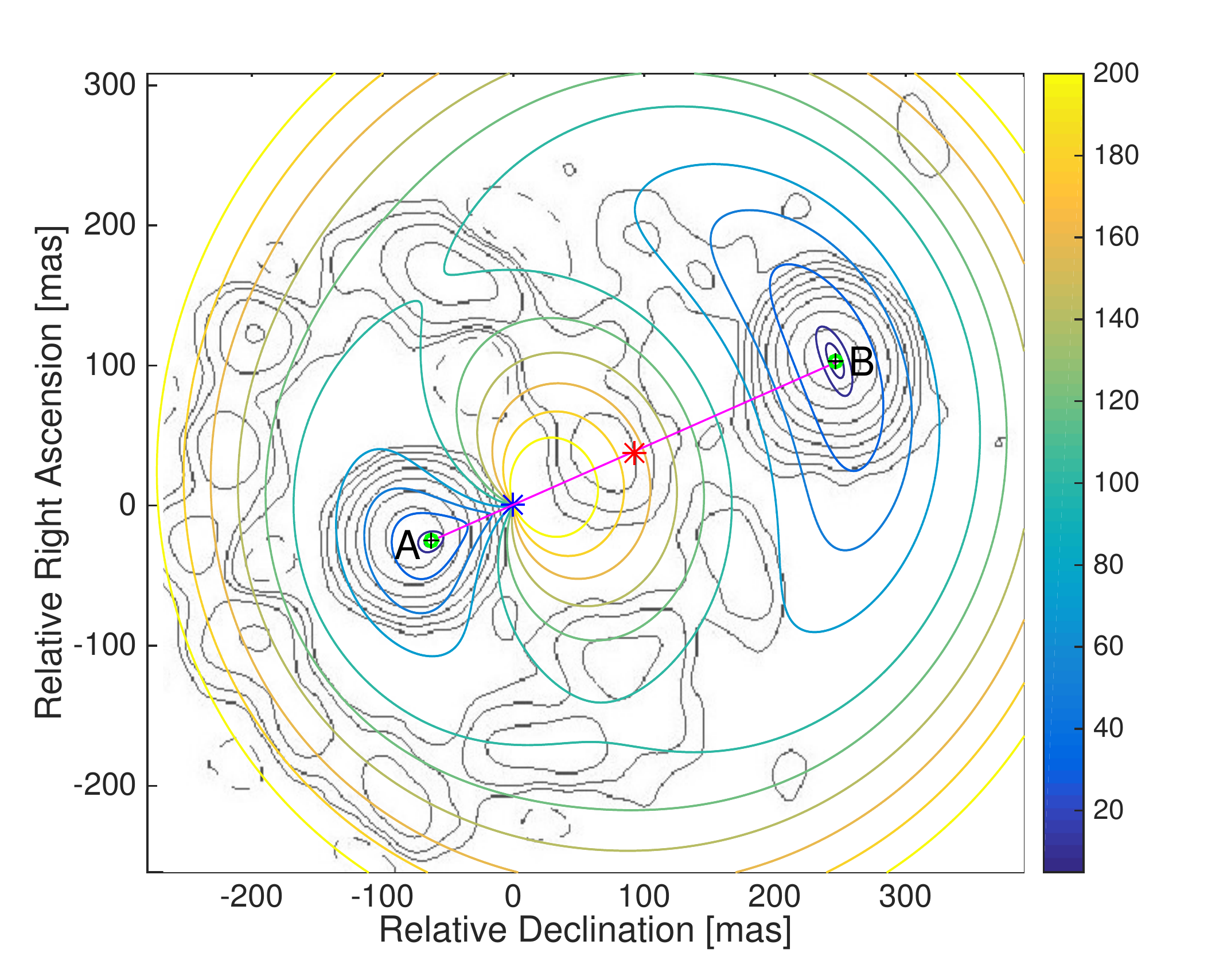}
\end{center}
\caption{\label{fig:Fermat}   Compilation of model and observation of the gravitationally-lensed blazar B2~0218+35. 
 					The structure of the lensed radio jet observed  at 1.687~GHz is shown as the gray contours.
					The Fermat surface is indicated as the color contours. 
					The lensed images form at the extrema of the Fermat surface. 
					The reconstructed positions of the lensed images are shown as black crosses.
					The observed positions of the 15~GHz lensed images of B2~0218+35 are shown as green open circles.
					The observed and reconstructed position of the lensed images show a perfect alignment.  
					The lens axis is shown as the magenta line. 
					The reconstructed positions of the lens and source as shown as the blue and red stars, respectively. 
					Figure from \cite{2016ApJ...821...58B}
					        }
\end{figure} 
%%%%%%%%%%%%%%%%%%%%%%%%%%%%%%%%%%%%%%%%%%%%%%%%%%%%
\subsection{Case Study: B2~0218+35 \label{sec:CaseB20218}}
%%%%%%%%%%%%%%%%%%%%%%%%%%%%%%%%%%%%%%%%%%%%%%%%%%%%

 B2~0218+35 is an example of a bright gravitationally-lensed blazar with variable emission detected from radio up to gamma rays. 
 Gamma-ray monitoring detected a series of bright gamma-ray flares. 
 This review will focus on two gamma-ray outbursts.
 Radio observations provide precise positions of the lensed images and gives strong constraints on the lens model. 
 Considering the above,  B2~0218+35 is a perfect target to apply the HPT method and to test if gamma-ray flares originate from the radio core. 
 
 Section~\ref{sec:propertiesB20218} presents a brief overview of B2~0218+35. 
Next in Section~\ref{sec:LensModel},  the approach to the lens modeling is outlined. 
Section~\ref{sec:B2Flare1} reviews gamma-ray time delay measurements, 
which, in Section~\ref{sec:OriginFlare1} is used as a base for the HPT approach 
that maps spatial offset between radio core and the origin of Flares~1. 
Section~\ref{sec:HPToffset} shows a possible interpretation of the offset. 
Section~\ref{sec:OriginFlare2} reveals the spatial origin of Flare~2. 
Section~\ref{sec:ConFlare12} examines the possibility that Flares~1 and~2 are connected events.
Section~\ref{sec:NotesH0} reviews the current measurements of H$_0$ and its implications for the HPT approach.  

%%%%%%%%%%%%%%%%%%%%%%%%%%%%%%%%%%%%%%%%%%%%%%%%%%%%
\subsubsection{Properties of the System \label{sec:propertiesB20218}}
%%%%%%%%%%%%%%%%%%%%%%%%%%%%%%%%%%%%%%%%%%%%%%%%%%%%

B2~0218+35 is a gravitationally-lensed system with the smallest known Einstein radius ($165\,$mas) \citep{1992AJ....104.1320O,1995MNRAS.274L...5P}. 
The system consists of a bright blazar  at redshift $z_S=0.944\pm 0.002$ \citep{2003ApJ...583...67C}.
This blazar is lensed by an apparently isolated galaxy at redshift $z=0.6847$ \citep{1993MNRAS.263L..32B}.
The lens bends the radio emission of the jet into two bright images of the core and extended structures, including an Einstein ring 
\citep{1992MNRAS.254..655P,1992AJ....104.1320O,1993MNRAS.261..435P,1995MNRAS.274L...5P,2000MNRAS.311..389J,2001MNRAS.322..821B,2003MNRAS.338..599B}.
These structures of the lensed radio jet observed at 1.687 GHz are shown in Figure~\ref{fig:Fermat} as gray contours. 

The radio time delay measurement  has been reported in the literature on three occasions. 
The first measurement of the radio time delay 
used VLA $15\,$GHz polarization observations and yielded a value of  $12\pm3\,$days \citep{1996IAUS..173...37C}.
The second time, high-precision VLA flux density measurements were used over the same epoch as \citep{1999MNRAS.304..349B},
 and resulted in the time delay of $10.1^{+1.5}_{-1.6}\,$days  \citep{2000ApJ...545..578C}. 
The third measurement resulted in  the time delay of $10.5 \pm 0.4\,$days and was based on a three-month VLA monitoring campaign at two frequencies \citep{1999MNRAS.304..349B}. 

Since discovery of B2~0218+35, the system has been considered as a ``golden lens" for Hubble constant measurement \citep{2004MNRAS.349...14W}.
However, despite precise measurements of the time delay, 
a clean lens environment without nearby companions or a surrounding cluster, 
and a negligible number of  structures along the line of sight which would complicate the modeling of the lens,
the H$_0$  values derived from this system are in the range 61-78~$\mbox{km\,s}^{-1}$Mpc$^{-1}$ 
\citep{2005MNRAS.357..124Y,2000ApJ...536..584L,2004MNRAS.349...14W}.
The most recent attempt to measure H$_0$ for B2~0218+35,  using a time delay of $11.46\pm0.16\,$days based on gamma-ray emission,
results in a Hubble constant of $64\pm4$~$\mbox{km\,s}^{-1}$Mpc$^{-1}$ \citep{2014ApJ...782L..14C}.
This large scatter in the H$_0$ values can indicate 
a complex structure of B2~0218+35 as anticipated in \citep{2015ApJ...799...48B}. 

%To investigate the detailed source structure, we review constraints on the lens model.
%%%%%%%%%%%%%%%%%%%%%%%%%%%%%%%%%%%%%%%%%%%%%%%%%%%%
\begin{figure*}
%\vskip 1cm
\begin{center}
\includegraphics[width=17.cm,angle=0]{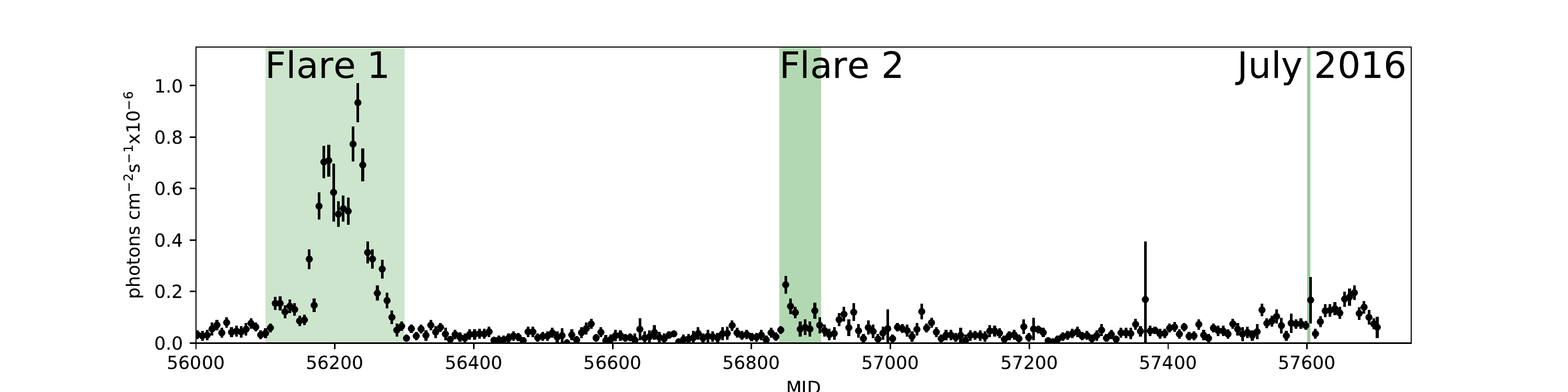}
\includegraphics[width=17.cm,angle=0]{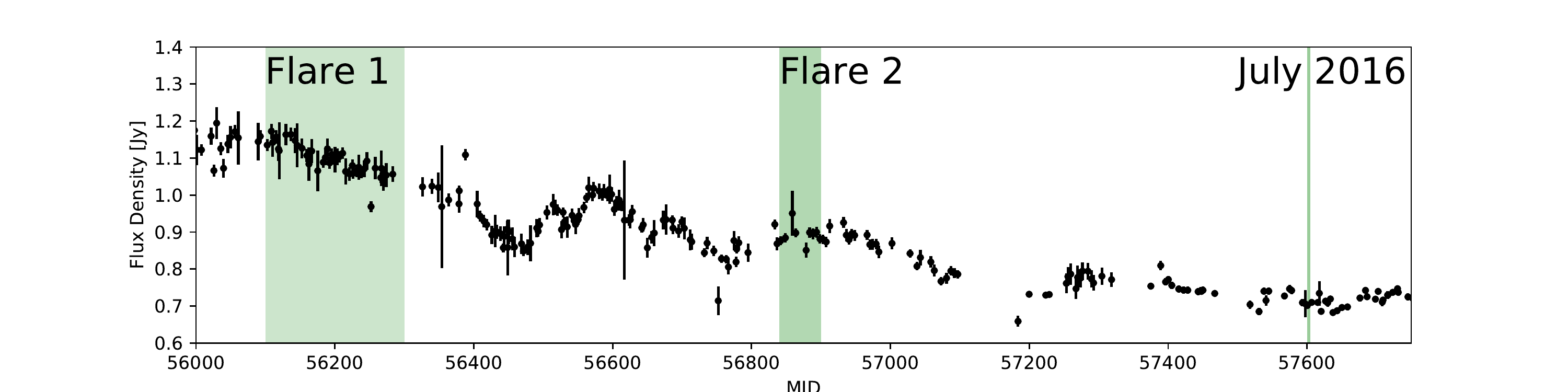}
\end{center}
\caption{\label{fig:lc_whole_B0218} 
                         {\bf Top: Gamma Rays} 
                         {\it Fermi}/LAT light curve of B2~0218+35. 
                           The light curve shows time period from March 2012 through November 2016.
                           The fluxes are based on seven-day binning.
                           The gamma-ray photons are selected in the energy range from 200 MeV to 300 GeV. 
                            {\bf Bottom: Radio}
                            15~GHz light curve of a gravitationally lensed blazar B2~0218+35 from the OVRO 40-m monitoring program. 
                           Time is represented in days since November 17, 1858 (Modified Julian Date: MJD).
                          % Figure from \cite{2016ApJ...821...58B}. 
                           }
\end{figure*}

%%%%%%%%%%%%%%%%%%%%%%%%%%%%%%%%%%%%%%%%%%%%%%%%%%%%
\subsubsection{The Lens Model \label{sec:LensModel}}
%%%%%%%%%%%%%%%%%%%%%%%%%%%%%%%%%%%%%%%%%%%%%%%%%%%%

Here, I give a brief overview how the parameters for the B2~0218+35 system are constrained following the analysis presented in \cite{2016ApJ...821...58B}. 
B2~0218+35 is an ideal system for lens modeling. 
The observations with the Hubble Space Telescope (HST) show that the lensing galaxy is isolated which indicates a clean and straightforward gravitational lens potential. 
Previous lens models confirm that the mass distribution of the lens is consistent with a Singular Isothermal Sphere (SIS) model \citep{2004MNRAS.349...14W,2011AstL...37..233L}.
Most of the previous studies of B2~0218+35 focused on the properties of the lensing galaxy and measurement of the Hubble parameter \citep{2005MNRAS.357..124Y}.

Here, the goal is to use the lens as a high-resolution telescope. 
The resolution of such gravitational telescope relies on precise reconstruction of the mass distribution of the lens.  
The accuracy of the reconstruction of the mass distribution of the lens can be defined as a difference between 
the observed and reconstructed positions of the lensed images giving observations of the lensed system. 
In the case of B2~0218+35, the goal is to reconstruct the mass distribution of the lens down to $\sim1\,$mas, 
which would allow us to find the origin of gamma-ray flares with such accuracy. 

The model of the lensed system includes parameters such as the Einstein radius, the lens and source positions and redshifts, the lens orientation and ellipticity, a slope for the mass distribution, etc.
The Einstein radius and the position of the source can be derived from the positions of the lensed images. 
The most precise positions of the lensed images of B2~0218+35 comes from radio VLBA observations at 15~GHz.
The positions of lensed images of the radio core are measured with 0.6~mas accuracy \citep[1994 Oct 3,][]{1995MNRAS.274L...5P}.
Figure~\ref{fig:Fermat} shows the positions of the lensed image B (brighter image located outside the Einstein ring; green circle) and image A (green circle). 
The Einstein radius is half the distance between the lensed images; 
$\theta_E = (\theta_A+\theta_B)/2 = 167.2\pm 0.6\,$mas  assuming a mass distribution close to a SIS. 
The Einstein radius gives us an estimate of the lens mass within one Einstein radius, 
which in the case of B2~0218+35 is  $\sim 2\times 10^{10}\,M_{\odot}$. 

The source is located at half the distance between the lensed images, $\theta_S = (\theta_A-\theta_B)/2$. 
The lensed images appear on the lens axis at distances of $\pm \theta_E$ from the source. 
The lens axis defined by the position of the source and the lens is shown as the magenta line in Figure~\ref{fig:Fermat}.
The source position may deviate from the lens axis if the mass distribution differs from an SIS. 
Thus the search for the true source position must allow a deviation from the estimates of source position.  

The major unknown in the modeling of B2~0218+35 is the location of the lens center. 
The optical center of the lens was derived with an accuracy of $\sim15\,$mas using the HST observations \citep{2005MNRAS.357..124Y}. 
However, it is unclear if the observed spiral galaxy is indeed the lens or the host galaxy of the blazar \citep{2017MNRAS.470.2814F}.  
The SIS lens geometry implies that the image axes and center of the Einstein ring are aligned. 
The center of the Einstein ring pinpoints the lens center. 
Thus, the image axis and the Einstein ring can constrain the center of the lens. 

The purpose of the lens modeling is to reconstruct the gravitational potential along with the source and lens locations that reproduce the observations.
The straightforward path to obtain the model of the lens is to compare the reconstructed positions of the lensed images with observed positions. 
The lens model is searched by repeating calculations of the parameters such as image positions for a range of lens parameters and comparing them to observation. 
The Monte Carlo simulations are used to investigate a range of complex models with parameters including  a core, a variable slope for the mass distribution, 
and a variable ellipticity and position angle of the lensing galaxy.

The best-reconstructed model was defined as the one that predicts the positions of the lensed images with the smallest offset
and predicting other parameters such as time delay and magnification ratio consistent with the observations.  
The simulations presented in \citep{2016ApJ...821...58B} yield an $\epsilon \sim 0$, confirming an isotropic SIS model of the lens. 
The best-reconstructed model of the system constrains the lens and source positions with an accuracy of $1\,$mas,
which corresponds to $8\,$pc at the redshift of B2~0218+35.

%%%%%%%%%%%%%%%%%%%%%%%%%%%%%%%%%%%%%%%%%%%%%%%%%%%%
\subsubsection{Time Delay of Flare~1 \label{sec:B2Flare1}}
%%%%%%%%%%%%%%%%%%%%%%%%%%%%%%%%%%%%%%%%%%%%%%%%%%%%

\begin{figure*}
%\vskip 1cm
\begin{center}
\includegraphics[width=5.7cm,angle=-90]{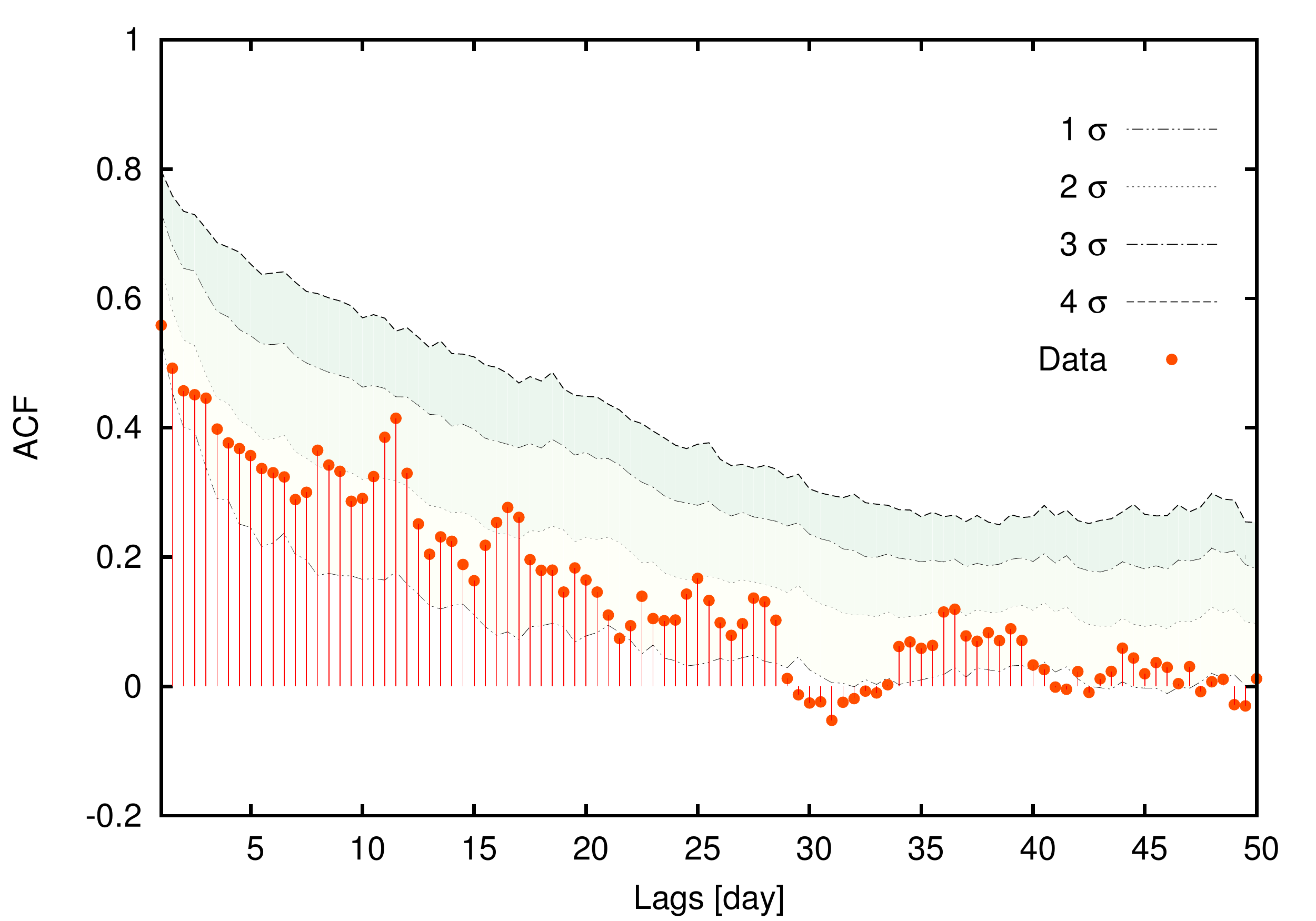}
\includegraphics[width=5.7cm,angle=-90]{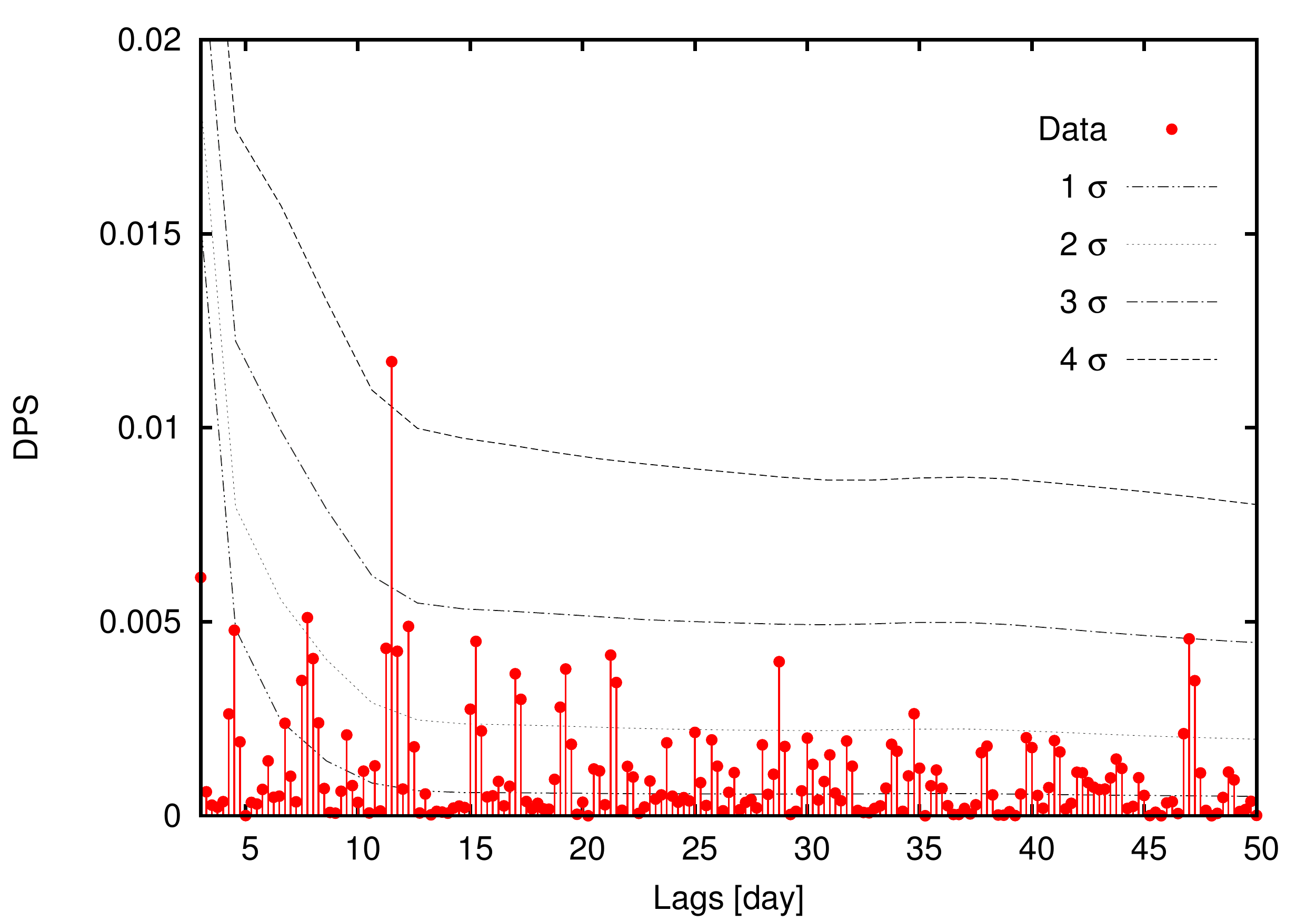}
\end{center}
\caption{\label{fig:dt_Flare1_all} 
			Autocorrelation Function ({\bf Left}) and Double Power Spectrum  ({\bf Right}) for Flare 1. % obtained from the sliced light curve. 
			Figure from \cite{2016ApJ...821...58B}.
				 }
\end{figure*}

On August of 2012, an exuberant outburst of gamma-ray activity was detected from B2~0218+35.
This event is shown in Figure~\ref{fig:lc_whole_B0218} as ``Flare 1".
The time delay estimated using an autocorrelation function shows a delay of $11.46\pm0.16\,$days  \citep{2014ApJ...782L..14C}.
The analysis of the same flare reported by \cite{2016ApJ...821...58B} resulted in  a time delay of $11.5\pm0.5\,$days using the autocorrelation function (Figure~\ref{fig:dt_Flare1_all}, Left), 
and $11.38\pm0.13\,$days using the DPS method (Figure~\ref{fig:dt_Flare1_all}, Right).
The results using the autocorrelation function and the DPS method are in agreement.
The DPS method provides measurement four times more precise as compared to the autocorrelation function.
However, the main advantage of the DPS method is manifested in the statistical significance of the measurement,
which establishes the sensitivity of the method and robustness of its results.
The autocorrelation function provided detection of a time delay  at $> 2\, \sigma$ level,
while the DPS method results in $> 4\, \sigma$ detection. 
The significance of the time delay detection was evaluated based on $10^6$ Monte Carlo realizations of artificial light curves (following \cite{2015ApJ...809..100B}). 
These Monte Carlo simulations were used to calculate the chance that a particular time delay will appear randomly in the simulated light curve which contains no intrinsic time delays. 
Four confidence levels evaluated based on these Monte Carlo simulations are shown in Figures~\ref{fig:dt_Flare1_all}. 

The temporal analysis of Flare~1 reveals a time delay at gamma rays measured with high significance and accuracy of 3~hours. 
The temporal accuracy of 3~hours in this lensed system corresponds to 1~mas spatial accuracy. 
The accuracy of the radio time delay of $10.5\,$days is of the order of 12~hours. 
Thus, despite a difference in the face value of radio and gamma-ray time delays, 
the time delays alone are insufficient to identify an offset between radio and gamma-ray emitting regions at high significance.  
I review application of the HPT method to find the spatial origin of Flare~1 and overcome shortcomings of the time delay approach. 

%%%%%%%%%%%%%%%%%%%%%%%%%%%%%%%%%%%%%%%%%%%%%%%%%%%%
\subsubsection{Origin of Flare~1  \label{sec:OriginFlare1}}
%%%%%%%%%%%%%%%%%%%%%%%%%%%%%%%%%%%%%%%%%%%%%%%%%%%%

\begin{figure}
%\vskip 1cm
\begin{center}
\includegraphics[width=10.5cm,angle=0]{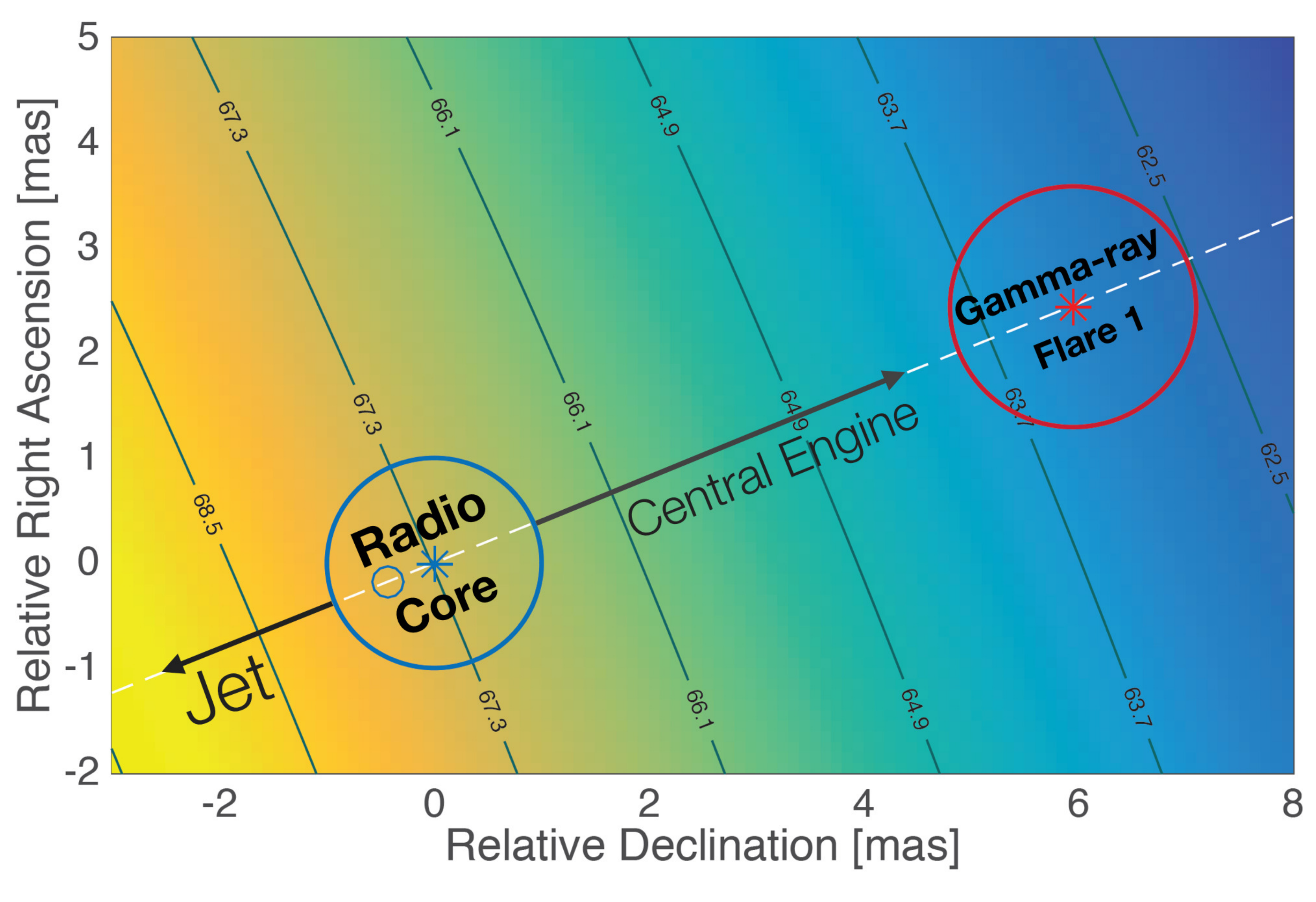}
\end{center}
\caption{\label{fig:HPT_Flare1} 
                          Hubble space.
                          The  blue star indicates the value of 
                          the Hubble parameter based on the  reconstructed position of the 15~GHz radio core. 
                          %The open blue point shows the Hubble parameter derived from the observed  positions of the 15~GHz radio images.
                          The  dashed line shows the  jet projection. 
                          Gray arrows show the direction from the radio core toward the central engine and toward the jet. 
                          The red circle locates the spatial origin of Flare~1.
                          The radius of the red circle corresponds to the uncertainty in the time delay.  
                          The spacing of the gray lines in Hubble space corresponds to $1.2$ km s$^{-1}$Mpc$^{-1}$, 
                          the  $1\,\sigma$ uncertainty in the Hubble parameter \citep{2013arXiv1303.5076P}.
                          Figure from \cite{2016ApJ...821...58B}.
                          }
\end{figure}

The lens model of B2~0218+35 is consistent with SIS.
Thus, the HPT method can be based on Equation~(\ref{eq:h}). % to find the spatial origin of gamma-ray Flare~1. 
%
%Here, the process of applying the HPT method is described. 
First, the positions of lensed images $\theta_A$ and $\theta_B$ are fixed to the positions of the resolved images of the 15 GHz radio core.
These image positions along with the model of the lens and the cosmological parameters are used 
to infer the expected time delay for the radio core.
The value of the reconstructed time delay of $\sim10.7\,$days agrees well with the measured radio time delay of $\sim10.5\,$days (see Table 3, \cite{2016ApJ...821...58B}).
Now, the reconstructed time delay and the positions of lensed images of the 15~GHz radio core are plugged into Equation~(\ref{eq:h}),
which results in the Hubble parameter of $67.3 \,\mbox{km\,s}^{-1}$Mpc$^{-1}$. 
Such reconstructed value of the Hubble constant is shown in Figure~\ref{fig:HPT_Flare1} as the blue star and indicates the position of the radio core. 
The radius of the blue circle indicates uncertainty of the position of the radio core.
This uncertainty is linked to the lens resolution of $1\,$mas.  
The calculation of the Hubble constant using a model of the lens acts as a consistency check. 
Such obtained ``true" values of the Hubble parameter are used as a reference point.
Once the reconstructed position of the radio core is set as the reference point, 
the lens model is used to calculate expected time delays within $\sim10\,$mas from the radio core. 
Equation~(\ref{eq:h}) allows us to convert these time delays to values of the Hubble constant at given distances from the radio core. 
Figure~\ref{fig:HPT_Flare1} shows such calculated Hubble parameters as a function of the position of the variable emitting region. 
The projection of Equation~(\ref{eq:h}) into a given region is called the Hubble space.

The Hubble space shows a range of locations with given value of Hubble parameters.
Additional information is required to limit the range of possible locations.  
In the case of B2~0218+35, knowledge of the projection of the jet allows us to limit possible origins of emission. 
The existence of the Einstein ring made of extended emission of the large-scale radio jet allows us to conclude that the radio jet is projected toward the center of the lens. 
This deduced projection of the jet is shown in Figure~\ref{fig:HPT_Flare1} as the arrow. 
The jet projection allows us to predict that if a flare is produced along the large-scale jet
then the site of emission will be closer to the lens center as compared to the radio core and shorter time delays would be observed. 
The Hubble parameter is inversely proportional to a time delay.  
The Hubble space predicts larger Hubble parameters for emitting regions closer to the lens center. 

The last step in finding the origin of Flare~1 in respect to the radio core is plugging the gamma-ray time delay along with the positions of the lensed images of the radio core into Equation~(\ref{eq:h}). 
This returns value of the Hubble parameter of $63 \,\mbox{km\,s}^{-1}$Mpc$^{-1}$, 
which is shown  in Figure~\ref{fig:HPT_Flare1}  as the red star along the extended jet axis.  
The Hubble space allows us to convert the offset in the Hubble parameters into a physical distance between the radio core and the site of Flare~1.
The reconstructed distance between the radio core and Flare~1 is $6.4\pm1.1\,$mas. % which corresponds to a projected distance of $51.2\pm8.8\,$pc. 
The offset between the radio core and Flare~1 is at $3.2\,\sigma$ level and is defined by the uncertainty of H$_0$ measured with Planck Collaboration \citep{2013arXiv1303.5076P}.

Figure~\ref{fig:lc_whole_B0218}  show the Owens Valley Radio Observatory (OVRO)\footnote{https://www.ovro.caltech.edu}  
monitoring of  B2-0218+35 at 15~GHz \citep{2011ApJS..194...29R}. 
Interestingly, the lack of radio variability from the 15~GHz core  during this huge outburst of gamma-ray emission \citep{2016MNRAS.457.2263S} 
supports the model that Flare~1 did not originate from the radio core. 

Intriguingly, the fact that the gamma-ray time delay is longer than the one expected from the radio core's places the origin of Flare~1 at the opposite site of the radio core from a large scale radio jet. 
Since B2~0218+35 is a blazar, only one side of the jet is observed due to the relativistic boosting effect. 
The counterpart jet cannot be observed in this configuration. 
As a result, the most distant observable structure is the supermassive black hole,
which implies that the supermassive black hole has to be located close to the site of Flare~1 or even beyond. 
As a consequence, the observed compact radio core is not tracing the position of the supermassive black hole. 
The supermassive black hole is located at the projected distance from the radio core of at least $51.2\pm8.8\,$pc. 
Moreover, the jet  is pointed toward us at a few degree angle. 
Thus, the true physical distance between the radio core and the supermassive black hole can be even as large as 1~kpc. 
Such a huge distance of the compact radio core and the supermassive black hole poses a challenge to our understanding of these extreme sources.  

%%%%%%%%%%%%%%%%%%%%%%%%%%%%%%%%%%%%%%%%%%%%%%%%%%%%
\subsubsection{Origin of Flare 2  \label{sec:OriginFlare2}}
%%%%%%%%%%%%%%%%%%%%%%%%%%%%%%%%%%%%%%%%%%%%%%%%%%%%

\begin{figure}
%\vskip 1cm
\begin{center}
\includegraphics[width=9.5cm,angle=0]{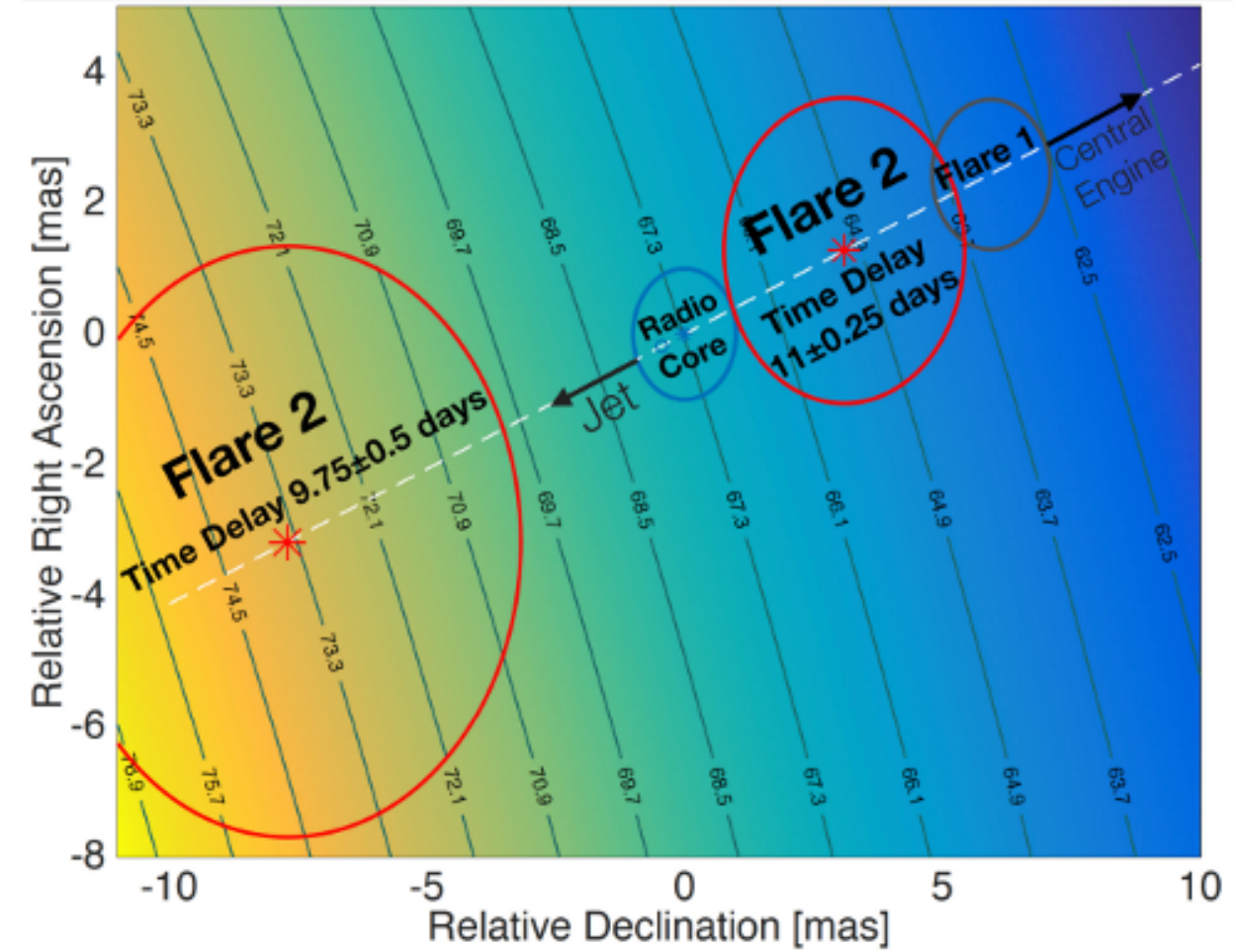}
\end{center}
\caption{\label{fig:HPT_Flare2} 
                          Hubble space calculated as in Figure~\ref{fig:HPT_Flare1}.
                          Red circles show the two possible sites for the  spatial origin of Flare 2 corresponding to
                          a time delay of $9.75\pm0.5\,$days or $11\pm0.25\,$days. }
\end{figure}

A short gamma-ray flare was detected by {\it Fermi}/LAT approximately two years after Flare~1. 
This event is highlighted in Figure~\ref{fig:lc_whole_B0218} as ``Flare 2". 
The advantage of having a single isolated flare like Flare~2 is the ease of a direct search for the echo flare using MPM.
The gamma-ray time delay constrained using MPM indicates values in one of two ranges:  $9.75\pm0.5\,$days or $11\pm0.25\,$days. 
For the first range, the reconstructed Hubble parameter is $69.85-77.4\,\mbox{km\,s}^{-1}$Mpc$^{-1}$.
The second range results in $63.64-66.6\,\mbox{km\,s}^{-1}$Mpc$^{-1}$.
The possible sites of the spatial origin of Flare~2 are shown in Figure~\ref{fig:HPT_Flare2}.
The HPT method reconstructs possible sites of emission to either $8.33\pm4.5\,$mas ($66.64\pm36.00\,$pc) in the direction of the jet, 
or $3.35\pm2.30\,$mas ($26.8\pm18.4\,$pc) from the core toward the central engine.

%%%%%%%%%%%%%%%%%%%%%%%%%%%%%%%%%%%%%%%%%%%%%%%%%%%%
\subsubsection{Connection Between Flares~1 and~2 \label{sec:ConFlare12}}
%%%%%%%%%%%%%%%%%%%%%%%%%%%%%%%%%%%%%%%%%%%%%%%%%%%%

%{\bf Motivation: are Flares~1 and~2 connected events? Findings: They might be!}
%Here, I review application of the HPT method to investigate  a physical connection between Flares~1 and~2. 

The spatial origin of Flare~2 is inconsistent with both the radio core and Flare~1. 
The Hubble parameter tuning  approach allows us to test a hypothesis if Flare~2 could result  
from a  moving blob of plasma launched somewhere in vicinity of the supermassive black hole,
and first  produced Flare~1 and then moved downstream along  the  jet to produce Flare~2.  

The time between Flare~1 and Flare~2, $\Delta t_{obs}$, is $690\,$days.
The projected distance between  Flares~1 and~2 is  $D_{projected}\sim24\,$pc and is constrained by the time delay of $11\pm0.25\,$days. 
In such scenario, the blob of plasma is moving relativistically with an apparent velocity of $\beta_{app}$
%
%\begin{equation}
%\label{eq:projected}
%\begin{split}
%\beta_{app} &   =  \frac{ D_{projected}(1+z_S)}{c\,\Delta t_{obs}}  \\
%                   &    \approx  70 \left(\frac{D_{projected}}{24\,\mbox{pc}}\right) \left(\frac{\Delta t_{obs}}{690\,\mbox{days}}\right)\,.
%\end{split}
%\end{equation}
\begin{equation}
\label{eq:projected}
\beta_{app}   =  \frac{ D_{projected}(1+z_S)}{c\,\Delta t_{obs}}    \approx  70 \left(\frac{D_{projected}}{24\,\mbox{pc}}\right) \left(\frac{\Delta t_{obs}}{690\,\mbox{days}}\right)\,.
\end{equation}

Very high superluminal apparent motions  are typical for gamma-ray blazars \citep{2013AJ....146..120L,2015ApJ...810L...9L}. 
For example, superluminal apparent motions of $\sim46\,c$  has been observed in the radio jet of the blazar PKS~1510-089 \citep{2005AJ....130.1418J}. 
Thus, the gamma-ray time delay of $11\pm0.25\,$days yields a reasonable physical model for B2~0218+35.

If the blob of plasma continues its motion with the same apparent velocity, $1.6\,$mas/year, 
it would pass through the stationary shock of  the 15~GHz core $\sim2\pm1\,$years after Flare~2 detected in July 2014.  
This model thus predicts impact of the blob of plasma with the radio core around July 2016. 
It is unclear what observational signatures should be expected when a blob of plasma impacts a stationary shock such as the radio core.
Interestingly, the {\it Fermi}/LAT instrument detected a long term increase in emission starting July 2016. 
The increase in gamma-ray emission is evident when emission in binned into 7-day bins as shown in Figure~\ref{fig:lc_whole_B0218}. 
The temporal analysis of gamma-ray light curve including the period of increased emission starting July 2016
results in time delay of $10.5\pm0.5\,$days using both the autocorrelation function and the DPS method. 
%exactly as expected from emission originating from the radio core,
Intriguingly, the OVRO monitoring did not show increase in radio emission simultaneous with gamma rays (see bottom of Figure~\ref{fig:lc_whole_B0218}). 

%Radio observations during this period could thus provide valuable insight into  the physical processes and plasma propagation along the jet. 

The second possible site of Flare~2, implied by the time delay of $\sim9.75\pm0.5$, is located at a projected distance of $\sim16\,$mas from Flare~1.
An apparent velocity of $350\,$c would be required to explain such a large projected distance.
Thus, these flares could not be produced by the same moving blob of plasma. 

There is no direct evidence that Flares~1 and~2 were caused by the same relativistic blob of plasma.
However, the longer time delay and the detection of the predicted increased emission after July 2016 implies a reasonable physical model for the source 
and  demonstrates the potential of the Hubble parameter tuning approach to predict observations of future events.

%%%%%%%%%%%%%%%%%%%%%%%%%%%%%%%%%%%%%%%%%%%%%%%%%%%%
\subsubsection{Interpretation of Radio Core vs Gamma-ray Flare Offset \label{sec:HPToffset}}
%%%%%%%%%%%%%%%%%%%%%%%%%%%%%%%%%%%%%%%%%%%%%%%%%%%%

It has been anticipated that the radio core, gamma-ray emission site, and the central engine are within parsecs of one another. 
The projected offset between the gamma-ray flare and the radio core of $\sim60\,$pc was unexpected. 
Our best insights on the connection between  gamma rays and radio emission comes from monitoring of the variability patterns. 
Figure~\ref{fig:lc_whole_B0218} illustrates the challenge of using the variability monitoring to infer the origin of the emission. 
During a period of an enormous gamma-ray activity marked as Flare~1, 
the radio emission not only did not show similar variability pattern as observed at gamma rays,
but the radio emission even decreased over this period \citep{2016MNRAS.457.2263S}.
A similar situation is observed for other gamma-ray flares of B2~0218+35.

Radio monitoring of a large sample of blazars shows that roughly 2/3 of the gamma-ray flares coincide  with 
the appearance of a new superluminal knot and/or a flare in the millimeter-wave core located parsecs from the central engine 
\citep{2010ApJ...710L.126M,2011ApJ...726L..13A,2012arXiv1204.6707M,2012arXiv1201.5402M,2014MNRAS.441.1899F,2014MNRAS.445.1636R,2015arXiv150503871C}.
Further study shows correlations and similarities  between multi-wavelength  and gamma-ray observations \citep{2012ApJ...754...23L,2012ApJ...749..191C,2014A&A...562A..79S}.
However, \cite{2014MNRAS.445..428M} modeled the light curves of blazars as red noise processes, and found  that only 1 of 41 sources with high-quality data in both the radio and gamma-ray bands shows correlations with a significance larger than $3\sigma$. 
They thus demonstrate the difficulties of measuring statistically robust multi-wavelength correlations even when the data span many years. 

The complex gamma ray and radio variability are also observed in the case of M87, 
where some of the gamma-ray flares are correlated with radio emission from the radio core, 
other with the HST-1 knot, and some of the gamma-ray flares appear without correlation with either the radio core or HST-1 
\citep{2012ApJ...746..151A,2010ApJ...716..819A,2006Sci...314.1424A,2008ApJ...685L..23A}. 

Now, let us consider the geometry of the innermost relativistic jets \citep{2016PASJ...68R...1H,2017ApJ...834...65A} and observations of M87 to demonstrate 
 one of possible scenarios to explain the presented findings on B2~0218+35.
The well-resolved radio observations of M87 revealed that 
 the jet is  maintaining a parabolic morphology from the base of the jet  up to
the HST-1 knot where it transitions to conical shape  \citep{2012ApJ...745L..28A,2013ApJ...775..118N}.

The parabolic part of the M87 jet close to the central engine (within $1\,$pc) is not well collimated  \citep{1999Natur.401..891J,2013ApJ...775...70H}.
The apparent  jet opening angle of  M87 at  the distance of $0.1\,$pc is $\theta_{jet}=33^\circ$. 
Jets became well collimated at larger distances.
At the distance of the HST-1 knot, the jet opening angle is $ \theta_{jet}\sim6^\circ$. 
The Lorenz factor can be approximated as $1/\theta_{jet}$ \citep{2014ApJ...794L...8B}. 
The viewing angle of M87 is no more than $\theta_{obs}\leq19^\circ$ from our line-of-sight \citep{1999ApJ...520..621B}.
In this configuration, the expected Doppler factor is $\mathcal{D}\sim3$, for both, the emission close to the central engine and the HST-1 knot. 

However, if one imagine the M87 jet pointed toward us at the viewing angle of $\theta_{obs}\sim3^\circ$, as in the case for blazars,
the radiation originating from the region close to SMBH would have the Doppler factor of $\mathcal{D}\sim3$.
However, the HST-1 knot, due to its collimation, would have the Doppler factor of $\mathcal{D}\sim16$. 
Taking into account that the radiation is enhanced by $\mathcal{D}^4$, 
and assuming similar intrinsic luminosities, the HST-1 knot would appear $\sim500$ times brighter that the emission close to the central engine.

The HST-1 knot is located at a projected distance of $\sim60\,$pc from the supermassive black hole \citep{1999ApJ...520..621B}.
If M87 were located at redshift equal to 1, the emission from the HST-1 knot would appear at an offset of $\sim7\,$mas from SMBH. 
Thus, relativistically boosted recollimation shocks are good candidates to explain offsets between radio and optical emission. 
This scenario also allows us to make a testable prediction. 
If indeed the observed radio core is one of the recollimation shocks, 
then the base of the jet should be present as a dozen to hundred times fainter feature upstream of the jet,
detectable with high contrast observations. 
Also, the effect depends on the viewing angle of a jet, and become prominent when the viewing angle is smaller. 
Thus, the model could be tested statistically by correlating viewing angle and the number of observed sources with an offset. 

%%%%%%%%%%%%%%%%%%%%%%%%%%%%%%%%%%%%%%%%%%%%%%%%%%%%
\subsection{Notes on the Hubble Constant \label{sec:NotesH0}}
%%%%%%%%%%%%%%%%%%%%%%%%%%%%%%%%%%%%%%%%%%%%%%%%%%%%

The reconstruction of the position of gamma-ray flares in respect to the radio core is based on the Hubble parameter H$_0=67.3\pm1.2\,$~km$\,$s$^{-1}\,$Mpc$^{-1}$ reported by  \cite{2013arXiv1303.5076P}. 
Many independent methods provide a measure of H$_0$.
For example  the Hubble Space Telescope Key Project provides  H$_0=72\pm$8~$\mbox{km\,s}^{-1}$Mpc$^{-1}$ \cite{2001ApJ...553...47F}, 
the Cepheid distance ladder gives  $73.8\pm$2.4~$\mbox{km\,s}^{-1}$Mpc$^{-1}$ \citep{2011ApJ...730..119R,2011ApJ...732..129R}
and $74.3\pm1.5\,(\mbox{stat})\pm 2.1\,(\mbox{sys})\, \mbox{km\,s}^{-1}$ Mpc$^{-1}$ \citep{2012ApJ...758...24F}.

The gamma-ray time delay combined with the position of the radio core gives a Hubble parameter of $H_0 = 63.64 \pm 0.67\,\mbox{km\,s}^{-1}$Mpc$^{-1}$.
Thus, even for the largest value ($74.3\pm1.5\,(\mbox{stat})\pm 2.1\,(\mbox{sys})\, \mbox{km\,s}^{-1}$ Mpc$^{-1}$) 
there would be a significant offset between the radio core and the gamma rays of at least $\sim3\sigma$.  
Therefore, the separation between the radio core and the gamma-ray emission is robust to the large spread in values of the Hubble constant. 

It is still under debate if the true H$_0$ is closer to  H$_0^{Planck+BAO}= 67.6\pm 0.6 \,\mbox{km\,s}^{-1}$Mpc$^{-1}$ \citep{2013arXiv1303.5076P}
inferred from  the cosmic microwave background (CMB) and large-scale structure (LSS) observations assuming the standard flat $\Lambda$CDM model,
or H$_0^{local}= 73.24\pm 1.74\,\mbox{km\,s}^{-1}$Mpc$^{-1}$ obtained from direct measurement in the local universe \citep{2016ApJ...826...56R}.  
However, the value of $63 \,\mbox{km\,s}^{-1}$Mpc$^{-1}$ from the gamma-ray time delay is not debatable and is an indication that Flare~1 did not originate from the radio core. 

%{\bf 72 expected time delay from radio core even shorter than 10.7 days. }

%%%%%%%%%%%%%%%%%%%%%%%%%%%%%%%%%%%%%%%%%%%%%%%%%%%%
\subsection{Discussion}
%%%%%%%%%%%%%%%%%%%%%%%%%%%%%%%%%%%%%%%%%%%%%%%%%%%%

Active galaxies are complex sources. 
Our inability to spatially resolve multi-wavelength emission originating from their inner regions led us to  
multiple assumptions on SMBHs and their connections to the compact radio core and variable emission.  
The Hubble parameter tuning approach allows us to convert temporal resolution into a spatial resolution 
to constrain physical distances between otherwise unresolved emitting regions. 
The approach allows us to take advantage of high-resolution observations of radio telescopes or optical facilities equipped in adaptive optics,
and combine it with observation of sources at higher energies where observations with sufficient angular resolution are impossible, 
however, source variability allows measuring time delays for gravitationally lensed sources. 

The case study of gravitationally lensed blazar B2~0218+35, reviewed here, 
demonstrates the potential of the HPT method to spatially resolve the origin of the emission and predict future flares. 
The case study focused on two gamma-ray flares. 
The HPT method combined time delay during Flare~1 of  $11.33\pm0.12\,$days and positions of well-resolved radio lensed, 
and revealed the distance between the radio core and the site of the gamma-ray flare is $6.4\pm1.1\,$mas, 
which corresponds to the projected distance of $51.2\pm8.8\,$pc.
The HPT method allowed us to take advantage of the well-resolved positions of the lensed images of the radio core, instead of an uncertain radio time delay.  

It is commonly assumed that the radio core traces the position of the supermassive black hole to within a few parsecs. 
A possible offset smaller that 1~mas may occur as a core shift effect. 
However, it is believed that the radio core is the closest structure to the supermassive black hole that we can resolve. 
It is also assumed that the radio core is a region where outbursts across the entire 
the electromagnetic spectrum is produced.
Thus, the results of the HPT method applied to B2~0218+35 challenge our assumptions. 

An accuracy in the time delay measurement of $0.12\,$days corresponds to a spatial position of $\sim1\,$milliarcsecond. 
The modern gamma-ray instruments resolve the spatial origin of the radiation down to $0.1\,$deg, at best. 
Therefore, the Hubble parameter tuning approach probes jet structure on a scale 
$>360000$ smaller than limited by spatial imaging with the world's best gamma-ray telescopes. 

Strong gravitational lensing allows us to constrain the spatial location of sources, 
but it does not constrain their sizes on small scales ($<pc$). 
The size of the source can be constrained using variability timescales and light crossing time. 
Size of the source can be also constrained using the microlensing effect \citep{2015NatPh..11..664N,2016A&A...586A.150V,2016MNRAS.459.1959S}. 
However, microlensing does not elucidate location of sources. 
The size of gamma-ray sources is comparable to the Einstein angle of a star located at cosmological distance. 
Thus, the effects of microlensing caustic-crossing on magnification and time delay \citep{2018MNRAS.473...80T} 
are reduced for gamma-ray emission region as compare to much smaller accretion disc.

Lensed high-energy sources monitored with detectors like Chandra, Swift 
and NuSTAR offer rich opportunities to extend the Hubble parameter tuning approach to other sources. 
There are more than 20 known lensed quasars with associated X-ray emission. 
Some of these systems already have enough observations to reconstruct the mass distribution of the lens. 
Further monitoring will enable measurement of time delays with X-ray detectors.
In the near future, SKA will resolve hundreds of thousands of radio images of gravitationally lensed quasars, 
and LSST will provide time delay measurements for thousands of these sources.

%%%%%%%%%%%%%%%%%%%%%%%%%%%%%%%%%%%%%%%%%%%%%%%%%%%%
\section{Caustics as Non-linear Amplifiers \label{sec:Caustics}}
%%%%%%%%%%%%%%%%%%%%%%%%%%%%%%%%%%%%%%%%%%%%%%%%%%%%

Comparisons of radio and optical positions of sources is severely limited 
by the astrometric accuracy and angular resolution of optical telescopes. 
The example of  B2~0218+35, reviewed in Section~\ref{sec:CaseB20218}, 
shows evidence for the offset between the radio core and central engine.
Such an offset implies that there should also be an offset between the radio core and optical emission primarily originating from the accretion disk surrounding the SMBH. 
Interestingly, recent observations by the Gaia satellite provide astrometry with $mas$ resolution of the centroid of the optical emission.
The first comparison of optical positions measured by Gaia and radio positions from VLBI  reveal that $\sim6\%$ of sources show significant offsets \citep{2016A&A...595A...5M}. 
Such offsets might be an indication of extreme phenomena including relativistic jets, SMBH binaries, or even recoiling SMBHs. 

The Gaia observations revolutionize our knowledge of the nearby universe. 
Gaia provides astrometric positions of the centroid of the emission for sources as faint as 20 mag.   
Resolving complex multi-wavelength emission of distant sources is still precluded 
by the limited resolution and astrometry of current and future instruments. 
To overcome these technological limitations, 
\cite{2017ApJ...846..157B} demonstrated that caustics of lensing galaxies can act as non-linear amplifiers
and can be used to enhance the performance of telescopes by orders of magnitude using only relative positions of lensed images.

When sources are located close to the caustic of a lensing galaxy, 
 even a small offset in positions of the source
results in a drastic difference in positions and magnification of lensed images. 
Thus, if optical and radio emissions originate from the same region, 
the position of the lensed images observed using optical and radio telescopes will coincide. 
However, if there is even a small offset between optical and radio emissions, then the positions of the lensed images will differ significantly. 

Section~\ref{sec:causticIdea} reviews the idea of using caustics of lensing galaxies as non-linear amplifiers. 
Section~\ref{sec:CausticChar} characterizes caustics using Monte Carlo simulations. 
Section~\ref{sec:CausticApp} discusses applications of the caustic approach. 

%%%%%%%%%%%%%%%%%%%%%%%%%%%%%%%%%%%%%%
\begin{figure}
%\vskip 1cm
\begin{center}
\includegraphics[width=8.cm,angle=0]{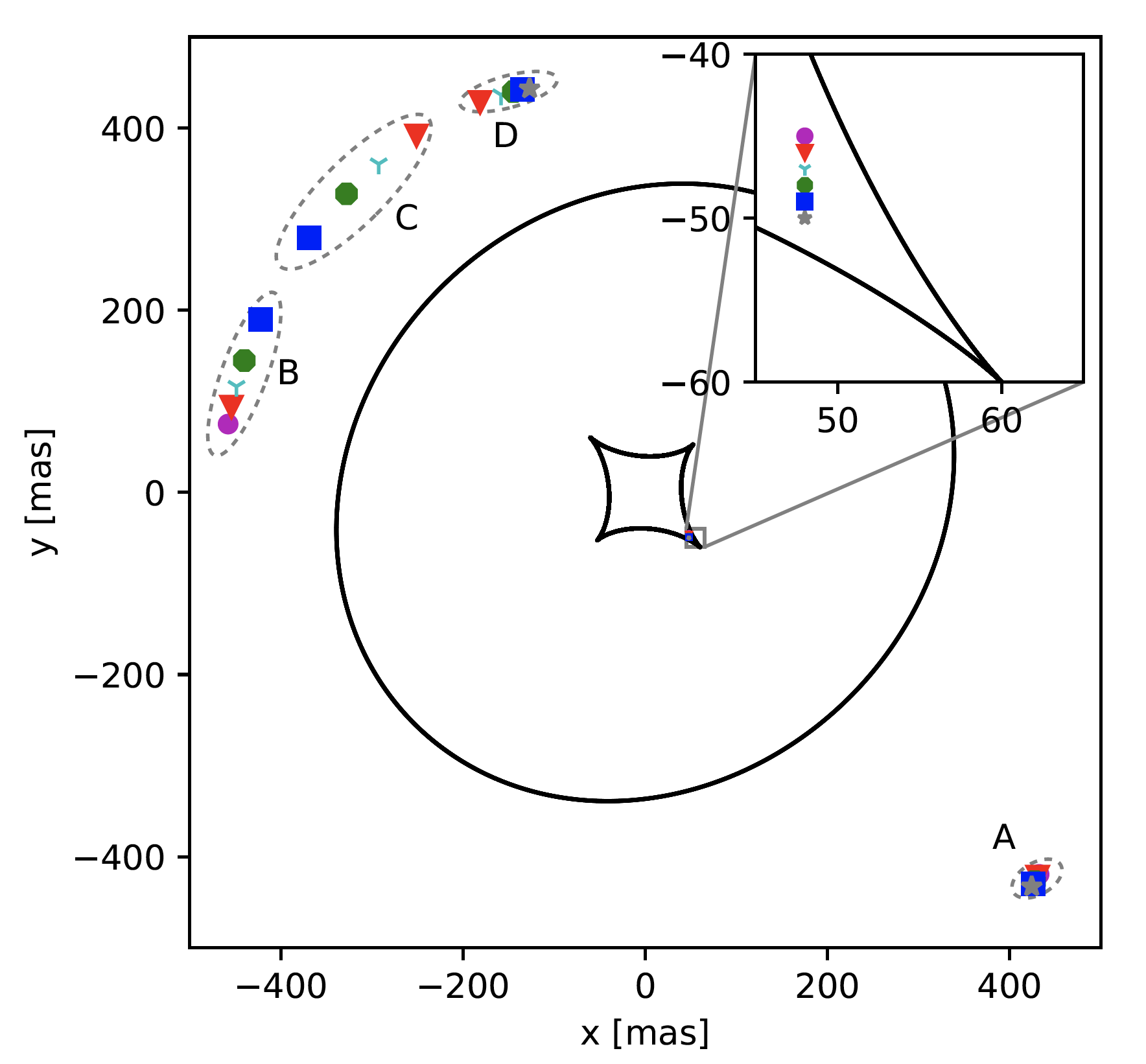}
% ProfilePlot.py 
\end{center}
\caption{\label{fig:ImageSeq} 
			Simulation of a lensed source located close to the caustic of lensing galaxy.
			%Toy Model lensed system demonstrating propagation of small offsets between emitting regions into large offsets between positions of lensed images.
		         Black lines show the inner and outer caustics of the lensing galaxy. 
                          Grey dotted ellipses group four classes of the lensed images to facilitate a comparison of the image positions.
                          The most drastic changes are visible for the C image class where the positions differ by $\sim50\,$mas 
                          for emitting regions offset by  $1\,mas$ in the source plane.
                          Figure from \cite{2017ApJ...846..157B}.
                          }
\end{figure}
%%%%%%%%%%%%%%%%%%%%%%%%%%%%%%%%%%%%%%%%%%%%%%%%%%%%
\subsection{Approach \label{sec:causticIdea}}
%%%%%%%%%%%%%%%%%%%%%%%%%%%%%%%%%%%%%%%%%%%%%%%%%%%%

Resolving the multi-wavelength emission of the inner regions of active galaxies requires 
angular resolution and astrometric accuracy below $1\,mas$, which is challenging to achieve with current and future facilities. 
 \cite{2017ApJ...846..157B}  proposed using caustics of lensing galaxies to study the multi-wavelength structure of sources. 
The caustic of the lensing galaxy is a place where lensed images merge or separate. 
Thus, the number of lensed images changes from two to four, when the source is located inside the inner caustic. 
The proximity of the caustic results in a drastic difference in the positions and magnifications of the lensed images 
when there is even a small offset in the positions of sources.  
The caustic turns a small offset between unresolvable sources into large offsets between lensed images of these sources. 
The offsets between lensed images can be resolved even with existing facilities. 
The relative position between lensed images can be used to understand the origin of the emission.
The properties of the caustic can be used to discover and investigate complex sources.

Here, I review a toy model used by  \cite{2017ApJ...846..157B} to demonstrate the idea of using the caustic of a lensing galaxy. 
The toy model system includes a source at redshift 2 
and a lensing galaxy at redshift 0.5 modeled as an Singular Isothermal Ellipsoid (SIE; see Section~\ref{sec:SIE}).
The SIE lens has an ellipticity of $e=0.2$, 
an angle $\phi=45^\circ$, and the velocity dispersion $\sigma =170\,km/s$, 
corresponding to the Einstein ring radius of 0.5". 

The toy model source consists of six aligned emitting regions separated by $1\,mas$
 ($1\,mas$ corresponds to $\sim10\,$pc at the source location). 
The configuration represents different offsets from 1 to $5\,mas$ for sources offset by  $1\,mas$,
or can simulate a jet consisting of six distinct emission components referred to as knots.
The toy model source is located close to the caustic. 
The knots separated by $1\,mas$ blend into one point. 
The top-right corner of Figure~\ref{fig:ImageSeq} shows the 18 times zoomed source region with resolved points.
Figure~\ref{fig:ImageSeq} shows the sequence of lensed images for six knots.
Interestingly, the positions of the lensed images differ by $20-50\,mas$  even for knots separated by only $1\,mas$. 

Figure~\ref{fig:ImageSeq}  demonstrates that the largest variation in the positions of the images occurs for the C image class. 
In the considered configuration, 
the angular offset in the source plane as small as $1\,mas$ can result in an offset between positions of the lensed images as large as $50\,$mas. 
The total change in the position of lensed images $\Delta \theta$ between two sources reaches $\Delta \theta \sim100\,mas$, 
and defines the offset amplification\footnote{Offset Amplification is defined as a ratio between an angular offset between lensed images, $\Delta\theta$,  
to the offset in the source plane $\Delta\beta$}. 
The change in the position of lensed images is accompanied by a significant change in flux magnification.
The magnification ratio between lensed images can be used as an additional way to constrain the spatial origin of the source. 
Thus, the offset amplification and magnification can allow us to determine the offset between unresolvable sources.

%%%%%%%%%%%%%%%%%%%%%%%%%%%%%%%%%%%%%%
  \begin{figure}
%\vskip 1cm
\begin{center}
\includegraphics[width=7.cm,angle=0]{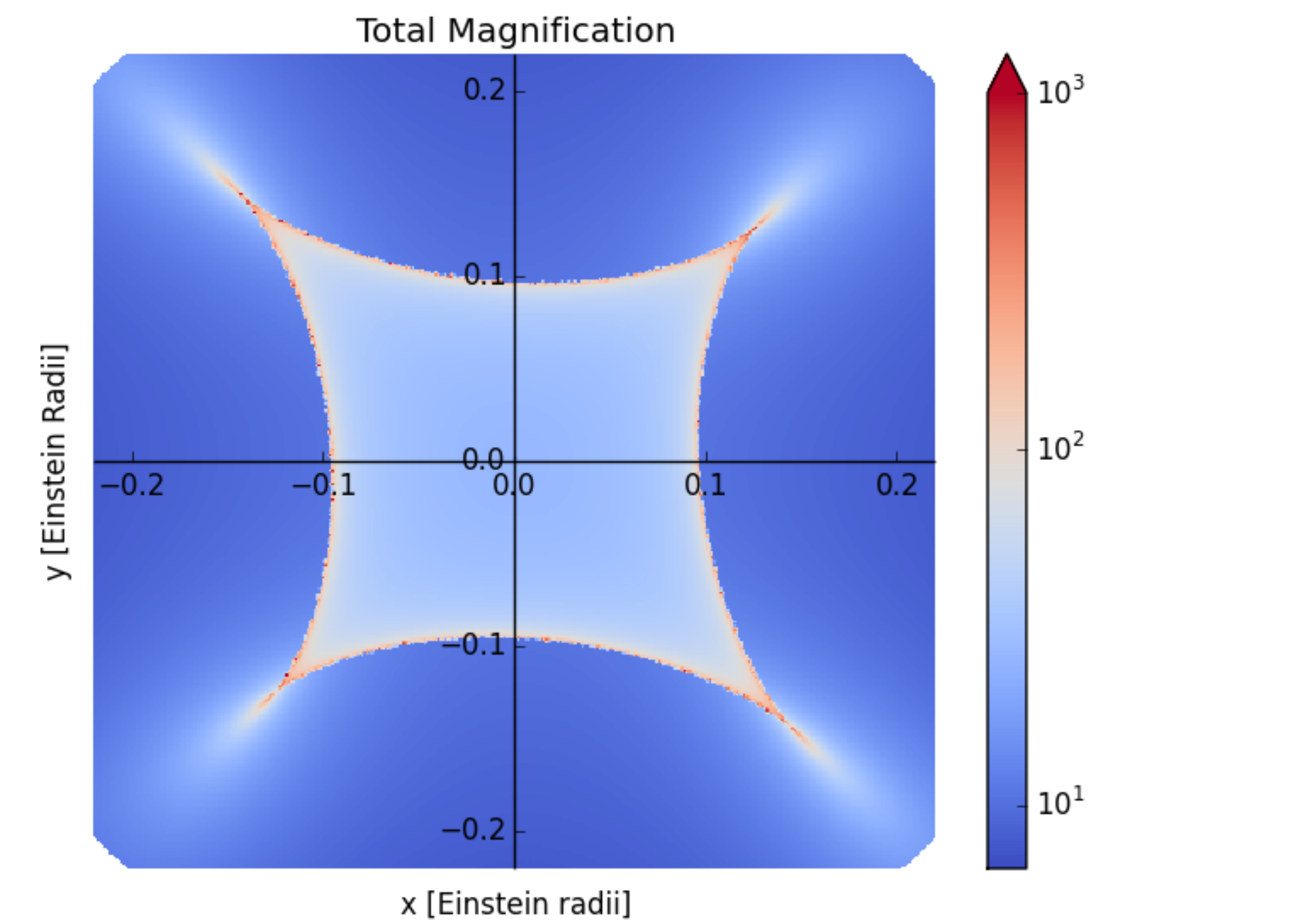}
\includegraphics[width=7.cm,angle=0]{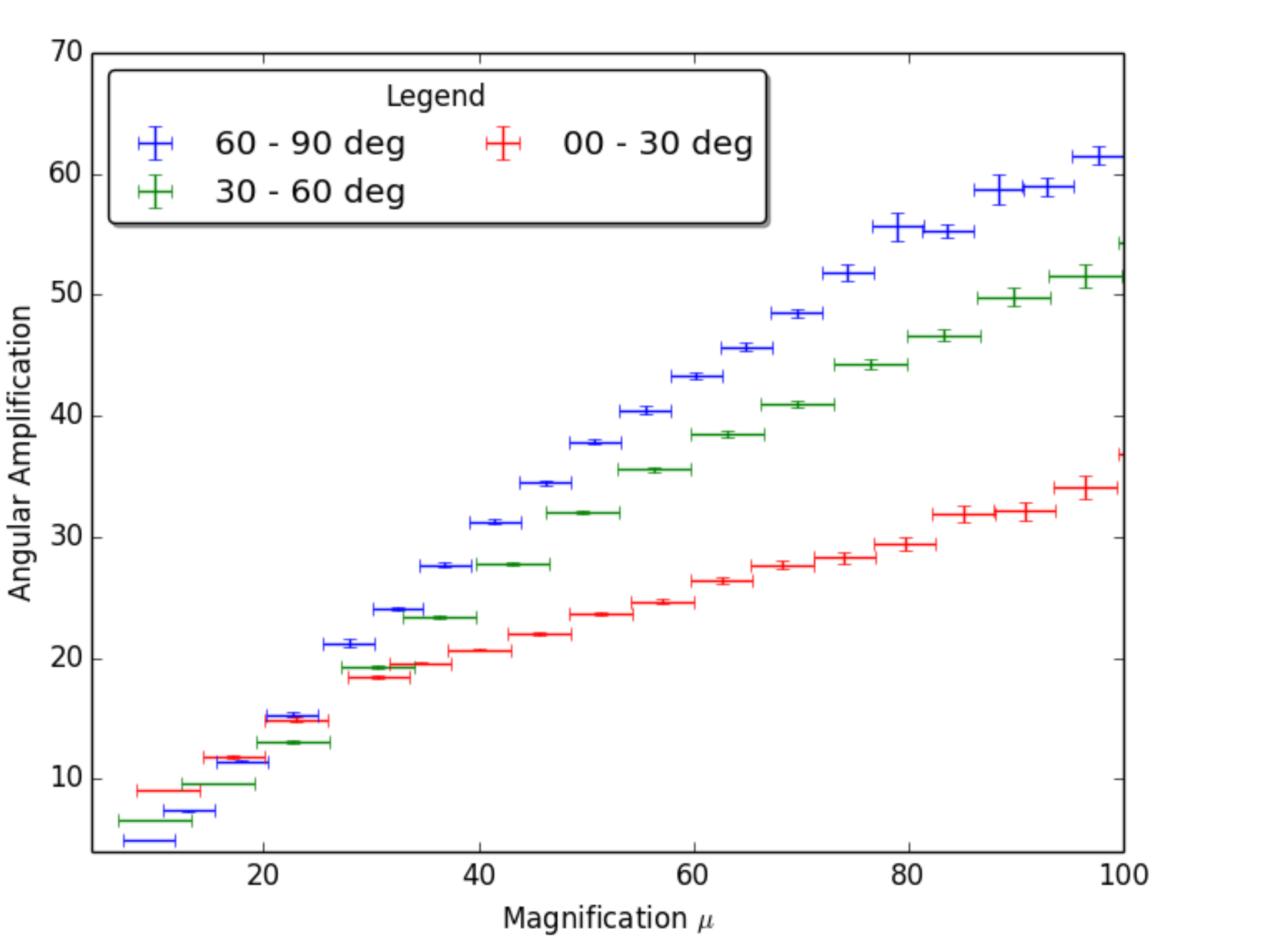}
% ProfilePlot.py 
\end{center}
\caption{\label{fig:TotalMag}  %PlotMagnification2.py 
                          {\it Left}: Total flux magnification defined as a sum of flux magnifications of all lensed images of  a source. 
                          Coordinates are shown relative to the lens center.
                          {\it Right}: Offset amplification as a function of total magnification for different  angles of pair of sources in relation to the caustic. 
                          Figures from \cite{2017ApJ...846..157B}.
                           }
\end{figure}

%%%%%%%%%%%%%%%%%%%%%%%%%%%%%%%%%%%%%%
\subsection{The Caustic Characteristics}
\label{sec:CausticChar}
%%%%%%%%%%%%%%%%%%%%%%%%%%%%%%%%%%%%%%

The performance of gravitational telescopes is defined by their ability to magnify fluxes and amplify offsets between emitting regions. 
Monte Carlo simulations show that the offset amplification in the caustic approximation increases drastically from 20 to even 60, 
with even sharper changes of magnification reaching 100. 
Here, I review  the Monte Carlo simulations performed to evaluate the offset amplification\footnotemark , flux magnification, and probability of a source being in the caustic configuration.

The Monte Carlo simulations included $10^6$  pairs of offset sources at different orientations and locations with respect to the  tangent to nearest caustic. 
The position and magnification of lensed images were obtained using {\tt glafic} code \citep{2010PASJ...62.1017O}.
The total magnification of the source as a function of the source position is shown in Figure~\ref{fig:TotalMag} (Left). 
The total magnification is the sum of magnifications of all lensed images for given source positions. 
Inside the inner caustic, four magnified images are created and the total magnification is greater than 10.
When a source approaches a side of the caustic, the total magnification can be greater than 100. 

The total offset between these positions of lensed images of pair of sources were calculated 
and normalized by the distance between the pair of sources to obtain the offset amplification.
The offset amplification as  a function of total magnification is shown in Figure~\ref{fig:TotalMag} (Right).
The largest offset amplification  is achieved when the pair of sources is located perpendicularly  to the caustic. 
When the pair of sources is located within $1\% \,r_E$ from the caustic, 
the flux magnification is $\sim 70$, and the offset amplification reaches 50. 

The image magnification is determined by the second derivative of the effective potential  (see Section~\ref{sec:magnification}). 
The positions of lensed images, and thus the offsets, are determined by the first derivative of the effective potential.
As a result, image magnification is changing faster than image position when a source is located close to the caustic.  
However, the advantage of using the image positions to reconstruct the origin of 
the emission is that the image positions are less sensitive to the substructures in the mass distribution of the lens and the effects of external shear.  
Substructures greater than $10^6\,M_\odot$ are required to breaks the symmetry of a smooth critical curve and can produce a shift in the position of lensed images greater than $mas$ \citep{2018arXiv180403149D,2017ApJ...850...49V}. 
Such substructures in the mass distribution could potentially complicate the lens modeling, 
however, if accounted for in the lens model would enhance the caustic properties as non-linear amplifiers.  

The last caustic characteristics considered here is a chance that the source will be located close enough to the caustic 
to experience the non-linear amplification. 
The caustic length in the toy model example is $\sim1.15\,r_E$. 
The significant offset amplification, $>20$, requires  the source to be located within an  $0.02\,r_E$ from the caustic.
The probability that the lensed source is located in the caustic configuration is $\sim0.005$. 

 However, in the caustic configuration, the source flux is magnified more than 20 times (see Figure~\ref{fig:TotalMag}). 
 The high magnification introduces magnification bias that increases a probability of observing gravitationally lensed systems in the caustic configuration
 \citep{1980ApJ...242L.135T,1984ApJ...284....1T,2003ApJ...583...58W,2002ApJ...577...57W}. 
The Monte Carlo simulations of magnification bias show  2\% probability that the source will be located within the region with magnification greater than 20. 
 However, due to the magnification bias, the probability of observing the lens system in the caustic configuration increases to 8\%. 
 Thus, a significant fraction of the observed gravitationally lensed quasars will be in the caustic configuration.
%and the brightest observed lensed sources are more likely to be in the caustic configuration.  

%%%%%%%%%%%%%%%%%%%%%%%%%%%%%%%%%%%%%%%%%%%%%%%%%%%%
\subsection{Applications \label{sec:CausticApp}}
%%%%%%%%%%%%%%%%%%%%%%%%%%%%%%%%%%%%%%%%%%%%%%%%%%%%

Sources located close to the caustic experience flux magnification and offset amplification up to two orders of magnitude. 
The flux magnification allows us to look for distant and faint sources of radiation including the first galaxies and quasars. 
This offset amplification allows us to investigate multi-wavelength complexity of the sources and search for multiple emission components such as binary black holes or relativistic jets. 
Here, I review example applications of the caustics of lensing galaxies as non-linear amplifiers. 

%%%%%%%%%%%%%%%%%%%%%%%%%%%%%%%%%%%%%%%%%%%%%%%%%%%%
\begin{figure*}
%\vskip 1cm
\begin{center}
%bb=0 0 612 792
\includegraphics[width=14.cm,angle=0]{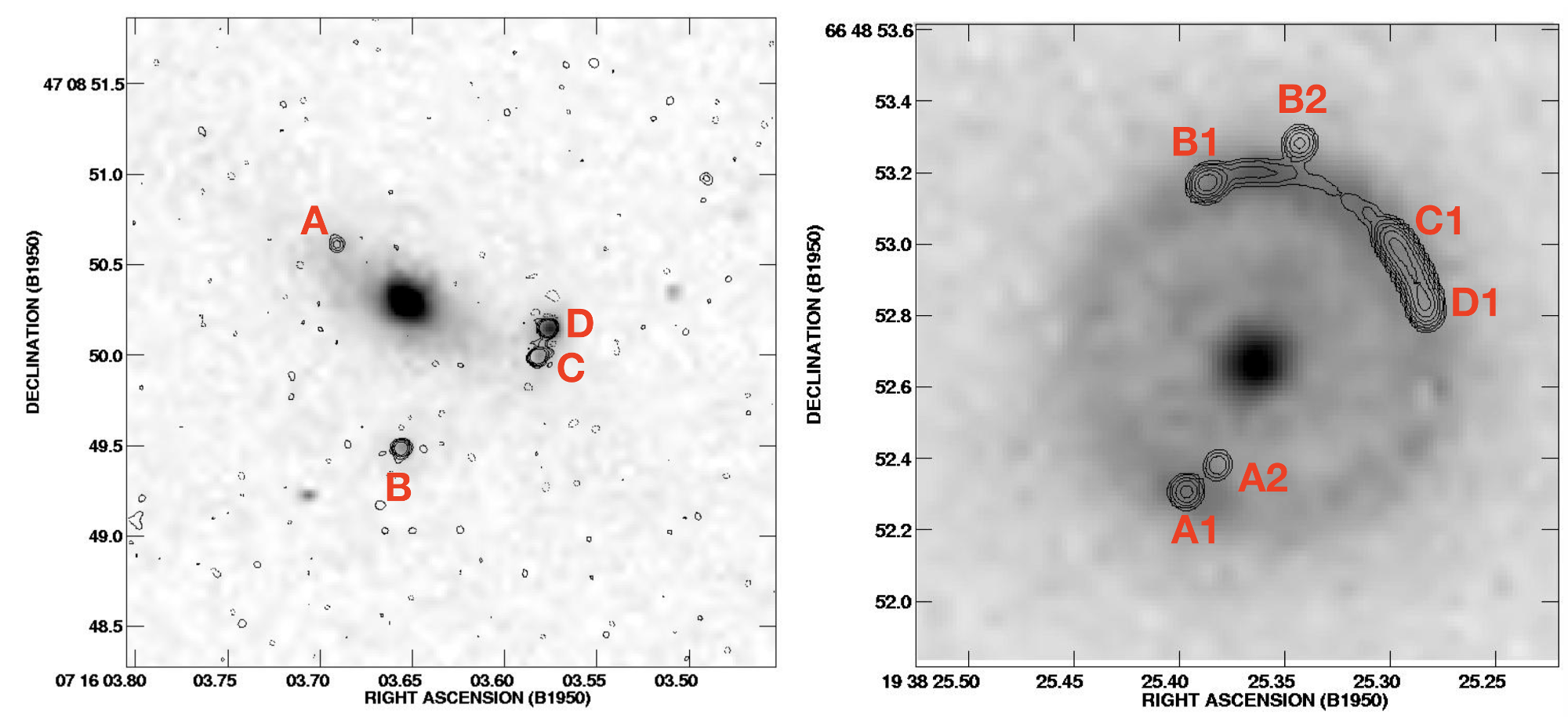}
\end{center}
\caption{\label{fig:offsets} 
			Gravitationally lensed systems with superimposed MERLIN radio contours on the HST observations.
			 {\it Left}:  The gravitational lens B0712+472 from \cite{1998MNRAS.296..483J}
			 {\it Right}: The gravitational lens B1938+666 from \cite{1998MNRAS.295L..41K}.
			%http://www.jb.man.ac.uk/research/gravlens/lensarch/lens.html
			%Figure from \cite{2016ApJ...821...58B}.
				 }
\end{figure*}
%%%%%%%%%%%%%%%%%%%%%%%%%%%%%%%%%%%%%%%%%%%%%%%%%%%%
\subsubsection{Offset Sources}
%%%%%%%%%%%%%%%%%%%%%%%%%%%%%%%%%%%%%%%%%%%%%%%%%%%%

Supermassive black holes powering AGN do not always reside at the centers of their host galaxies  \citep{2012MmSAI..83..925B}.
Such systems could be offset from the galaxy center after receiving a kick from binary coalescence, or they could host a supermassive black hole binary. 
Offset systems are challenging to identify and study directly due to limiting astrometry and resolution of telescopes.
For example, a source consisting of a single emitting region observed with optical and radio facilities can be resolved down to $mas$, 
but limited astrometric accuracy precludes investigation of the offset between such sources. 
Historically, such observations would be interpreted as co-spatial. 

Offset systems could be constrained if observed in the caustic configuration.
If, for example, optical and radio emission originate from the same region, 
then positions of lensed images resolved using optical and radio observations should be well aligned. 
Figure~\ref{fig:offsets} shows superimposed radio contours on the HST observation of gravitationally lensed source  B0712+472 (Left). 
The center of the image shows the optical emission of the lensing galaxy. 
The existence of four lensed images indicates that the source is located inside the inner caustic. 
The close separation between lensed images C and D indicates that the source is located close to the caustic. 
Moreover, the optical and radio positions of the lensed images are very well aligned suggesting a co-spacial origin of optical and radio emission (Spingola and Barnacka, in preparation).

However, if the source located close to the caustic is complex, then more than four lensed images will be observed, and optical, and radio emission might be misaligned. 
Figure~\ref{fig:offsets} shows superimposed radio contours on optical observation of the gravitationally lensed source B1938+666 (Right). 
The radio observations reveal the existence of six lensed images. 
Four of these lensed radio images marked as A1, B1, C1, and D1 belong to an emitting region located inside the inner caustic. 
While, the lensed radio images A2 and B2 are part of the radio source located outside inner caustic. 
These six lensed radio images point to a complex radio source consisting of two emitting regions. 
The tangential extent of the radio images points to extended radio emission suggesting the relativistic jet origin. 
Interestingly, the HST observations of this system show very different morphology with emission forming an Einstein ring 
suggesting that the resolved optical emission does not coincide with the radio emission. 

Detailed analysis of lensed systems in the caustic configuration has a potential to provide a precise measurement of the offsets between the optical and radio emission 
and will give us an insight into the size of the emitting region based on the tangential extent of lensed images. 

%%%%%%%%%%%%%%%%%%%%%%%%%%%%%%%%%%%%%%%%%%%%%%%%%%%%
\begin{figure*}
%\vskip 1cm
\begin{center}
%bb=0 0 612 792
\includegraphics[width=14cm,angle=0]{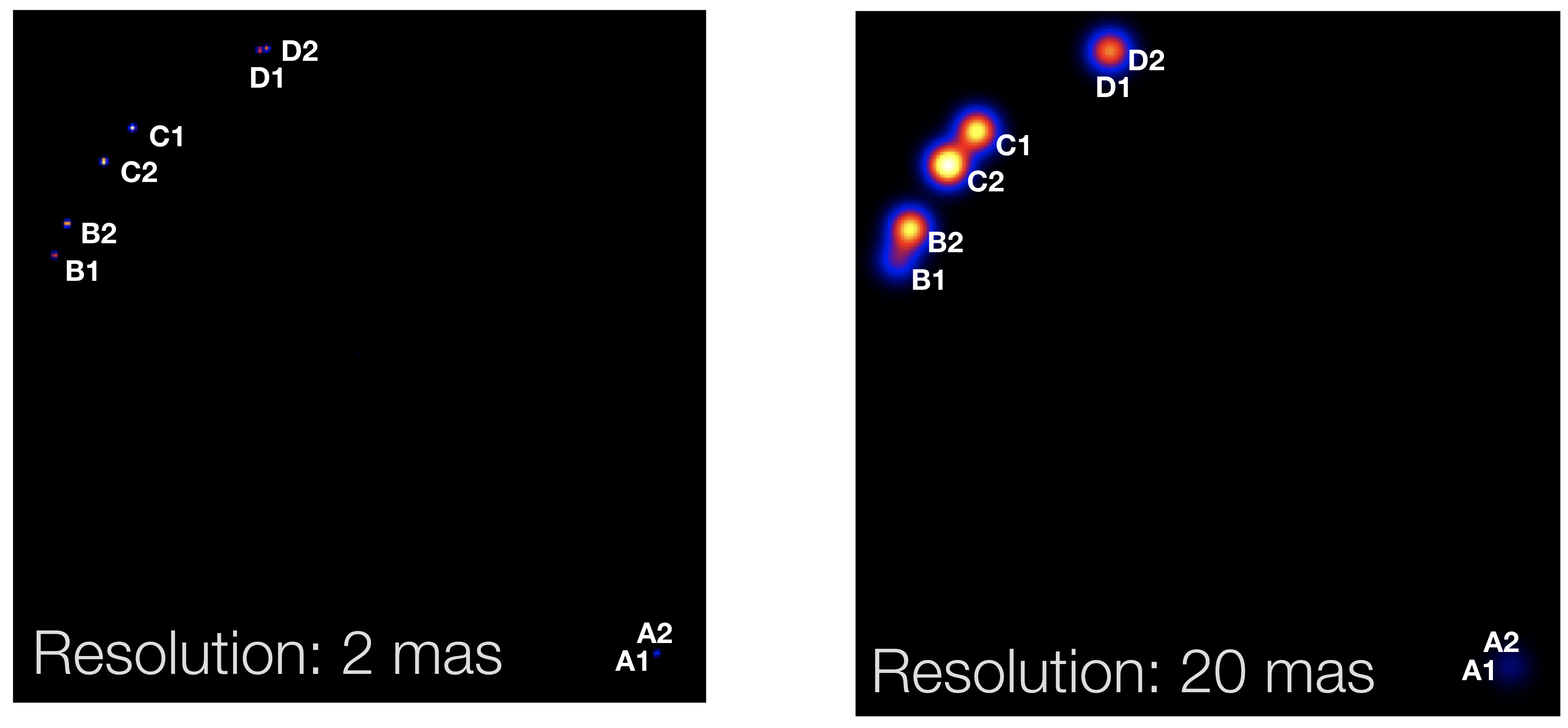}
\end{center}
\caption{ \label{fig:0402Model}          Simulation of a lensed supermassive black hole binary based on observations of the  0402+379  system. 
			The simulation is smoothed to demonstrate SKA resolution of $2\,mas$  ({\it Left})
			and $20\,mas$ ({\it Right}).	
			%http://www.jb.man.ac.uk/research/gravlens/lensarch/lens.html
			%Figure from \cite{2016ApJ...821...58B}.
				 }
\end{figure*}
%%%%%%%%%%%%%%%%%%%%%%%%%%%%%%%%%%%%%%%%%%%%%%%%%%%%
\subsubsection{Supermassive Black Hole Binaries}
%%%%%%%%%%%%%%%%%%%%%%%%%%%%%%%%%%%%%%%%%%%%%%%%%%%%

The evolution of supermassive black holes binaries is crucial to our understanding of galaxy formation.
%The evolution of galaxies and their SMBHs is inextricably connected. 
Galaxies evolve through mergers which can be directly observed \citep[e.g.,][]{1992AJ....104.1039S}.
Most galaxies in the universe harbor SMBHs at their centers \citep{1998Natur.395A..14R}.
Thus, the formation of gravitationally bound SMBH binaries is inevitable. 
Such SMBH binaries can merge and produce outbursts of gravitational wave emission.  

The separation of binaries with gravitational wave emission affecting their evolution begins at the binary separation smaller than 0.1$\,pc$. 
The exact temporal evolution of SMBHBs is uncertain from both a theoretical and observational perspective. 
In principle, less than 10$^{-3}$ of active galaxies at redshift z$<$0.7 may host SMBHBs \citep{2009ApJ...703L..86V}. 
Consequently, observational search for SMBHBs must involve a large sample of AGN. 

The study of the black hole coalescence has been challenging due to the required resolution to monitor systems separated by less than one parsec  \citep{2013ApJ...777...44J}.  
One of the most compelling observations of  SMBHB is the radio galaxy 0402+379 with two compact-core sources separated by a projected distance of  $7.3\,pc$ \cite{2017ApJ...843...14B}.
The elliptical morphology of the 0402+379 host galaxy \citep{2016ApJ...826...91A} suggests that this object is the result of merging of two massive and roughly equal mass galaxies (a major merger).
The 0402+379 system is located at redshift z=0.05. 
Thus, the system is well resolved, which allowed to constrain the relative motion of the two cores of $\beta = v/c=0.0054\pm0.0003$. 

Let us imagine an analog of the binary 0402+379 system located at redshift of 1 using the toy model described in Section~\ref{sec:causticIdea}. 
The separation between the binaries would correspond to $\sim 1\,mas$. 
Thus, SKA will be unable to resolve even systems like 0402+379, with relatively large separation. 
However, gravitationally lensed systems similar to the binary 0402+379 will be easy resolvable with SKA. 
Figure~\ref{fig:0402Model} shows a simulation of the analog of the binary 0402+379 system  gravitationally lensed in the caustic configuration. 
The images simulate SKA resolution of $2\,mas$ (Left) and $20\, mas$ (Right). 
The caustic configuration creates four lensed images per source. 
Even with a resolution of $20\,mas$ it is evident that there are more than four lensed images, 
which indicates more than one source of emission.

The future surveys including SKA and Euclid will detect $\sim10^5$ gravitationally lensed compact flat-spectrum AGN \citep{2004NewAR..48.1085K,2015aska.confE..84M}.
At least $\sim10^4$ of these sources will be in the caustic configuration following estimation discussed in Section~\ref{sec:CausticChar}.
These lensed sources will be at large redshifts, with an average z=2. 
 \cite{2009ApJ...703L..86V} predicts that the number of binaries increases rapidly with redshift. 
The number of detectable binaries increases by a factor $\sim$5-10 from z = 0.7 to z = 1.
Thus, more  than 100 of the lensed quasars may host SMBHB assuming predictions for the number of binaries at higher redshift are correct.  
As a result, there will be a considerable number of lensed quasars with SMBHB. 

However, our ability to identify these sources will be limited by the angular resolution of telescopes.
The best SKA resolution, which is planned to be achieved at 10~GHz, will resolve sources with $2\,mas$ resolution. 
At redshift of 1, a resolution of $1\,mas$ corresponds to $8\,pc$. 
Thus, SKA will be able to select binaries with separation down to $16\,pc$. 
The sub-parsec resolution is required to find candidates SMBHB emitting gravitational waves.
Such sub-parsec resolution can be achieved with SKA if SMBHBs are gravitationally lensed in the caustic configuration. 
The caustic configuration introduces angular amplification, 
which on average can boost angular resolution 20 times,
and in the best case scenario even 60 times. 
Thus, gravitational lensing will allow us to identify SMBHBs with separation down to $0.2\,pc$. 
The additional orders of magnitude in separating binaries can be achieved by taking advantage of having multiple images of the lensed source, 
which provide a frame of reference to measure relative positions of lensed images. 
The gain from relative astrometry of lensed images will depend on image brightness. 
In the caustic configuration, source flux will be magnified 30 times in average.
In the best case scenario, the magnification can reach even 100 for sources located less than dozens of parsecs from the caustic. 
The flux magnification will facilitate observation and detection of the population of the faint sources.  

The magnification ratio between lensed images is the second derivative of the lens gravitational potential, 
and as such is very sensitive to lens substructures. 
However, the position of the lensed images depends on the first derivative of the lens gravitational potential. 
As a result, the position of the lensed images can be modeled with a smooth potential and are not sensitive to the lens substructures.   
Thus, the relative positions of lensed images or distance ratio between lensed images
will allow us to resolve even very tight systems, with separation $<0.01\,pc$, and detect changes due to the orbital motion of  binary systems. 
Detail analysis of positions of lensed images will allow us to select candidate SMBHB systems, 
which can be further followed with VLBI to investigate the nature of sources and to constrain orbital motion of these systems.

%%%%%%%%%%%%%%%%%%%%%%%%%%%%%%%%%%%%%%%%%%%%%%%%%%%%
%\subsubsection{Gravitational Waves}
%%%%%%%%%%%%%%%%%%%%%%%%%%%%%%%%%%%%%%%%%%%%%%%%%%%%

%%%%%%%%%%%%%%%%%%%%%%%%%%%%%%%%%%%%%%%%%%%%%%%%%%%%
\subsubsection{Astrometry}
%%%%%%%%%%%%%%%%%%%%%%%%%%%%%%%%%%%%%%%%%%%%%%%%%%%%

Astrometry is one of the most limiting factors in revealing complex multi-wavelength structures of sources. 
The precise determination of the offsets between the radio and optical emission relies on accurate astrometry. 
For example, the HST absolute astrometry has a typical uncertainty of 0.2"-1", 
which is greater than the angular resolution. 
The accuracy of absolute astrometry may be considerably improved by matching multiple objects visible on images
to deep ground-based astrometric catalogs, like in the case of HST \citep{2016AJ....151..134W}.
Absolute astrometry, which requires comparison of the position of sources to reference frame, can be significantly improved for sources located in crowded fields with a large field of view.
However, this is not the case for quasars, which are usually observed as a compact isolated source. 
Inaccurate astrometry can introduce a systematic offset between optical and radio emission.

Despite limiting absolute astrometry, observations can provide excellent relative astrometry for bright sources.
For example, the relative positions of point-like sources on HST images can be measured with sub-mas accuracy \citep{2011PASP..123..622B}.
In the caustic configuration, four lensed images per source are produced.

The multiple lensed images provide a reference frame that can be used to find radio and optical offsets independent of the coordinate system. 
For example, the positions of lensed images can be measured in relation to the brightest image.
The angular offset can be determined with even sub-mas accuracy in relation to the caustic of the lensing galaxy 
and can be converted into physical units knowing the redshift of the source.
The position of the source can be reconstructed based on a model of the lens,
which allows mapping the positions of lensed images into the source plane. 
The source position can be defined with relation to the center of the lens or in respect to the caustic,
which eliminates the need for absolute astrometry. 
The caustic configuration will reduce possible systematics in the offset measurement arising from imprecise astrometry. 

%%%%%%%%%%%%%%%%%%%%%%%%%%%%%%%%%%%%%%%%%%%%%%%%%%%%
\section{Large Ensemble of Sources \label{sec:Esemble}}
%%%%%%%%%%%%%%%%%%%%%%%%%%%%%%%%%%%%%%%%%%%%%%%%%%%%

The upcoming transition to the petabit astronomy propelled by the sky surveys will transform the way in which astronomy is done \citep{2013pss2.book..223D}.
These surveys will enable a very wide range of science and will open a new window of discoveries. 
The synoptic all-sky surveys  including SKA, {\it Euclid}, and LSST are 
set to increase the number of gravitationally-lensed quasars from $\sim10^2$ known today 
to $\sim10^5$, in the next decade. 
This three orders of magnitude increase in number of lensed quasars will turn the next decade into the era of strong gravitational lensing. 

In Section~\ref{sec:SSSS}, I briefly review the synergy between future facilities. 
In Section~\ref{sec:ML}, I stress a need for a machine learning approach in studies of strong gravitational lensing. 
Then, in Sections~\ref{sec:CosmicEvolution} and~\ref{sec:Cosmology}, 
I present perspectives on investigating cosmic evolution of complex sources and their implications for cosmology. 

%%%%%%%%%%%%%%%%%%%%%%%%%%%%%%%%%%%%%%%%%%%%%%%%%%%%
\subsection{Synergy of Synoptic Sky Surveys \label{sec:SSSS}}
%%%%%%%%%%%%%%%%%%%%%%%%%%%%%%%%%%%%%%%%%%%%%%%%%%%%

The reviewed methods of turning galaxies into high resolution telescopes rely on gravitationally-induced time delays, positions of lensed images, and lens models. 
The  future surveys will provide a set of observations to apply the methods to a large ensemble of sources.  
SKA will provide observations with a resolution of $\sim2\,mas$ at $10\,$GHz, and $\sim20\,mas$
at $1\,$GHz \citep{2009IEEEP..97.1482D,2012PASA...29...42G}. 
The well-resolved radio positions of lensed images of quasars 
will provide precise positions of lensed images and relative astrometry which 
will set a foundation for reconstructing the mass distribution of lenses, 
and it will give a reference frame for comparison with other observations.

{\it Euclid} will map three-quarters of the extragalactic sky with  resolution  comparable to 
the HST telescope for objects below $\sim24\,$mag \citep{2011arXiv1110.3193L,2013LRR....16....6A}. 
These observations will provide constraints on relative positions of lensed images, lens and source redshifts, and mass distribution of the lens.
The Large Synoptic Survey Telescope \citep[LSST;][]{2002SPIE.4836...10T,2008arXiv0805.2366I} will open a new chapter in time domain astronomy, 
and will deliver thousands of gravitationally-induced time delays \citep{2013arXiv1310.4830D,2015ApJ...800...11L,2018ApJ...855...22G},
which will be used to investigate origins of variable emission and constrain cosmological parameters. 

In addition, the combination of offsets measured using Gaia and gravitational lensing will give us complementary insight into inner regions of galaxies at all redshifts. 
It is expected that Gaia will detect more than 500 000 quasars  \citep{2015A&A...574A..46P},
and among them  about 3000 gravitationally lensed quasars  \citep{2012MmSAI..83..944F}. 
Gaia astrometry for these gravitationally lensed quasars will provide an excellent frame for comparison of the positions of lensed images. 

Potentially interesting sources selected from  the surveys can be a target for follow-up observations with multiple facilities allowing more in-depth studies. 
The follow-up observations with the James Webb Space Telescope\footnote{https://jwst.nasa.gov} (JWST),  
the Wide Field Infrared Survey Telescope\footnote{https://wfirst.gsfc.nasa.gov} (WFIRST),
or ground facilities like the Extremely Large Telescope\footnote{http://www.eso.org/public/teles-instr/elt/} (ELT) equipped with adaptive optics 
 will allow us to detect a population of faint quasars
and will enhance our capability to search for the most distant quasars. 

The observations of lensed quasars in the caustic configuration will have the potential to provide unique insights into the origin of their X-ray emission.
The angular resolution of the {\it Chandra} satellite\footnote{http://chandra.harvard.edu} is $0.5"$. 
At redshift of 1, the Chandra X-ray telescope resolution corresponds to  $\sim4\,$kpc.  
For comparison, the M87 jet spreads throughout a projected distance of $\sim1.6\,$kpc. 
Consequently, if M87 were located at redshift  $\sim1$, 
the {\it Chandra} satellite would observe the M87 jet as a point source. 
However, if an M87-like source was gravitationally lensed in the caustic configuration, 
the offset amplification of 50 in combination with the advantage of relative astrometry, and combined with observations from the sky surveys,  
would allow us not only to resolve the jet, but also separate HST-1-like structures from the supermassive black hole. 
The future X-ray missions including Lynx\footnote{https://wwwastro.msfc.nasa.gov/lynx/} and {\tt ATHENA}\footnote{http://www.the-athena-x-ray-observatory.eu}
will not provide an improvement in angular resolution. 
Thus, gravitational lensing is the only way to resolve the origin of the X-ray emission at scales smaller than $0.5"$.

%%%%%%%%%%%%%%%%%%%%%%%%%%%%%%%%%%%%%%%%%%%%%%%%%%%%
\subsection{Machine Learning in Gravitational Lensing \label{sec:ML}}
%%%%%%%%%%%%%%%%%%%%%%%%%%%%%%%%%%%%%%%%%%%%%%%%%%%%

The future surveys will increase the number of lensed sources by three orders of magnitude. 
Such increases will open a new window for discovery, 
but will also force us to develop new approaches to data handling. 
Today, the reconstruction of the mass distribution of a lens is performed individually for every system based on maximum likelihood modeling of observations. 
Such approach is time and resource intensive, and for more complicated systems as in the caustic configuration,
 obtaining a satisfactory model of the lens by a skilled person may take even months for a single system. 
Today's approach to lens modeling is not scalable to the next decade of strong gravitational lensing. 

\citep{2017Natur.548..555H} show that fast, automated analysis of strong gravitational lenses can be obtained with a convolutional neural network.
Their analysis focused on extended images of lensed galaxies, and show that the light from the lens can be removed quickly using automatic procedures. 
The parameters of the lens can be reconstructed with an accuracy comparable to that obtained using sophisticated models, but about ten million times faster. 

Modeling of lensed quasars in the caustic configuration
using traditional approaches is even more challenging
due to the non-linear properties of the caustic.
In principle, the lensing potential near a critical curve where images form when the source is close to the caustic
can be extracted based on the generic properties of images using the model-independent approach proposed by \citep{2016A&A...590A..34W,2016arXiv161201793W}.
Moreover, these non-linear properties and the two-fold symmetry of the
caustic of an elliptical galaxy result in positions of lensed images in specific patterns. 
The presence of these patterns makes reconstruction of the caustic configuration parameters a suitable 
task for supervised machine learning algorithms.
Thus, the analysis of these sources will give a foundation for using gravitationally lensed quasars in the caustic configuration 
to understand the origin of emission.

%%%%%%%%%%%%%%%%%%%%%%%%%%%%%%%%%%%%%%%%%%%%%%%%%%%%
\subsection{Structure of Sources and Their Cosmic Evolution \label{sec:CosmicEvolution}}
%%%%%%%%%%%%%%%%%%%%%%%%%%%%%%%%%%%%%%%%%%%%%%%%%%%%

The co-evolution of super massive black hole with the dark matter halo and the star-formation history of a galaxy
is supported by many observational and theoretical studies. 
Relativistic outflows and jets
reveal  their presence during the active galactic nucleus phase
in galaxy evolution.
AGN feedback appears to be important in regulating star formation in galaxies \citep{2012MNRAS.427.2998I,2015ApJ...811...73L}. 

A consistent picture of  quasar cosmic evolution has been built over the last decades \citep{2004MNRAS.355.1010P,2007ApJ...658..721H}. 
However, the cosmic evolution of jets, 
the scales and growth of jets as a function of time, 
and the energy and the amount of power released,
%the cosmic particle acceleration history, 
have not been explored. 
The picture of  cosmic evolution of jets includes
 magnetic fields, jet compositions, 
particle acceleration mechanisms and  environmental photon fields.

Extragalactic jets interact with the environment  by
releasing high-energy radiation into the intergalactic medium.
Accelerated particles interact with photon fields, 
including 
the cosmic microwave background (CMB), 
the infrared-to-optical extragalactic background light (EBL), 
and local sources of  low-energy photon fields within a host galaxy.

The story of the cosmic evolution of jets can be revealed through 
 analysis of a large sample of sources at different redshifts. 
Current instruments, such as HST and Chandra, 
provide large samples of well-resolved jets at low redshift.
However, resolving the emission of distant sources is challenging. 
Strong gravitational lensing magnifies the emission from distant jets,
and the caustic,  time delay, and  HPT methods  applied to the large ensemble of data 
can be used to measure the offset between the emitting regions at different energies and redshifts.
The identification of the nature of offset sources at high redshift could potentially help to  
distinguish different models of SMBH growth.

%%%%%%%%%%%%%%%%%%%%%%%%%%%%%%%%%%%%%%%%%%%%%%%%%%%%
 \begin{figure}
%\vskip 1cm
\begin{center}
\includegraphics[width=10.2cm,angle=0]{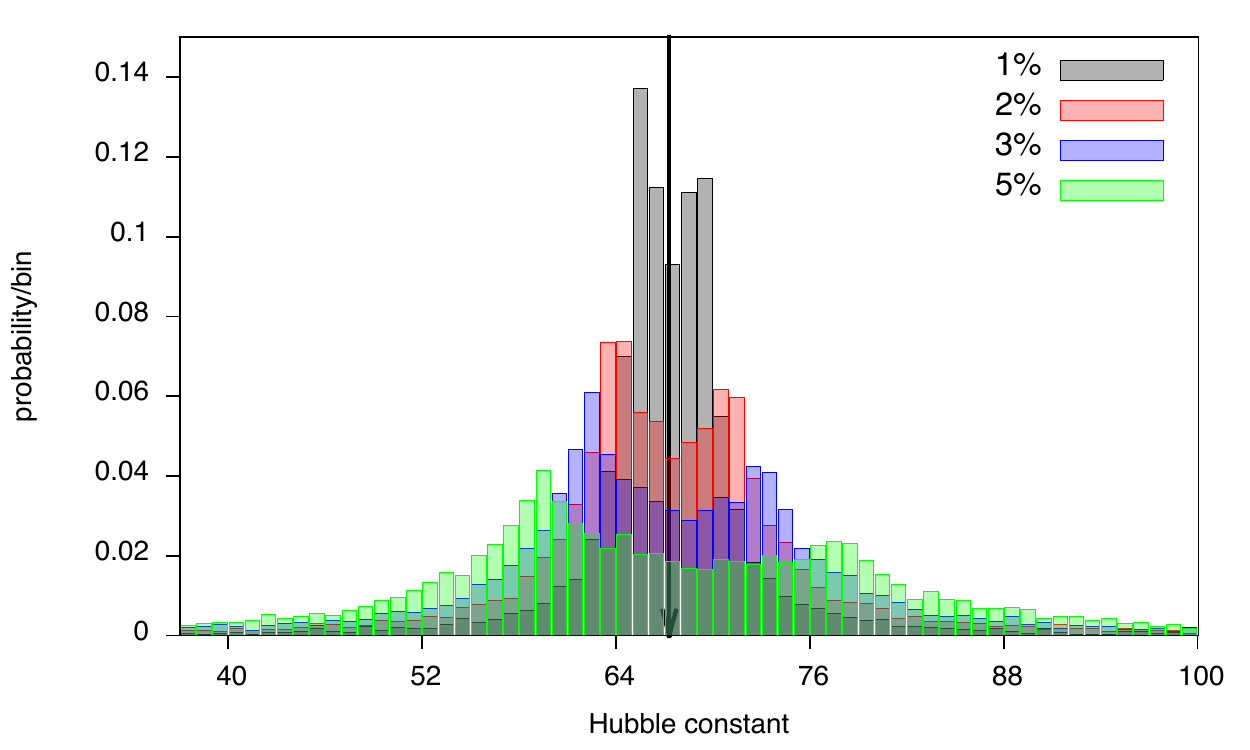}
\end{center}
 \caption{\label{fig:h_MC} Distributions of the Hubble constant obtained with the Monte Carlo simulation,
 				   with randomly selected positions of the core and the jet alignments.
				   The only free parameter in this model is the separation between the variable emission region and the core. 
                                         The distribution shows separations of 1\%, 2\%, 3\% and 5\% of the Einstein radius. 
                                         In the toy model, 1\% of the Einstein radius  corresponds to 30 pc.
                                          A black line indicates the Planck satellite measurement, H$_0=67.3$.
                                          Figure from \cite{2015ApJ...799...48B}. 
                                         }
\end{figure}
%%%%%%%%%%%%%%%%%%%%%%%%%%%%%%%%%%%%%%%%%%%%%%%%%%%%
\subsection{Implications for Cosmology \label{sec:Cosmology}}
%%%%%%%%%%%%%%%%%%%%%%%%%%%%%%%%%%%%%%%%%%%%%%%%%%%%

Time delays are the most straightforward route to obtaining  the Hubble parameter.
Gravitationally-lensed quasars have been extensively monitored, 
at all accessible wavelengths, since 1979,
when the first gravitationally-lensed quasar  was discovered.
A significant fraction of  monitored gravitationally-lensed quasars harbor powerful jets, 
including the first lensed quasar, Q0957+561, whose large-scale jet appears in radio and X-ray images.

Figure~\ref{fig:h_MC} shows the distribution of  Hubble constants obtained with 
 Monte Carlo simulations assuming different offsets between the quasar central engine and emitting region along the jet. 
 In general, the offset can be understood as a projected distance between the persistent components used to measure the positions of lensed images 
 and the variable components used to determine the time delays.
The Hubble constant  distributions for observed gravitationally lensed systems  at different energies
can be used to measure an average offset between  the central part of the quasar and the large scale jet as a function of energy.

There will be thousands of time delay measurements,
opening a new window for exploring the universe.  
Understanding  the structure of  lensed sources 
will allow us to use this large ensemble of time delays as a cosmological probe. 
The large ensemble of time delays will also allow us to determine the Hubble parameter with improved accuracy. 

As demonstrated by \cite{2015ApJ...799...48B}, 
the shape of the distribution of  Hubble constants obtained 
through strong gravitational lensing carries not only information about the geometry and expansion of the universe, 
but also contains the imprint of the complex structure of the sources,
which  may appear  as very characteristic bimodal distribution 
with a dip at the true value (see Figure~\ref{fig:h_MC}). 
Comparison of these Monte Carlo simulations 
with current measurements of H$_0$ 
shows a similar feature in the H$_0$ distributions \cite[][Figure 5]{2015ApJ...799...48B}, 
demonstrating the potential of this approach. 

Other  systematics may manifest their presence in the distribution of the Hubble constant in different ways.
They may shift or broaden the distribution.
Thus, in the near future, when  large samples of H$_0$ measurements become available, 
the shape of the observed H$_0$ distribution will be a useful tool for identifying and quantifying these systematics,
and will provide a deeper insights into geometry of the universe. 

%%%%%%%%%%%%%%%%%%%%%%%%%%%%%%%%%%%%%%%%%%%%%%%%%%%%
\section{Outlook and Perspectives} 
%%%%%%%%%%%%%%%%%%%%%%%%%%%%%%%%%%%%%%%%%%%%%%%%%%%%

Gravitational lenses magnify fluxes of distant objects.
This magnification allows us to detect sources which are too faint or too distant to be observed without  lensing effects. 
In addition, gravitational lenses provide insight into source properties and complex multi-wavelength structures 
far below the resolution or sensitivity limit of current and future facilities. 
Thus, gravitational lenses act as high-resolution telescopes.
However, as we don't have the ability to point these ``telescopes"  and the chance of a galaxy-galaxy alignment is low, 
so far strong gravitational lensing has been used to study the properties of individual sources,
and the efforts have been focused on increasing the number of lensed objects. 

Strong gravitational lensing effect requires close alignment of a distant source with a foreground galaxy.
According to estimations by \cite{2010MNRAS.405.2579O}, 
there are only $\sim0.2$ lenses per square degree for an i-band limiting magnitude of 21.
Consequently, discovery of gravitationally lensed systems requires observations of a large number of sources with angular resolution sufficient to identify multiple lensed images. 
Significant number of sources have been recently discovered with wide-field surveys, 
such as the Sloan Digital Sky Survey \citep[SDSS][]{2000AJ....120.1579Y} and the Dark Energy Survey \citep[DES][]{2017ApJS..232...15D}. 
For example, the SDSS-III BOSS quasar lens survey resulted in discovery of thirteen gravitationally lensed quasars \cite{2016MNRAS.456.1595M}. 
Recently, 24 lensed quasars were discovered using Gaia data\citep{2018MNRAS.tmp..893L}.

The search for lensed quasars is mostly limited to radio and optical surveys. 
Around 100 gravitationally lensed sources discovered at radio have been followed with HST observations \citep{1999AIPC..470..163K}.
Gravitationally lensed quasars can be  followed up with observations at higher energies. 
The {\it Chandra} X-ray Observatory has observed $\sim20$ lensed quasars. 
At X-ray,  gravitationally lensed quasars have been extensively monitored in search 
of microlensing effects that provides a way to constrain the size of the X-ray emitting part of the accretion disc \citep{2012ApJ...755...24C}. 

At gamma rays, there are two known gravitationally lensed sources; B2~0218+35 and PKS~1830-211.
The {\it Fermi} satellite has provided a unique set of observations by monitoring the entire sky, thereby providing uniform gamma-ray lightcurves since 2008. 
For the very first time, almost uninterrupted multi-year monitoring of lensed sources has become possible,
imparting a new precision for measuring gravitationally induced time delays. 
The study cases presented in this review rely on the {\it Fermi}/LAT lightcurves of gravitationally lensed blazars 
to demonstrate the potential of strong gravitational lensing in order to test assumptions on the origins of multi-wavelength emissions.  

Observations of the high energy universe are limited by the angular resolutions of detectors and the number of sources.
Thus, future discoveries of gravitationally lensed quasars will rely on radio and optical surveys 
such as SKA, LSST, and Euclid, which are expected to increase the number of gravitationally lensed quasars in the next decade  from $\sim10^2$ known today 
to $\sim10^5$. 
The three orders of magnitude increase in the number of lensed quasars will open an era of strong gravitational lensing and will enable applying strong gravitational lensing as a tool to resolve emissions for a large ensemble of sources. 

%%%%%%%%%%%%%%%%%%%%%%%%%%%%%%%%%%%%%%%%%%%%%%%%%%%%
\section{Conclusions} 
%%%%%%%%%%%%%%%%%%%%%%%%%%%%%%%%%%%%%%%%%%%%%%%%%%%%

Strong gravitational lensing allows us to improve the angular resolution of current and future telescopes and eliminates the need for absolute astrometry. 
Gravitational lenses induce time delays and produce multiple images of sources. 
Both time delays and positions of lensed images can be used to infer spatial origins of emissions on scales not accessible with the current and future facilities. 
Therefore, gravitational lenses act as ``high-resolution telescopes." 
 
The case studies of PKS~1830-211 and B2~0218+35 demonstrated applications of  strong gravitational lensing as high-resolution telescopes. 
In the case of PKS~1830-211, the time delay approach improved angular resolution at gamma rays 10,000 times, which resulted in evidence 
that variable emissions can be produced in 
 multiple emitting regions along the relativistic jet.
Observation of  variable emission at large distances from the SMBH  challenges our understanding of particle acceleration mechanisms, 
and introduces a possible source of systematics for using gravitationally-induced time delays to measure the Hubble parameter. 

Today, the Hubble parameter is constrained with many independent methods. 
Thus, the problem can be inverted and the Hubble parameter can be used to find the spatial origin of an emission. 
The Hubble parameter tuning approach applied to the case of B2~0218+35 improved angular resolution at gamma rays 1,000,000 times,
and illuminated $\sim60\,pc$ offset between the radio core and gamma-ray flare. 
Such a large offset questions our understanding of connections between multi-wavelength emission and SMBHs. 
 
The caustics of lensing galaxies acting as non-linear amplifiers will further allow us to resolve the origin of emission and investigate possible offsets in multi-wavelength emission of the inner regions of active galaxies.
The caustics of lensing galaxies provide flux magnification and amplify offsets between sources in the positions of their lensed images. 
These positions of lensed images will allow us
to resolve complex sources, including relativistic jets, and uncover the population of supermassive black hole binaries at high redshifts.

Future surveys,  including SKA, LSST, and Euclid will monitor large fractions of the sky 
and are set to increase the number of gravitationally lensed quasars from the $\sim10^2$ known today 
to $\sim10^5$, in the next decade. 
This three orders of magnitude increase in the number of lensed quasars will turn the next decade into the era of strong gravitational lensing. 
These synoptic all-sky surveys will provide sets of resolved positions of lensed images, time delays and distances;
all essential ingredients to apply strong gravitational lensing to resolve the physical nature of the inner regions of active galaxies.  

%%%%%%%%%%%%%%%%%%%%%%%%%%%%%%%%%%%%%%
\section{Acknowledgments}
%%%%%%%%%%%%%%%%%%%%%%%%%%%%%%%%%%%%%%

I would like to thank Marc Kamionkowski for the invitation to write this review. 
I thank the referee for providing very valuable comments which greatly helped to improve the manuscript. 
I would like to also thank Dan Schwartz, Martin Elvis, Jan Kansky, Chiara Mingarelli, Robert Kirshner, Michal Ostrowski, Bronek Rudak for comments and useful discussions.

The author is  supported by NASA through Einstein Postdoctoral Fellowship.
This research was supported in part by PLGrid Infrastructure.
This research has made use of data from the OVRO 40-m monitoring program \citep{2011ApJS..194...29R}
which is supported in part by NASA grants NNX08AW31G, NNX11A043G, and NNX14AQ89G and NSF grants AST-0808050 and AST-1109911.

\newpage
{\bf Bibliography}
\bibliography{ReviewGravitationalTelescopes}

\end{document}